\documentclass[%
  reprint,
superscriptaddress,
showpacs,preprintnumbers,
nofootinbib,
 amsmath,amssymb, 
 aps,
 prd,longbibliography,
]{revtex4-1}

\usepackage[dvipsnames]{xcolor}

\usepackage[normalem]{ulem}
\usepackage{tikz-feynman}

\usepackage{cancel}
\usepackage{accents}
\usepackage{mciteplus,slashed}
\usepackage{amssymb,cancel,amsmath}
\usepackage{dcolumn}% Align table columns on decimal point
\usepackage{bm}% bold math
\usepackage[caption=false]{subfig}
\usepackage{appendix}
\usepackage{feynmp-auto}
\usepackage[T1]{fontenc}	
\usepackage{csvsimple}
\usepackage[colorlinks=true,citecolor=blue,linkcolor=blue, allcolors=blue]{hyperref}
\usepackage[capitalise]{cleveref}
\usepackage{booktabs}
\usepackage{graphicx}
\usepackage{mathrsfs}
\usepackage[utf8]{inputenc}
\usepackage{siunitx}
\makeatletter
\def\l@subsubsection#1#2{}
\makeatother

\DeclareSIUnit\electronvolt{e\kern-.05em V}
\DeclareSIUnit\parsec{pc}
\DeclareSIUnit \h {\ensuremath{\mathit{h}}}

\begin{document}

% PREPRINT NUMBERS
\preprint{\hfill FERMILAB-PUB-20-560-AE-PPD-T}
\preprint{\hfill MIT-CTP/5252}

\title{New Pathways to the Relic Abundance of Vector-Portal Dark Matter}

\author{Patrick J. Fitzpatrick}
\email{fitzppat@mit.edu}
\affiliation{Center for Theoretical Physics, Massachusetts Institute of Technology, Cambridge, Massachusetts 02139, U.S.A.}

\author{Hongwan Liu}
\email{hongwanl@princeton.edu}
\affiliation{Center for Cosmology and Particle Physics, Department of Physics, New York University, New York, NY 10003, U.S.A.}
\affiliation{Department of Physics, Princeton University, Princeton, New Jersey, 08544, U.S.A.}

\author{Tracy R. Slatyer}
\email{tslatyer@mit.edu}
\affiliation{Center for Theoretical Physics, Massachusetts Institute of Technology, Cambridge, Massachusetts 02139, U.S.A.}

\author{Yu-Dai Tsai}
\email{ytsai@fnal.gov}
\affiliation{Fermilab, Fermi National Accelerator Laboratory, Batavia, IL 60510, U.S.A.}

\date{\today}

\begin{abstract}

We fully explore the thermal freezeout histories of a vector-portal dark matter model, in the region of parameter space in which the ratio of masses of the dark photon $A^{\prime}$ and dark matter $\chi$ is in the range $1 \lesssim m_{A^{\prime}}/m_{\chi} \lesssim 2$. In this region $2 \rightarrow 2$ and $3 \rightarrow 2$ annihilation processes within the dark sector, as well as processes that transfer energy between the dark sector and the Standard Model, play important roles in controlling the thermal freezeout of the dark matter. We carefully track the temperatures of all species, relaxing the assumption of previous studies that the dark and Standard Model sectors remain in thermal equilibrium throughout dark matter freezeout. Our calculations reveal a rich set of novel pathways which lead to the observed relic density of dark matter, and we develop a simple analytic understanding of these different regimes. The viable parameter space in our model provides a target for future experiments searching for light (MeV-GeV) dark matter, and includes regions where the dark matter self-interaction cross section is large enough to affect the small-scale structure of galaxies.

\end{abstract}
 
\maketitle 

\tableofcontents

\newpage

\section{Introduction}

The particle nature of dark matter (DM) remains a mystery, whose solution requires us to search beyond the Standard Model (SM). There are a great many suggestions for new physics particles that might solve the DM puzzle. One well-studied class of DM candidates is the weakly-interacting massive particles (WIMPs). WIMPs are theoretically attractive because they naturally arise in various Beyond-Standard Model (BSM) theories of new weak-scale physics, and because the thermal production of WIMPs, through their $2 \rightarrow 2$ annihilations to SM particles, naturally leads to the correct DM relic abundance. However, with increasingly strong experimental constraints being placed on the WIMP scenario, we are also motivated to consider alternative scenarios where other interactions control the final DM abundance.

There has been considerable recent interest in exploring thermal relic scenarios that naturally produce DM at light (sub-GeV) masses (see for example Ref.~\cite{Battaglieri:2017aum}), as existing direct detection constraints are much less sensitive to sub-GeV mass DM (e.g.\ Refs.~\cite{Akerib:2016vxi,Aprile:2018dbl,Agnese:2017njq,Wang:2020coa}). Existing beam dump experiments are sensitive to sub-GeV DM but leave much of the parameter space unconstrained~\cite{Bergsma:1985is,Bergsma:1985qz,Konaka:1986cb,Bjorken:1988as,Davier:1989wz,Blumlein:1990ay,Blumlein:1991xh,Banerjee:2018vgk,Batley:2015lha,Tsai:2019mtm}. New accelerator and direct detection experiments will soon explore the parameter space of light DM with unprecedented sensitivity (see Ref.~\cite{Battaglieri:2017aum} and references therein); consequently, it is important to understand the landscape of models which naturally populate this sub-GeV region.

Previous studies have identified a mechanism for thermally producing sub-GeV DM in which strong $3 \rightarrow 2$ self-annihilations among DM particles control the thermal relic abundance. This strongly-interacting-massive-particle (SIMP) scenario naturally leads to strongly-coupled DM ($\alpha_D \sim 1$) with mass similar to the QCD scale ($m_\chi \sim 10$--\SI{100}{\mega\eV}) \cite{Hochberg_2014}. The natural emergence of the strong scale in the thermal SIMP scenario makes it a particularly attractive framework. In this scenario, the DM and SM sectors remain in thermal equilibrium throughout freezeout via elastic scattering between DM and SM particles. 

An alternative thermal production mechanism for light DM arises when this condition is relaxed; in the elastically decoupling relic (ELDER) scenario the DM and SM sectors thermally decouple through the elastic DM-SM scattering while strong $3 \rightarrow 2$ self-annihilations are still active \cite{Kuflik:2015isi, Kuflik:2017iqs}. In the ELDER scenario, although thermal freezeout proceeds through the $3 \rightarrow 2$ DM self-annihilations, the DM relic abundance is nevertheless determined by the decoupling of DM-SM elastic scattering. This is achieved through a dark sector process called ``cannibalization'' \cite{1992ApJ...398...43C}, which occurs immediately after elastic decoupling and proceeds until $3 \rightarrow 2$ freezeout. During cannibalization, while the DM and SM sectors are thermally secluded, $3 \rightarrow 2$ DM self-annihilations convert mass to kinetic energy and heat the dark sector. As a result, the dark sector temperature evolves slowly (logarithmically as a function of SM temperature) during cannibalization, and likewise, the DM abundance evolves slowly. This leads to a DM relic abundance that is primarily determined by its value at kinetic decoupling. The ELDER scenario also naturally leads to MeV-GeV mass DM.

Distinctive thermal production mechanisms for light DM have also been realized in the well-studied vector-portal DM model of a Dirac fermion DM particle $\chi$ charged under a hidden U(1) gauge symmetry with dark gauge boson $A^{\prime}$, which is coupled to the SM photon through kinetic mixing. In the region of parameter space in which the dark photon is more massive than the DM ($r \equiv m_{A^{\prime}}/m_{\chi} > 1$), the kinematically suppressed $2 \rightarrow 2$ annihilations of DM to heavier $A^{\prime}$s ($\chi \bar{\chi} \rightarrow A^{\prime} A^{\prime}$) can control the relic abundance. In this ``forbidden DM'' (FDM) mechanism \cite{PhysRevD.43.3191, D_Agnolo_2015} the exponential suppression of the $2 \rightarrow 2$ process setting the relic abundance of DM naturally gives rise to DM exponentially lighter than the weak scale. The FDM mechanism was shown to be a viable mechanism for producing sub-GeV DM.

More recently, Ref.~\cite{Cline_2017} showed that in the region of parameter space of the dark photon model in which $1.5 \lesssim r \lesssim 2$, the kinematic suppression of the $\chi \bar{\chi} \rightarrow A^{\prime} A^{\prime}$ annihilation process is compensated for by a kinematically allowed $3 \rightarrow 2$ ($\chi \chi \bar{\chi} \rightarrow \chi A^{\prime}$) annihilation channel, which can then play a dominant role in setting the thermal relic abundance of DM. This ``not-forbidden dark matter'' (NFDM) scenario is analogous to the thermal SIMP scenario in that $3 \rightarrow 2$ processes can determine the DM relic abundance, realized in the simple and well-studied vector-portal DM model. The NFDM scenario was also demonstrated to be a viable mechanism for naturally producing sub-GeV DM. In both the FDM and NFDM scenarios, the DM and SM sectors were assumed to remain thermally coupled throughout the freezeout of DM. 

\begin{figure*}
\centering
    \includegraphics[width=0.25\textwidth]{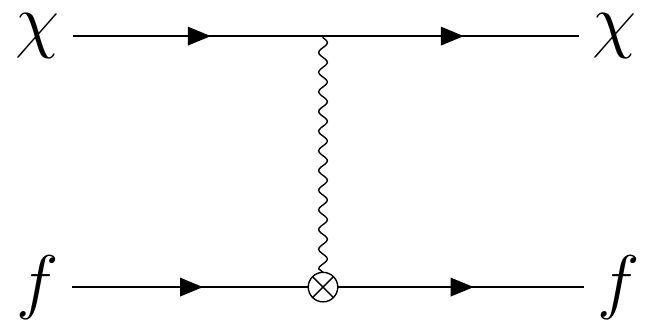} \quad \quad \quad
    \includegraphics[width=0.21\textwidth]{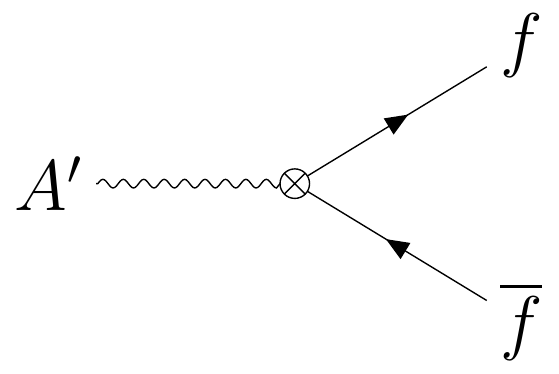} \quad \quad \quad
    \includegraphics[width=0.25\textwidth]{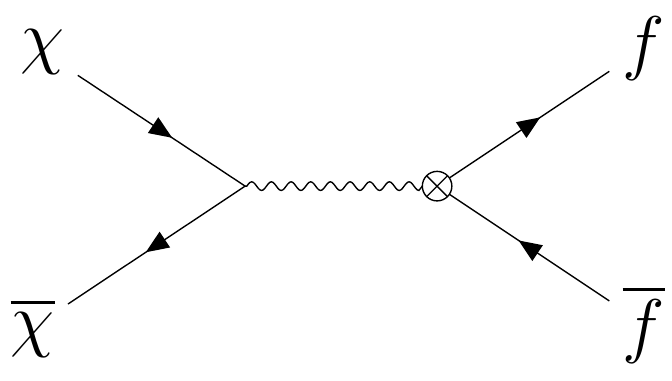}
    \caption{Tree-level interactions between dark sector and SM particles, including \textit{(left)} $\chi f \to \chi f$ elastic scattering, \textit{(center)} $A' \to f \overline{f}$ decay, and \textit{(right)} $\chi \overline{\chi} \to f \overline{f}$ annihilation into SM fermions.} 
    \label{fig:kinetic}
\end{figure*}

In this paper, we extend both of these frameworks to consider cases in which the DM and SM sectors are allowed to kinetically decouple during thermal freezeout of the DM. We fully explore the $1 \lesssim r \lesssim 2$ region of parameter space of the dark photon model, in which the kinematically suppressed $2 \rightarrow 2$ ($\chi \bar{\chi} \rightarrow A^{\prime} A^{\prime}$) channel and the kinematically allowed $3 \rightarrow 2$ ($\chi \chi \bar{\chi} \rightarrow \chi A^{\prime}$) channel play important roles in controlling thermal freezeout, and relax the condition that kinetic equilibrium is maintained between the two sectors throughout the freezeout process. We find a rich set of novel cosmological histories leading to a range of different mechanisms for obtaining the correct DM relic density. Among these, we identify a general class of mechanisms in which the DM relic abundance is determined by processes controlling the kinetic decoupling of the DM and SM sectors (which we call the KINetically DEcoupling Relic -- KINDER). This KINDER scenario in the dark photon model generalizes the ELDER scenario to cases in which multiple processes control the thermal coupling between dark and SM sectors, and in which a $3 \rightarrow 2$ annihilation process among multiple dark sector species supports heating of the dark sector.

The outline of our paper is as follows. In Section~\ref{sec:model}, we describe the dark photon model in the $1 \lesssim r \lesssim 2$ region we consider, including the primary interactions controlling chemical equilibrium in the dark sector, and those between the dark sector and SM particles. In Section~\ref{sec:dark_sector_freezeout} we discuss general features of dark sector freezeout in our model, including the relevant interaction processes, the Boltzmann equations, which describe the thermodynamic evolution of the system, and the freezeout conditions of relevant processes. In this section, we also classify three thermodynamic phases (A, B, and C), which generally describe the various stages of the thermal histories realized in our model. 

In Sections~\ref{sec:NFDM_regime} and~\ref{sec:FDM_region}, we characterize the thermal freezeout histories for $1.5 \lesssim r \lesssim 2$ and $1 \lesssim r \lesssim 1.5$ respectively. In each case, we identify a rich set of freezeout histories and analytically determine the parameter space regions where they occur. These different histories are naturally classified into specific regions in the $\epsilon$--$\alpha_D$ plane, where $\epsilon$ describes the mixing between the dark photon and the SM photon, and $\alpha_D$ is the dark sector coupling. In Section~\ref{sec:NFDM_regime} we study the $1.5 \lesssim r \lesssim 2$ region of our model, where the $2 \leftrightarrow 2$ process freezes out before the $3 \leftrightarrow 2$ process; the possible histories can be classified into the WIMP, NFDM and KINDER regimes. In Section~\ref{sec:FDM_region}, we examine the $1 \lesssim r \lesssim 1.5$ region of our model, where the $3 \rightarrow 2$ process freezes out prior to the $2 \rightarrow 2$ process, and find four distinct regimes in addition to the WIMP regime (Regimes I -- IV). In Section~\ref{sec:constraints} we discuss the relevant experimental and cosmological constraints; finally, in Section~\ref{sec:conclusion} we summarize our conclusions.

Throughout this paper, we make use of Planck 2018 cosmological parameters~\cite{Aghanim:2018eyx}, using the TT,TE,EE+lowE+lensing results; we take the DM abundance to be the central value of $\Omega_\chi h^2 = 0.12$, with $h = 0.6736$. All quantities are expressed in natural units, with $\hbar = c = k_B = 1$. Finally, we use many different symbols for approximations in this paper, and have attempted to keep them consistent with the following definitions: \textit{(i)} we use ``$\simeq$'' when the approximation is a physical limit, e.g., \ a nonrelativistic limit; \textit{(ii)} we use ``$\approx$'' for statements that are true within an order of magnitude, but which we will take to be an equality for the purpose of analytic results; \textit{(iii)} finally, we use ``$\sim$'' for statements that are true within an order of magnitude, but we do not use the fact either analytically or numerically.

\section{Model}

\label{sec:model}

\begin{figure*}[!t]
\centering
    \includegraphics[width=0.2\textwidth]{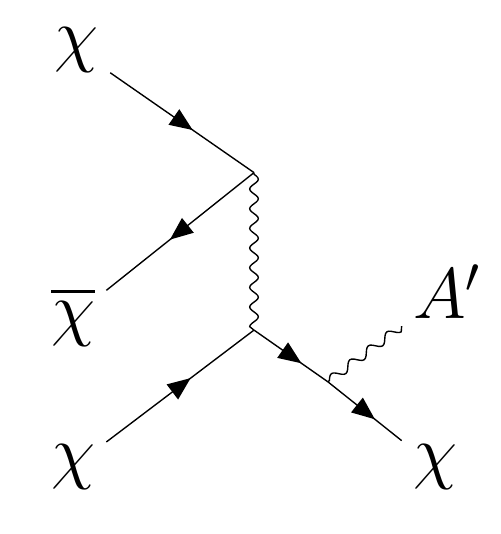} \qquad \qquad \qquad
    \includegraphics[width=0.25\textwidth]{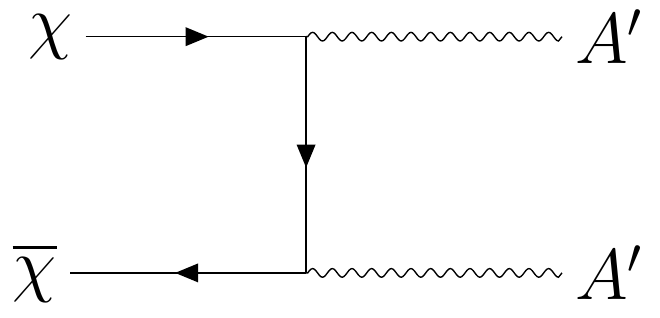}
    \caption{The dark-sector-only \textit{(left)} $\chi \overline{\chi} \chi \leftrightarrow A' \chi$ ($ 3 \leftrightarrow 2$) process and \textit{(right)} the $\chi \overline{\chi} \leftrightarrow A'A'$ ($2 \leftrightarrow 2$) process.} 
    \label{fig:number_change}
\end{figure*}

In the mass basis, the Lagrangian of the dark photon model we consider is:

\begin{alignat}{1}
	\mathcal{L} \supset - \frac{1}{4} F_{\mu\nu} F^{\mu\nu} - \frac{1}{4} F_{\mu\nu}' F'^{\mu\nu} + \frac{1}{2} m_{A'}^2 A_\mu' A'^\mu \nonumber \\ 
	\overline{\chi}(i \slashed{D} - m_\chi) \chi + \epsilon A_\mu' J^\mu_{\text{EM}}
	\label{eq:Lagrangian}
\end{alignat}
where the gauge coupling is $\alpha_D = g_D^{2}/4 \pi$, and $\slashed{D} \equiv \slashed{\partial} - i g_D \slashed{A}^{\prime}$. The dark photon $A^{\prime}$ kinetically mixes with the SM photon, giving rise to a small coupling between the dark photon and the SM electromagnetic current $J_\text{EM}^\mu$, set by the kinetic mixing parameter $\epsilon$. The value of $\epsilon$ can naturally range from as small as $10^{-13}$ up to $10^{-1}$~\cite{Gherghetta:2019coi}. The hidden U(1) symmetry can be spontaneously broken through a Higgs-like mechanism with the dark Higgs taken to be heavy enough to be excluded from this low-energy effective description, since we will always be considering energies $\lesssim m_\chi, m_{A'}$. The kinetic mixing generates the tree-level interactions between dark and SM particles shown in Fig.~\ref{fig:kinetic}: $\chi f \rightarrow \chi f$ , $\chi \bar{\chi} \rightarrow f \bar{f}$, $A^{\prime} \rightarrow f \bar{f}$.

We are primarily interested in scenarios in which the dominant DM-number-changing interactions are the $\chi \chi \bar{\chi} \leftrightarrow \chi A^{\prime}$ ($3 \leftrightarrow 2$) process and the kinematically suppressed $\chi \bar{\chi} \leftrightarrow A^{\prime} A^{\prime}$ ($2 \leftrightarrow 2$) process, shown in Fig.~\ref{fig:number_change}. This restricts us to the region of parameter space in which $1 \lesssim r \lesssim 2$.

At lower values of $r$, where $r < 1$, the dominant process controlling thermal freezeout is $\chi \bar{\chi} \rightarrow A^{\prime} A^{\prime}$ (which is then kinematically allowed), and the $A^{\prime}$ decays promptly to SM particles. This regime is strongly ruled out by cosmic microwave background (CMB) constraints on the annihilation cross section of DM into SM particles for $m_{\chi} \lesssim \SI{10}{\giga\eV}$~\cite{Cirelli_2017}.

At higher values of $r$, where $r > 2$, the $s$-channel annihilation of $\chi \bar{\chi} \rightarrow f \bar{f}$ via an off-shell $A^{\prime}$ dominates the DM-number-changing interactions: the $\chi \overline{\chi} \to A'A'$ process is very kinematically suppressed and the $\chi \chi \bar{\chi} \rightarrow A^{\prime} \chi$ process reduces to a scattering process among the $\chi$s as the final-state $A^{\prime}$ promptly decays back to $\chi \bar{\chi}$. Dark sector freezeout proceeds via the classic WIMP freezeout scenario, which also runs into stringent CMB constraints on the $s$-wave annihilation of Dirac fermion dark matter below 10 GeV~\cite{Aghanim:2018eyx}.

In the intermediate ($1 \lesssim r \lesssim 2$) region of interest to us, which of the $2 \rightarrow 2$ or $3 \rightarrow 2$ processes dominates during thermal freeze-out depends on the ratio $r$. The $ 2 \rightarrow 2$ process receives a kinematic suppression from $\chi$ particles annihilating into heavier $A^{\prime}$ particles (with an exponential factor of the form $e^{-2(r-1)m_\chi/T}$), while the $3 \to 2$ receives a Boltzmann suppression from an extra factor of $\chi$ number density in the initial state (with an exponential factor of the form $e^{-m_\chi/T}$). In the lower half of the range in $r$ we consider ($1 < r \lesssim 1.5$), the Boltzmann suppression is more severe for the $3 \to 2$ process, and therefore the $2 \rightarrow 2$ process dominates during thermal freezeout. In this regime, and for the case in which the DM and SM sectors remain thermally coupled throughout thermal freezeout, the $2 \rightarrow 2$ process determines the relic abundance; this is the FDM scenario described in the introduction.

In the upper half of the range in $r$ we consider ($1.5 \lesssim r < 2$), in contrast, the large kinematic suppression of the $2 \rightarrow 2$ process renders it subdominant to the $3 \rightarrow 2$ process at freezeout. In this regime, and for the case in which the DM and SM sectors remain thermally coupled throughout thermal freezeout, the $3 \rightarrow 2$ process determines the thermal relic abundance; this is the NFDM scenario described in the introduction.

\section{Dark Sector Freezeout}
\label{sec:dark_sector_freezeout}

Before we detail all of the different regimes in which the dark sector can evolve to obtain the final DM relic density, we will begin by discussing some general features of the dark sector freezeout in this model. By ``dark sector freezeout'', we mean the cosmological evolution from the initial state, when both sectors are in thermal equilibrium, to the point where the DM has attained its final comoving relic abundance.

\subsection{Thermodynamic Variables}
\label{subsec:thermodynamic_variables}

Throughout freezeout, for the parameter space we consider, $\chi$ and $A'$ remain in thermal equilibrium with each other through $\chi$--$A^{\prime}$ scattering. The dark sector can therefore be described by a single dark sector temperature $T^{\prime}$. The general expressions for the number densities of the particles in the nonrelativistic limit are:
\begin{alignat}{1}
    n_\chi(T') &\equiv \frac{2 g_\chi m_\chi^2 T'}{2 \pi^2} K_2 \left( \frac{m_\chi}{T'} \right) \nonumber \\
    &\simeq 2g_\chi \left(\frac{m_\chi T'}{2\pi}\right)^{3/2} e^{-m_\chi/T'} e^{\mu_\chi(T')/T'} \label{eq:n_chi} \,, \\
    n_{A'}(T') &\equiv \frac{g_{A'} m_{A'}^2 T'}{2 \pi^2} K_2 \left( \frac{m_{A'}}{T'} \right) \nonumber \\ &\simeq g_{A'} \left(\frac{r m_\chi T'}{2 \pi} \right)^{3/2} e^{-r m_\chi/T'} e^{\mu_{A'}(T')/T'} \label{eq:n_Ap} \,,
\end{alignat}
where $g_\chi = 2$ and $g_{A'} = 3$ are the numbers of degrees of freedom associated with each particle. The factor of two in Eq.~\eqref{eq:n_chi} accounts for the fact that we are including both $\chi$ and $\overline{\chi}$ in the definition of $n_\chi$. We have also included effective chemical potentials $\mu_\chi$ and $\mu_{A'}$ which are in general nonzero; we denote the number densities of $\chi$ and $A'$ with zero chemical potential as $n_{\chi,0}(T')$ and $n_{A',0}(T')$ respectively. We will also frequently use the inverse dimensionless temperatures, $x \equiv m_\chi/T$ and $x' \equiv m_\chi/T'$. 

The energy densities and pressures of $\chi$ and $A^{\prime}$ are related to their number densities by:
\begin{alignat}{1}
    \rho_{\chi} \left(T^{\prime} \right) = \frac{n_{\chi} \left( T^{\prime} \right)}{n_{\chi,0} \left( T^{\prime} \right) } \rho_{\chi, 0} \left( T^{\prime} \right) \, 
    \label{eq:rho_chi}
\end{alignat}
and 
\begin{alignat}{1}
    P_{\chi} \left( T^{\prime} \right) = \frac{n_{\chi} \left( T^{\prime} \right)}{n_{\chi,0} \left( T^{\prime} \right) } P_{\chi, 0} \left( T^{\prime} \right) \,  ,
    \label{eq:P_chi}
\end{alignat}
where the Maxwell-Boltzmann distributions with zero chemical potential for $\rho_{\chi, 0} \left( T^{\prime} \right)$ and $P_{\chi, 0} \left( T^{\prime} \right)$ are given by
\begin{alignat}{1}
    \rho_{\chi,0}(T') &\equiv \frac{2 g_\chi m_\chi^2 T'}{2 \pi^2} \left(m_\chi K_1\left( \frac{m_\chi}{T'} \right) + 3 T' K_2 \left( \frac{m_\chi}{T'} \right) \right) \,, \nonumber \\
    P_{\chi,0}(T') &\equiv \frac{2 g_\chi m_\chi^2 T'^2}{2 \pi^2} K_2 \left(\frac{m_\chi}{T'} \right) = n_{\chi,0}(T')T'\,.
\end{alignat}
Similar relations hold for $A^{\prime}$. The entropy of the dark sector is conserved when no heat is transferred between the dark sector and the SM through processes that involve both dark sector and SM particles. The entropy density of the dark sector is
\begin{alignat}{1}
    s_D \equiv \frac{(\rho_\chi + \rho_{A'}) + (P_\chi + P_{A'}) - \mu_\chi n_\chi - \mu_{A'} n_{A'}}{T'} \,.
    \label{eq:dark_sector_entropy_density}
\end{alignat}
Entropy conservation of the dark sector in the limit where heat transfer processes are inefficient is a useful fact that we will use extensively in obtaining an analytic understanding of our results. When the dark sector entropy is conserved, $d(s_Da^3)/dt = 0$, where $a$ is the expansion scale factor.

Since we will be discussing the time evolution of $n_\chi$ and $n_{A'}$ frequently in the context of analytic estimates, we will derive here several expressions related to $\dot{n}_\chi$ and $\dot{n}_{A'}$ that will be useful throughout the paper. First, taking the time derivative of $n_\chi$ gives
\begin{alignat}{1}
    \dot{n}_\chi &\simeq -\left[\frac{3}{2T'} + \frac{m_\chi}{T'^2} + \frac{d}{dT'} \left(\frac{\mu_\chi}{T'}\right)\right] \frac{dT'}{dT} HT n_\chi \,,
    \label{eq:n_chi_dot}
\end{alignat}
where we have used $dT/dt \simeq -HT$. We will often make the approximation that $m_\chi/T' \gg 1$ during freezeout, and so the term $3/(2T')$ can often be neglected, unless $\mu_\chi \sim m_\chi$. Similarly, 
\begin{alignat}{1}
    \dot{n}_{A'} \simeq - \left[\frac{3}{2T'} + \frac{r m_\chi}{T'^2} + \frac{d}{dT'} \left(\frac{\mu_{A'}}{T'}\right)\right] \frac{dT'}{dT} HT n_{A'} \,.
    \label{eq:n_Ap_dot}
\end{alignat}

We will often be interested in comparing the final number density of the dark matter after the dark sector completely freezes out to the number density required to achieve the relic abundance of dark matter today. Defining $Y_\chi \equiv n_\chi / s_\text{SM}$, where $s_\text{SM}$ is the entropy density of the SM sector after the dark sector has completely decoupled, the correct relic abundance is obtained when~\cite{Edsjo:1997bg}
\begin{alignat}{1}
    Y_\chi \equiv \frac{n_\chi}{s_\text{SM}} = 4.32 \times 10^{-10} \left(\frac{\SI{}{\giga\eV}}{m_\chi}\right) \,.
    \label{eq:relic_abundance_requirement}
\end{alignat}

\subsection{Relevant Processes}
\label{subsec:relevant_processes}

In the conventional WIMP regime, DM freezes out through the process $\chi \overline{\chi} \to f \overline{f}$, where $f$ is a SM fermion. Once the mixing parameter $\epsilon \lesssim 10^{-5}$--$10^{-4}$, however, $\chi \overline{\chi} \to f \overline{f}$ freezes out while other dark sector processes are still active, and these processes play a significant role in the freezeout of the dark sector~\cite{PhysRevLett.115.061301,Cline:2017tka}. 

Outside the WIMP regime, there are four main processes that play important roles during the freezeout of the dark sector when $1 \lesssim r \lesssim 2$: 

\begin{enumerate}

    \item \textit{The} $2 \leftrightarrow 2$ \textit{dark sector process, }$\chi \overline{\chi} \leftrightarrow A' A'$. This process was shown in Ref.~\cite{PhysRevLett.115.061301} to be responsible for the freezeout of the dark sector for $1 \lesssim r \lesssim 1.5$, under the assumption that the dark sector was in full thermal equilibrium with the SM. As we described in the introduction, this process is kinematically forbidden for $r > 1$ for stationary $\chi$ particles, leading to a velocity-averaged annihilation cross section that is exponentially suppressed as a function of the dark sector temperature $T'$. Explicitly, the annihilation cross section is given by~\cite{PhysRevLett.115.061301}
    \begin{alignat}{1}
        \langle \sigma v \rangle_{\chi \overline{\chi} \to A'A'} &= \frac{n_{A',0}^2}{n_{\chi,0}^2} \langle \sigma v \rangle_{A' A' \to \chi \overline{\chi}} \nonumber \\
        &= \frac{9}{16} r^3 e^{2(1-r)m_\chi/T'} \langle \sigma v \rangle_{A' A' \to \chi \overline{\chi}} \,.
    \end{alignat}
    We provide the expression for $\langle \sigma v \rangle_{A' A' \to \chi \overline{\chi}}$ in App.~\ref{sec:Cross}; to make our analytic estimates more convenient, however, we parametrize this annihilation cross section as follows: 
     \begin{alignat}{1}
         \langle \sigma v \rangle_{A' A' \to \chi \overline{\chi}} \equiv \frac{\alpha_D^2}{m_\chi^2}g(r) 
         \label{eq:definition_of_g(r)}
    \end{alignat}
    where $g(r)$ is a function of $r$ that captures the nontrivial $r$-dependence. Typical values of $g(r)$ are shown in Table~\ref{tab:f_and_g}. 

    \renewcommand{\arraystretch}{1.5}

    \setlength{\tabcolsep}{6pt}

    \begin{table}[t!]
    \begin{tabular}{c c c c c c c c}

    \toprule
    \hline
    $r$ & 1.2 & 1.3 & 1.4 & 1.5 & 1.6 & 1.7 & 1.8 \\
    \hline
    $f(r)$ & 9.47 & 14.1 & 23.7 & 45.9 & 105.7 & 312.9 & 1427\\
    $g(r)$ & 4.44 & 5.49 & 5.90 & 5.94 & 5.77 & 5.50 & 5.19\\
    \botrule
    \end{tabular}
    \caption{List of $f(r)$ and $g(r)$ values, as defined in Eqs.~\eqref{eq:definition_of_g(r)} and~\eqref{eq:definition_of_f(r)}, evaluated at typical $r$-values of interest in this paper.}
    \label{tab:f_and_g}
    \end{table}

    As $r$ increases, the rate of the forward process becomes exponentially more suppressed as the mass difference between $\chi$ and $A'$ increases. Note that in the forward direction, $\chi \overline{\chi} \to A'A'$ removes kinetic energy from the dark sector; the rate of the forward reaction also becomes exponentially suppressed as $T'$ decreases, since less kinetic energy is available to $\chi$ particles for conversion into the rest mass of $A'$ particles.

    \item \textit{The $3 \leftrightarrow 2$ dark sector process,} $\chi \overline{\chi} \chi \leftrightarrow A' \chi$. For $1.5 \lesssim r \lesssim 2$, the freezeout of the dark sector is mainly controlled by this process, as examined in Ref.~\cite{Cline:2017tka}, once again under the assumption of a dark sector in thermal equilibrium with the SM. The forward process is a $3 \to 2$ process, with velocity-averaged annihilation cross section given by
    \begin{alignat}{1}
        \langle \sigma v^2 \rangle_{\chi \overline{\chi} \chi \to A' \chi} \equiv \langle \sigma v^2 \rangle \equiv \frac{\alpha_D^3}{m_\chi^5} f(r) \,,
        \label{eq:definition_of_f(r)}
    \end{alignat}
    where $f(r)$ encodes the nontrivial $r$-dependence of the cross section; once again, the full expression for $\langle \sigma v^2 \rangle_{\chi \overline{\chi} \chi \to \chi A'}$ is given in App.~\ref{sec:Cross}. For ease of notation, we will drop the subscript on the thermally averaged cross section from here on, unless it is needed to avoid ambiguity. Typical values for $f(r)$ across the range of $r$ considered in this paper are shown in Table~\ref{tab:f_and_g}. Note that the forward reaction converts rest mass to kinetic energy, and heats the dark sector, similar to other $3 \to 2$ processes found in cannibal dark matter models~\cite{Carlson:1992fn,Kuflik:2015isi,Pappadopulo:2016pkp,Farina:2016llk,Kuflik:2017iqs}.

    \item $A' \leftrightarrow f \overline{f}$. The dark photon kinetically mixes with the SM photon, and can decay into a pair of SM fermions. This process is an important number-changing process for $A'$ particles, and is one of two important processes responsible for transferring energy between the two sectors. The decay width $\Gamma$ of $A'$ is given in full in App.~\ref{sec:Cross}.

    \item $\chi f \leftrightarrow \chi f$. This elastic scattering process, and all possible processes related by conjugation, allows $\chi$ to directly transfer energy to or from the SM. This process as well as $A' \leftrightarrow f \overline{f}$ together determine how efficiently energy gets transferred between the two sectors. Once both $\chi f \leftrightarrow \chi f$ and $A' \leftrightarrow f \overline{f}$ become sufficiently inefficient, the dark sector and the SM can lose thermal contact and kinetically decouple, falling out of thermal equilibrium.

\end{enumerate}

There are additional $3 \leftrightarrow 2$ dark-sector-only processes that we do not consider, such as $\chi \bar{\chi} A^{\prime} \rightarrow A^{\prime} A^{\prime}$ and $A^{\prime} A^{\prime} A^{\prime} \rightarrow \chi \bar{\chi}$. Since we are only considering $m_{A^{\prime}} > m_{\chi}$, these $3 \to 2$ processes have rates that are parametrically suppressed by at least one power of $\exp(-(r-1) m_\chi/T')$ compared to $\chi \overline{\chi} \chi \to A' \chi$, and $\alpha_D$ times at least one power of $\exp(-m_{A'}/T')$ compared to $\chi \overline{\chi} \to A'A'$. These slower processes are therefore relatively unimportant compared to the much faster $3 \leftrightarrow 2$ and $2 \leftrightarrow 2$ processes shown here. 

We also neglect the processes $A' f \leftrightarrow \gamma f$ and $A' \gamma \leftrightarrow f \overline{f}$: these processes are suppressed by an additional factor of the electromagnetic fine structure constant $\alpha_\text{EM}$ relative to $A' \to f \overline{f}$, and are also Boltzmann suppressed by $n_{A'} \ll n_\chi$ relative to $\chi f \to \chi f$. Consequently, they never control when thermal decoupling between the two sectors occurs. They also do not play any important role based on the analytic understanding that we will develop below; they may only appear as terms proportional to $n_{A'} - n_{A',0}(T)$ in the Boltzmann equations, and can therefore be treated as small corrections to energy transfer rate arising from decays.

\subsection{Boltzmann Equations}
\label{subsec:boltzmann_equations}

The evolution of the system is governed by the coupled Boltzmann equations for the number densities of $\chi$ and $A^{\prime}$, $n_{\chi}$ and $n_{A^{\prime}}$, respectively, along with their energy densities $\rho_{\chi}$, $\rho_{A^{\prime}}$ and pressures $P_\chi$, $P_{A'}$:
\begin{alignat}{2}
\frac{d n_\chi}{dt} + 3 H n_\chi  &=&& \,\, -\frac{1}{4} \langle \sigma v^{2} \rangle \left[n_\chi^3 - \frac{ n_{\chi,0}(T')^2 }{ n_{A',0}(T') } n_\chi n_{A'} \right] \nonumber \\
& &&+  \langle \sigma v \rangle_{A^{\prime} A^{\prime} \rightarrow \bar{ \chi} \chi} \left[ n_{A^{\prime}}^{2}  - \frac{n_{A',0}(T')^2}{ n_{\chi,0}(T')^2 }    n_\chi^2 \right]   \nonumber \\
& &&- \frac{1}{2} \langle \sigma v \rangle_{\chi \bar{\chi} \rightarrow f \bar{f} } \left[ n_\chi^2 - n_{\chi,0}^2(T) \right] \,, 
 \label{eq:Boltz_chi}
\end{alignat}
\begin{alignat}{2}
    \frac{d n_{A^{\prime}}}{dt} + 3 H n_{A^{\prime}} &=&& \,\, \frac{1}{8} \langle \sigma v^{2} \rangle \left[ n_\chi^3 - \frac{ n_{\chi,0}(T')^2 }{ n_{A',0}(T') } n_\chi n_{A'} \right] \nonumber \\
	& &&- \langle \sigma v \rangle_{A^{\prime} A^{\prime} \rightarrow \bar{ \chi} \chi} \left[ n_{A^{\prime}}^{2} - \frac{n_{A',0}(T')^2}{ n_{\chi,0}(T')^2 }    n_\chi^2 \right] \nonumber \\
	& &&- \Gamma \left[ n_{A^{\prime}} - n_{A',0}(T) \right] \, ,
	\label{eq:Boltz_Ap}
\end{alignat}
and
\begin{alignat}{1}
    &\frac{d(\rho_\chi + \rho_{A'})}{dt} + 3 H(\rho_\chi + \rho_{A'} + P_\chi + P_{A'}) \nonumber \\
    & \qquad = \, - \langle \sigma v \delta E \rangle_{\chi f \to \chi f} n_\chi n_f  - m_{A'} \Gamma \left[n_{A'} - n_{A',0}(T)\right] \nonumber \\
    & \qquad \quad \,\, - \frac{1}{2} m_{\chi} \langle \sigma v \rangle_{\chi \overline{\chi} \to f \overline{f}} \left[n_\chi^2 - n_{\chi,0}^2(T) \right] \,,
    \label{eq:Boltz_rho}
\end{alignat}
where $n_f$ is the number density of charged SM particles, which for simplicity we assume to consist only of electrons and positrons. This assumption is justified because we are considering sub-GeV dark matter and dark photons, so thermal equilibrium between the SM and dark sector typically holds down to temperatures of $T \lesssim \SI{100}{\mega\eV}$, at which point all other SM particles have annihilated or decayed away. Note that all dark sector (SM) variables are evaluated at the dark sector temperature $T'$ (SM temperature $T$) unless otherwise stated. The prefactors for each term account for our convention of including both $\chi$ and $\overline{\chi}$ in $n_\chi$, and for initial state symmetry factors. Our convention, as well as the derivation of the dark sector annihilation cross sections, can be found in Ref.~\cite{Cline:2017tka}. We take the limit of nonrelativistic $\chi$ and $A^{\prime}$ for the $A^{\prime} \rightarrow f \bar{f}$ and $\chi \bar{\chi} \rightarrow f \bar{f}$ energy transfer rates. Details on the energy transfer rate for elastic scattering $\chi f \to \chi f$ can be found in App.~\ref{sec:Rate}; in particular, we highlight the fact that we have calculated $\langle \sigma v \delta E \rangle_{\chi f \to \chi f}$ analytically without assuming that $f$ is relativistic, which is to our knowledge a new result. This result is important when $m_\chi \sim \mathcal{O}(\SI{}{\mega\eV})$.
 
Eqs.~\eqref{eq:Boltz_chi}--\eqref{eq:Boltz_rho} contain three unknowns: $n_\chi$, $n_{A'}$ and $T'$, and can be solved numerically for the coupled evolution of these variables as a function of the SM temperature $T$. The numerical solution of these equations is used for all of the results throughout the paper. 

We will also rely significantly on analytic approximations to gain some intuition for these results. To this end, it is useful to write the energy density Boltzmann equation Eq.~\eqref{eq:Boltz_rho} in the nonrelativistic limit. Expanding the energy densities to first order in $1/x'$, which is a small parameter once $T,T' \ll m_\chi$, we find
\begin{alignat}{1}
    \rho_\chi &\simeq m_\chi n_\chi \left(1 + \frac{3}{2x'}\right) \,, \nonumber \\
    \dot{\rho}_\chi &\simeq m_\chi \dot{n}_\chi \left(1 + \frac{3}{2x'}\right) - \frac{3x}{2 x'^2} \frac{dx'}{dx} H m_\chi n_\chi \,,
\end{alignat}
and similarly for $\rho_{A'}$. We have also made the approximation $dx/dt \simeq Hx$. With these expansions, we obtain
\begin{multline}
    (\dot{n}_\chi + r\dot{n}_{A'}) \left(1 + \frac{3}{2x'} \right) \\
    + 3 H (n_\chi + r n_{A'}) \left(1 + \frac{5}{2x'} - \frac{x}{2x'^2} \frac{dx'}{dx} \right) \\
    \simeq -\langle \sigma v \delta E \rangle \frac{n_\chi n_f}{m_\chi} - r \Gamma\left[n_{A'} - n_{A',0}(T) \right] \,.
    \label{eq:nonrel_Boltzmann_energy_density}
\end{multline}
We have neglected DM annihilation into SM fermions in this analytic estimate for simplicity, since this process is typically not important in the regions of parameter space we will be interested in. We also find numerically that $(x/2x'^2)dx'/dx \ll \mathcal{O}(1)$ in all scenarios, and thus can be neglected in Eq.~\eqref{eq:nonrel_Boltzmann_energy_density}. The simplified Boltzmann energy density equation to leading order then reads
\begin{multline}
    \dot{n}_\chi + r \dot{n}_{A'} + 3 H (n_\chi + r n_{A'}) \simeq -\langle \sigma v \delta E \rangle \frac{n_\chi n_f}{m_\chi} \\
    - r \Gamma \left[ n_{A'} - n_{A'0}(T) \right] \,.
    \label{eq:simplified_energy_density_Boltzmann}
\end{multline}
Comparing this with the sum of the $\chi$ and $A'$ number density Boltzmann equations, Eqs.~\eqref{eq:Boltz_chi} and~\eqref{eq:Boltz_Ap}, which is given by
\begin{multline}
    \dot{n}_\chi + \dot{n}_{A'} + 3 H (n_\chi + n_{A'}) \\
    = - \frac{1}{8} \langle \sigma v^2 \rangle \left[ n_\chi^3 - \frac{n_{\chi,0}^2}{n_{A',0}} n_\chi n_{A'} \right] \\
    - \Gamma [ n_{A'} - n_{A',0}(T) ] \,,
\end{multline}
we finally obtain the following compact expression for the $\chi$ number density evolution:
\begin{multline}
    \dot{n}_\chi + 3 H n_\chi \simeq -\frac{n_\chi n_f}{(1-r)m_\chi} \langle \sigma v \delta E \rangle \\
    + \frac{r}{8(1-r)} \langle \sigma v^2 \rangle \left[ n_\chi^3 - \frac{n_{\chi,0}^2}{n_{A',0}} n_\chi n_{A'} \right] \,.
    \label{eq:n_chi_evolution}
\end{multline}
Comparing this expression with the number density Boltzmann equation for $\chi$, we find
\begin{multline}
    \langle \sigma v \rangle_{A'A' \to \chi \overline{\chi}} \left[n_{A'}^2 - \frac{n_{A',0}^2}{n_{\chi,0}^2} n_\chi^2 \right] \\
    \simeq  \frac{2-r}{8(1-r)}  \langle \sigma v^2 \rangle \left[n_\chi^3 - \frac{n_{\chi,0}^2}{n_{A',0}} n_\chi n_{A'} \right] \\
    - \frac{n_\chi n_f}{(1-r) m_\chi} \langle \sigma v \delta E \rangle \,.
    \label{eq:2_to_2_and_3_to_2_relation}
\end{multline}
With this relation, we can also reformulate the number density Boltzmann equation for $A'$ as
\begin{multline}
    \dot{n}_{A'} + 3 H n_{A'} \simeq - \frac{1}{8(1-r)} \langle \sigma v^2 \rangle \left[n_\chi^3 - \frac{n_{\chi,0}^2}{n_{A',0}} n_\chi n_{A'} \right] \\
    + \frac{n_\chi n_f}{(1-r)m_\chi} \langle \sigma v \delta E \rangle - \Gamma \left[n_{A'} - n_{A',0}(T) \right]    \,.
    \label{eq:n_A_evolution}
\end{multline}
These equations show that in the nonrelativistic limit, the Boltzmann equations establish certain relations between the rates of the various processes, determined ultimately by number and energy conservation. These equations will prove to be extremely useful for gaining analytic understanding of our numerical results.

\subsection{Fast Reactions and Freezeout}
\label{subsec:fast_reactions_and_freezeout}

To gain an understanding of the freezeout behavior of our dark sector, it is useful to understand when processes are occurring at rates fast enough to influence the freezeout process, and when they cease to be important. For temperatures $T \gtrsim m_\chi$, the rates of all of the process are generally fast, i.e.\ the rates of all processes in one direction are all much larger than the Hubble rate. For example, the $3 \leftrightarrow 2$ process is considered fast when 
\begin{alignat}{1}
    \frac{1}{4}  n_\chi \langle \sigma v^2 \rangle \gg H(T) \,.
\end{alignat}
While a process is fast, the corresponding terms in square brackets in Eqs.~\eqref{eq:Boltz_chi} and~\eqref{eq:Boltz_Ap} will generically be small, e.g.\ for the $3 \leftrightarrow 2$ process,
\begin{alignat}{1}
    n_\chi^3 \approx \frac{n_{\chi,0}^2}{n_{A',0}} n_\chi n_{A'} \quad (\text{fast } 3 \leftrightarrow 2)
    \label{eq:fast_3_to_2_n_relations}
\end{alignat}
such that 
\begin{alignat}{1}
    \frac{1}{4} \langle \sigma v^{2} \rangle \left[n_\chi^3 - \frac{ n_{\chi,0}^2 }{ n_{A',0} } n_{A'} n_\chi \right] \approx H n_\chi \quad (\text{fast } 3 \leftrightarrow 2) \,;
    \label{eq:fast_3_to_2}
\end{alignat}
otherwise, the $3 \leftrightarrow 2$ process can change the number densities of both $\chi$ and $A'$ within a time much faster than the Hubble time, until Eq.~\eqref{eq:fast_3_to_2} is satisfied. 

Similarly, the $2 \leftrightarrow 2$ process is fast when
\begin{alignat}{1}
    n_\chi \langle \sigma v \rangle_{\chi \overline{\chi} \to A'A'} \gg H(T) \,,
\end{alignat}
with
\begin{alignat}{1}
    n_{A'}^2 \approx \frac{n_{A',0}^2}{n_{\chi,0}^2} n_\chi^2 \quad (\text{fast } 2 \leftrightarrow 2)
    \label{eq:fast_2_to_2_n_relations}
\end{alignat}
such that
\begin{alignat}{1}
    \langle \sigma v \rangle_{\chi \overline{\chi} \to A'A'} \left[n_{A'}^2 - \frac{n_{A',0}^2}{n_{\chi,0}^2} n_\chi^2 \right] \approx H n_\chi \quad (\text{fast } 2 \leftrightarrow 2) \,.
\end{alignat}
Once $T \ll m_\chi$, the number densities of both $\chi$ and $A'$ are Boltzmann suppressed and rapidly decrease. At some point, the forward rates of these processes become comparable to the Hubble rate, and the process freezes out. For the $3 \to 2$ process, this happens when
\begin{alignat}{1}
    \frac{1}{4} n_\chi \langle \sigma v^2 \rangle \approx H(T) \quad (3 \leftrightarrow 2 \text{ freezeout})
    \label{eq:3_to_2_freezeout_condition}
\end{alignat}
and for the $2 \leftrightarrow 2$ process, 
\begin{alignat}{1}
     n_\chi \langle \sigma v \rangle_{\chi \overline{\chi} \to A' A'} \approx H(T) \quad (2 \leftrightarrow 2 \text{ freezeout}) \,.
    \label{eq:2_to_2_freezeout_condition}
\end{alignat}
Similar results hold for $\chi \overline{\chi} \to f \overline{f}$, just like in the conventional WIMP scenario. 

The approximate relations found in Eqs.~\eqref{eq:fast_3_to_2_n_relations} and~\eqref{eq:fast_2_to_2_n_relations} when the $3 \leftrightarrow 2$ and $2 \leftrightarrow 2$ processes are fast can be rewritten in terms of the effective chemical potential $\mu_\chi$ and $\mu_{A'}$ as
\begin{alignat}{1}
    2 \mu_\chi \approx \mu_{A'} \qquad\qquad (\text{fast } 3 \leftrightarrow 2) \,,
    \label{eq:mu_relation_3_to_2}
\end{alignat}
and
\begin{alignat}{1}
    \mu_\chi \approx \mu_{A'} \qquad\qquad (\text{fast } 2 \leftrightarrow 2) \,
    \label{eq:mu_relation_2_to_2}
\end{alignat}
respectively. Note that when both processes are fast, these relations together enforce $\mu_\chi \approx \mu_{A'} \approx 0$. 

For processes that are responsible for transferring heat between the SM and dark sector, the criterion for when these processes are ``fast'' depend on how much heat is generated/removed due to the $2 \leftrightarrow 2$ and $3 \leftrightarrow 2$ processes described above. Since the energy density of the dark sector for $T \ll m_\chi$ is dominated by the $\chi$ as $n_\chi \gg n_{A'}$, the rate of change of dark sector energy density per dark sector particle is given approximately by $m_\chi \dot{n}_{\chi}/n_{\chi}$; processes are considered ``fast'' if they can transfer heat between the sectors at a comparable rate. 
 
As discussed in Sec.~\ref{subsec:relevant_processes}, the two most important processes transferring energy between the two sectors are $A' \leftrightarrow f \overline{f}$ and $\chi f \leftrightarrow \chi f$. Let us first focus on the process $A' \leftrightarrow f \overline{f}$. In scenarios where both the $3 \leftrightarrow 2$ and $2 \leftrightarrow 2$ processes are fast, the number densities of the dark sector particles are given by $n_{\chi,0}(T')$ and $n_{A',0}(T')$. When $T \gg m_\chi$, $A' \leftrightarrow f \overline{f}$ is generally fast enough to maintain thermal equilibrium between the two sectors, so that $T' = T$. However, once $m_\chi > T$, $n_{A',0}(T)$ drops rapidly, and the number densities of the dark sector $n_{\chi,0}(T)$ and $n_{A',0}(T)$ evolve to a point where
\begin{alignat}{1}
    \frac{n_{A',0}}{n_{\chi,0}} r m_\chi \Gamma \approx \frac{m_\chi \dot{n}_{\chi,0}}{n_{\chi,0}} \,.
    \label{eq:Ap_decay_kinetic_decoupling}
\end{alignat}
After this point, the term on the left-hand side starts to become small relative to the right-hand side, and $A' \leftrightarrow f \overline{f}$ becomes ineffective at maintaining both sectors in thermal equilibrium. Similarly, the dark sector number densities can evolve to a point where
\begin{alignat}{1}
    n_f \langle \sigma v \delta E \rangle_{\chi f \to \chi f} \approx \frac{m_\chi \dot{n}_{\chi,0}}{n_{\chi,0}} \frac{T - T'}{T} \,,
    \label{eq:elastic_scatter_kinetic_decoupling}
\end{alignat}
after which $\chi f \leftrightarrow \chi f$ is too slow to maintain thermal equilibrium. Once both Eq.~\eqref{eq:Ap_decay_kinetic_decoupling} and~\eqref{eq:elastic_scatter_kinetic_decoupling} have been met, kinetic decoupling occurs, and the dark sector temperature $T'$ starts to diverge from the SM temperature $T$. Keep in mind that $\langle \sigma v \delta E \rangle_{\chi f \to \chi f}$ is proportional to $(T' - T)/T$; the comparison made in Eq.~\eqref{eq:elastic_scatter_kinetic_decoupling} is therefore between the heat transfer rate when $|T' - T|/T \sim \mathcal{O}(1)$ and the energy lost due to $n_{\chi,0}$ decreasing.

We are now ready to understand the broad features of the thermodynamic evolution of the dark sector. There are three thermodynamic phases that the dark sector in our model may go through:
\begin{enumerate}
    \item \textit{Thermodynamic phase A: dark sector in thermal equilibrium with the SM.} Interactions between the dark sector and the SM allow the two sectors to exchange heat. If these interactions are sufficiently fast, the dark sector stays in thermal equilibrium with the SM with $T' = T$, and the number densities of $\chi$ and $A'$ are simply given by $n_{\chi,0}(T)$ and $n_{A',0}(T)$; 
    \item \textit{Thermodynamic phase B: $T' \neq T$ with zero chemical potential.} Once $A' \to f \overline{f}$ and $\chi f \to \chi f$ become too slow, the dark sector kinetically decouples, and develops a temperature different from $T$. The $2 \leftrightarrow 2$ and $3 \leftrightarrow 2$ dark sector processes can inject or remove heat from the dark sector. While both processes are fast, Eqs.~\eqref{eq:mu_relation_3_to_2} and~\eqref{eq:mu_relation_2_to_2} enforce $\mu_\chi \approx \mu_{A'} \approx 0$. 
    \item \textit{Thermodynamic phase C: $T' \neq T$, with nonzero chemical potential.} If either the $3 \to 2$ or the $2 \to 2$ process freezes out after the SM-dark sector processes become slow, $\chi$ and $A'$ develop a chemical potential $\mu_\chi(T')$ and $\mu_{A'}(T')$ respectively, according to either Eqs.~\eqref{eq:mu_relation_3_to_2} and~\eqref{eq:mu_relation_2_to_2}. 
\end{enumerate}
In some parts of parameter space in the models we study, the dark sector goes through all three phases sequentially; in other parts of parameter space, a nonzero chemical potential develops once $T'$ starts diverging from $T$, leading to a direct transition from phase A to C without spending any significant time in phase B. 

Previous studies investigating this model~\cite{PhysRevLett.115.061301,Cline:2017tka} have assumed that the dark sector only stays in thermodynamic phase A, with Ref.~\cite{PhysRevLett.115.061301} making the further assumption that $n_{A'} = n_{A',0}(T)$ throughout in their thermally coupled model. However, we shall see that for values of $\epsilon$ as large as $10^{-5}$, the dark sector does not stay in thermodynamic phase A throughout the process of freezeout, changing the dependence of the relic abundance on the model parameters drastically.

Throughout this paper, we will mostly be interested in $\epsilon$ values that are small, of order $10^{-5}$ or smaller. However, if $\epsilon$ is too small, the dark sector and the SM sector need not have been in thermal contact at any point, calling into question the basic assumption we make that the two sectors start out in thermal equilibrium. To obtain an estimate for the minimum value of $\epsilon$ above which we are guaranteed to have the dark sectors in thermal equilibrium at $T \sim m_\chi$, we follow Ref.~\cite{Evans:2017kti}, and set this minimum value of $\epsilon$ to be when the $f \overline{f} \to A'$ rate exceeds the Hubble rate at $T = m_{A'}$. When this condition is met, $A'$ particles can be produced at a rate much faster than Hubble at $T \sim m_\chi$, allowing the whole dark sector to come into chemical equilibrium with the SM prior to the onset of the Boltzmann suppression from $A'$ and $\chi$ going nonrelativistic. This condition can be written as~\cite{Evans:2017kti}
\begin{alignat}{1}
    \frac{\pi^2}{12 \zeta(3)} \frac{m_{A'}}{T} \Gamma \sim \frac{T^2}{M_\text{pl}} \,,
\end{alignat}
where $\zeta$ is the Riemann zeta function, and $M_\text{pl}$ is the Planck mass. Using the expression for $\Gamma$ in App.~\ref{sec:Cross} and setting $T = m_{A'}$, we obtain the following estimate for $\epsilon_\text{eq}$, the minimum value of $\epsilon$ at which thermal equilibrium is guaranteed by $T \sim m_\chi$: 
\begin{alignat}{1}
    \epsilon_\text{eq} \sim 7 \times 10^{-9} \left(\frac{m_{A'}}{\SI{}{\giga\eV}}\right)^{1/2} \,.
    \label{eq:epsilon_thermal_equilibrium}
\end{alignat}
In practice, experimental constraints will limit us to values of $\epsilon \gtrsim 10^{-8}$; we can therefore safely assume the dark sector to be thermally coupled to the SM at $T \sim m_\chi$ throughout this paper. 

\section{\texorpdfstring{$1.5 \lesssim r \lesssim 2$}{1.5 ~< r ~< 2}}
\label{sec:NFDM_regime}

We begin our discussion of the freezeout of the vector-portal dark matter model with $1.5 \lesssim r \lesssim 2$. For these values of $r$, the $2 \leftrightarrow 2$ process freezes out before the $3 \leftrightarrow 2$ process. Under the assumption that the dark sector stays in thermodynamic phase A with $T' = T$, this regime --- which we call the ``classic not-forbidden dark matter (NFDM)'' regime --- was studied in Ref.~\cite{Cline:2017tka}, and was found to be a viable model for sub-GeV dark matter with appreciable self-interaction rates and thus the potential to affect the small-scale structure of galaxies. Here, we explore $1.5 \lesssim r \lesssim 2$ including the temperature evolution of the dark sector.

\subsection{``Classic Not-Forbidden'' Regime}
\label{subsec:classic_NFDM}

For sufficiently small values of $\epsilon$ with $r \gtrsim 1.5$, the $3 \leftrightarrow 2$ process eventually freezes out later than $\chi \overline{\chi} \to f \overline{f}$ --- the process that controls conventional WIMP freezeout --- and starts to become the main process that controls the final abundance of $\chi$. This transition occurs when
\begin{alignat}{1}
    n_\chi \langle \sigma v \rangle_{\chi \overline{\chi} \to f \overline{f}} \approx H \approx \frac{1}{4} n_\chi^2 \langle \sigma v^2 \rangle \,,
\end{alignat}
i.e.\ when both processes freeze out at roughly the same time. Using the analytic expressions for the quantities above, we obtain an estimate for $\epsilon_{\text{N/W}}$, the value of $\epsilon$ that sets the boundary between the `classic NFDM' regime and the WIMP regime:
\begin{multline}
    \epsilon_{\text{N/W}} \sim 2 \times 10^{-5} (4-r^2) \left(\frac{g_*(x_f)}{10.75}\right)^{1/8}  \\
    \times \left(\frac{\alpha_D}{1.0}\right)^{1/4} \left(\frac{20}{x_f}\right)^{1/2} \left(\frac{m_\chi}{\SI{}{\giga\eV}}\right)^{1/4} \left(\frac{f(r)}{105.7}\right)^{1/4} \,,
    \label{eq:eps_condition_WIMP_NFDM}
\end{multline}
where $x_f \equiv m_\chi/T_f$, and $T_f$ is the temperature at which freezeout of either of these two processes occur. $g_*$ is the effective number of relativistic degrees of freedom that enters into the Hubble parameter, $H(T) = 1.66 g_*^{1/2}(T) T^2 / M_\text{pl}$. Further requiring that the final relic abundance of DM is equal to the observed one today gives a relation between $\alpha_D$, $\epsilon$ and $m_\chi$. In the WIMP regime, where freezeout is controlled by $\chi \overline{\chi} \to f \overline{f}$, the correct relic abundance is obtained when Eq.~\eqref{eq:relic_abundance_requirement} is satisfied. This allows us to predict:
\begin{multline}
    \epsilon_{\text{N/W}} \sim 10^{-5} (4-r^2) \left(\frac{\alpha_D}{1.0}\right)^{1/2} \left(\frac{g_{*,s}(x_f)}{10.75}\right)^{1/6} \\
    \times\left(\frac{g_*(x_f)}{10.75}\right)^{1/12} \left(\frac{20}{x_f}\right)^{5/6} \left(\frac{f(r)}{105.7}\right)^{1/3}
    \label{eq:eps_condition_relic_abundance_NFDM}
\end{multline}
as the boundary between the conventional WIMP-like regime and the ``classic NFDM'' regime when the correct relic abundance is achieved. 

For $\epsilon < \epsilon_{\text{N/W}}$, the freezeout of the $3 \leftrightarrow 2$ process determines the abundance of DM, and the parameters that generate the correct relic abundance become virtually independent of $\epsilon$, provided that $\epsilon$ is large enough that the system remains in thermodynamic phase A (i.e.\ $T'=T$) throughout freezeout.

\subsection{Kinetically Decoupling Relic (KINDER) Regime}
\label{subsec:KINDER}

\begin{figure*}[tbp]
    \centering
    \includegraphics[width=0.47\textwidth]{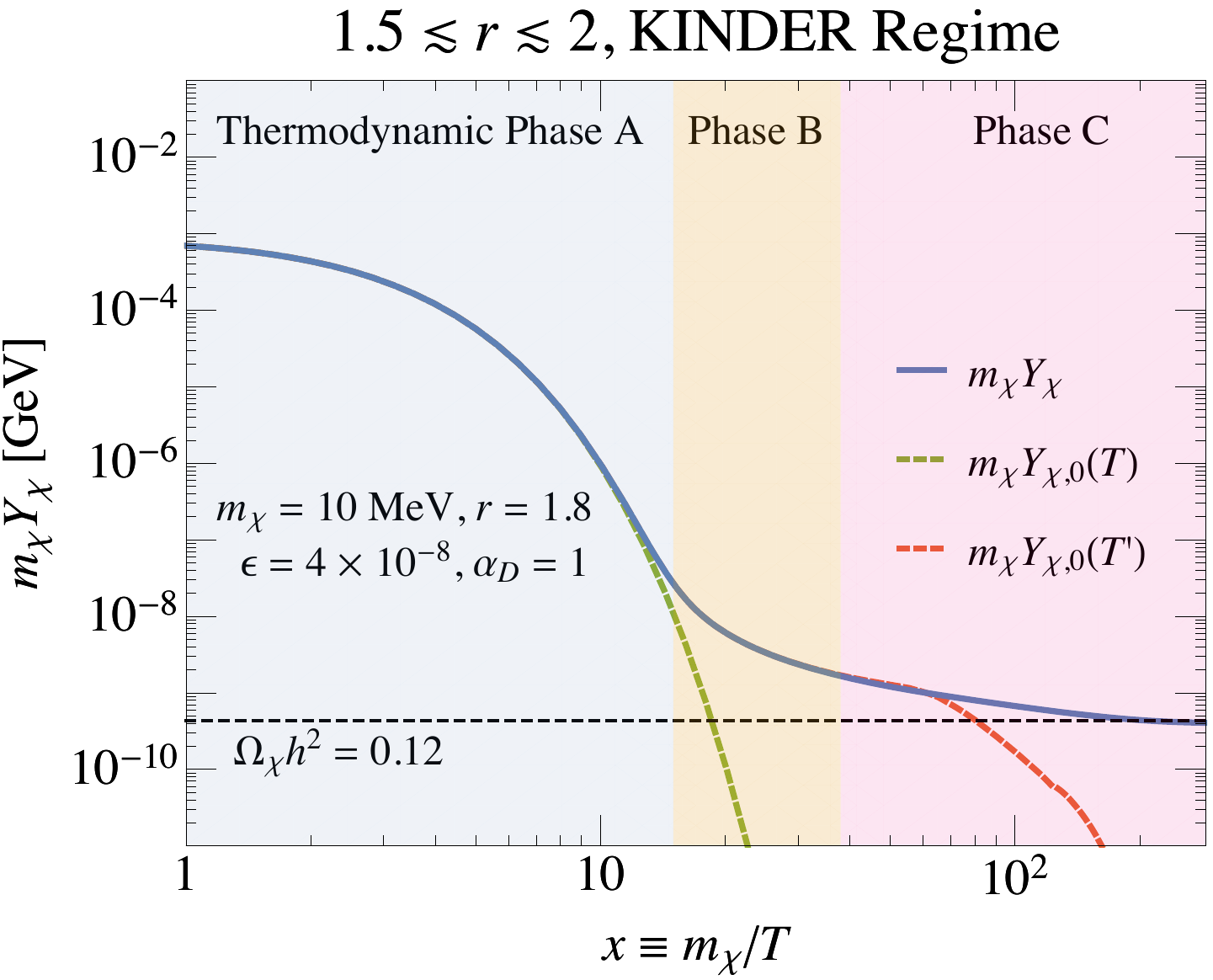}
    \qquad
    \includegraphics[width=0.47\textwidth]{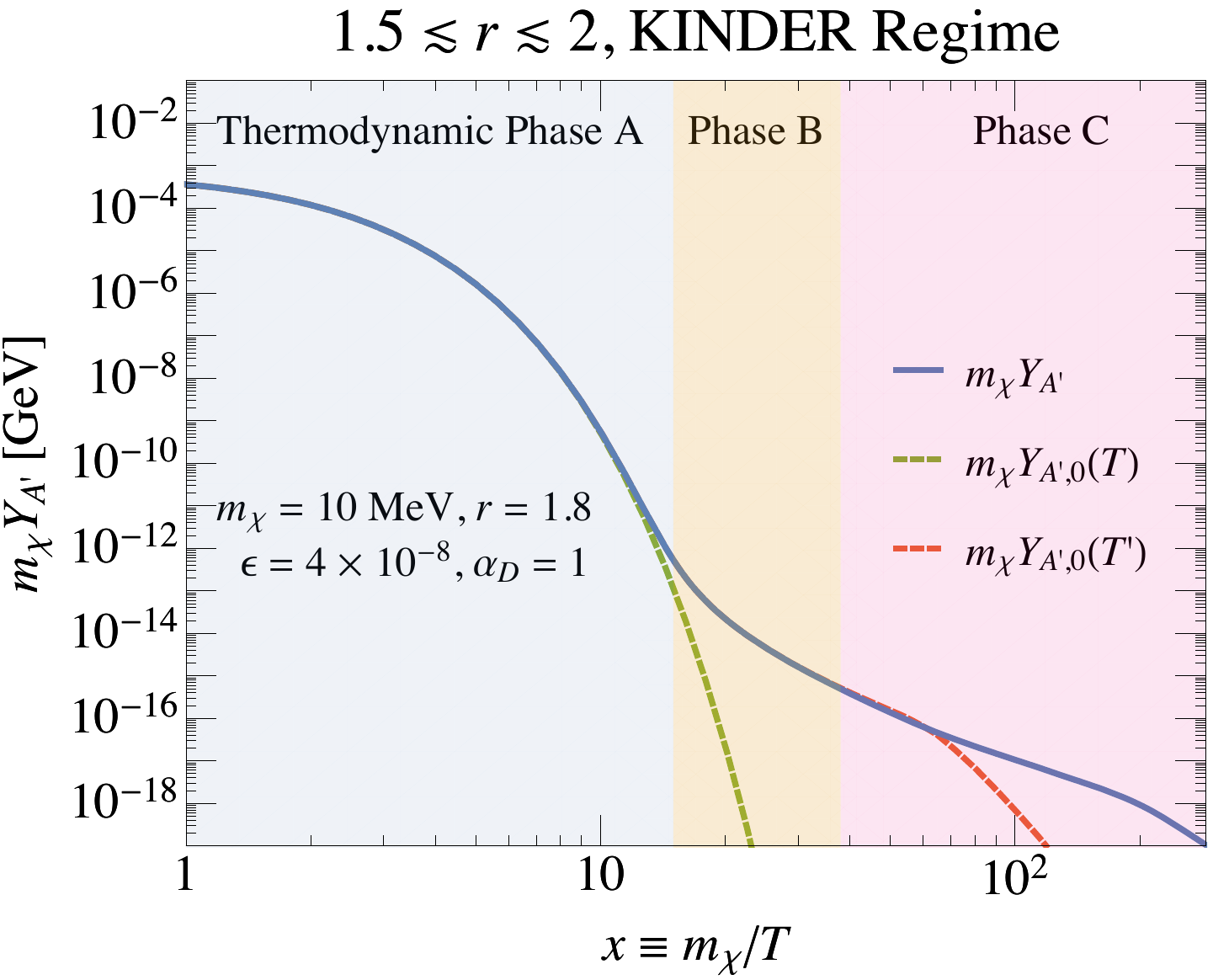}
    \includegraphics[width=0.47\textwidth]{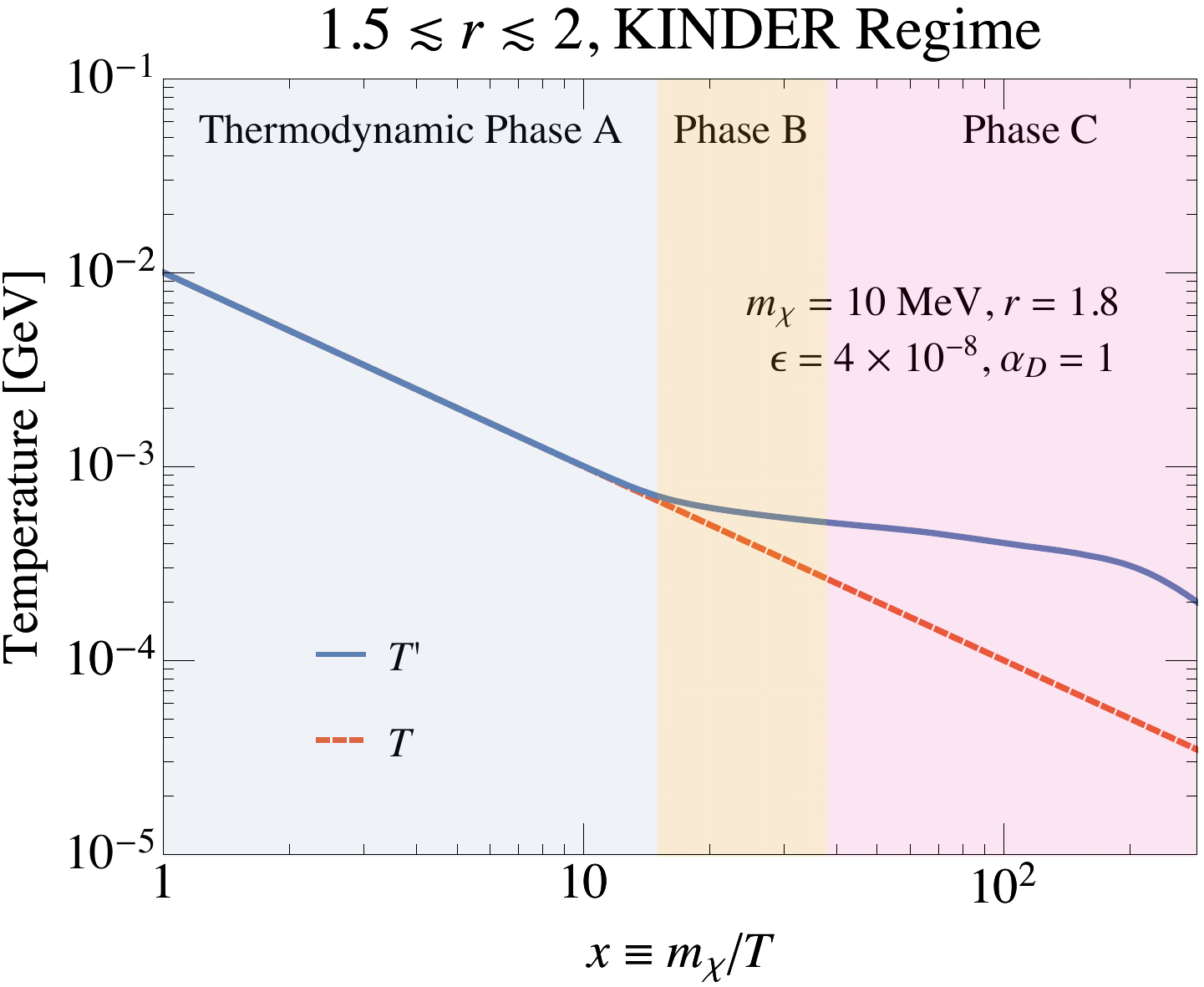}
    \caption{Dark sector evolution in the KINDER regime for $1.5 \lesssim r \lesssim 2$, with parameters $m_\chi = \SI{10}{\mega\eV}$, $r = 1.8$, $\epsilon = 4 \times 10^{-8}$ and $\alpha_D = 1$. In all three plots, thermodynamic phases A, B and C as defined in Sec.~\ref{subsec:fast_reactions_and_freezeout} are shown in light blue, yellow and pink respectively. (\textit{Top left}) $\chi$ abundance (given as $m_\chi Y_\chi$) as a function of $x$ (blue line), with the zero chemical potential abundance at the SM temperature $m_\chi Y_{\chi,0}(T)$ (green dashed line) and the dark sector temperature $m_\chi Y_{\chi,0}(T')$ (red dashed line) shown for reference. The observed DM abundance is indicated by the horizontal black dashed line, as defined in Eq.~\eqref{eq:relic_abundance_requirement}. (\textit{Top right}) $A'$ abundance (given as $m_\chi Y_{A'}$) as a function of $x$ (blue line), with $Y_{A',0}(T)$ (green dashed line) and $Y_{A',0}(T')$ (red dashed line) once again given for reference. (\textit{Bottom}) The dark sector temperature $T'$ (blue line), as a function of the SM temperature (red dashed line).} 
    \label{fig:abundances_NFDM}
\end{figure*}
\begin{figure*}[tbp]
    \centering
    \includegraphics[width=0.47\textwidth]{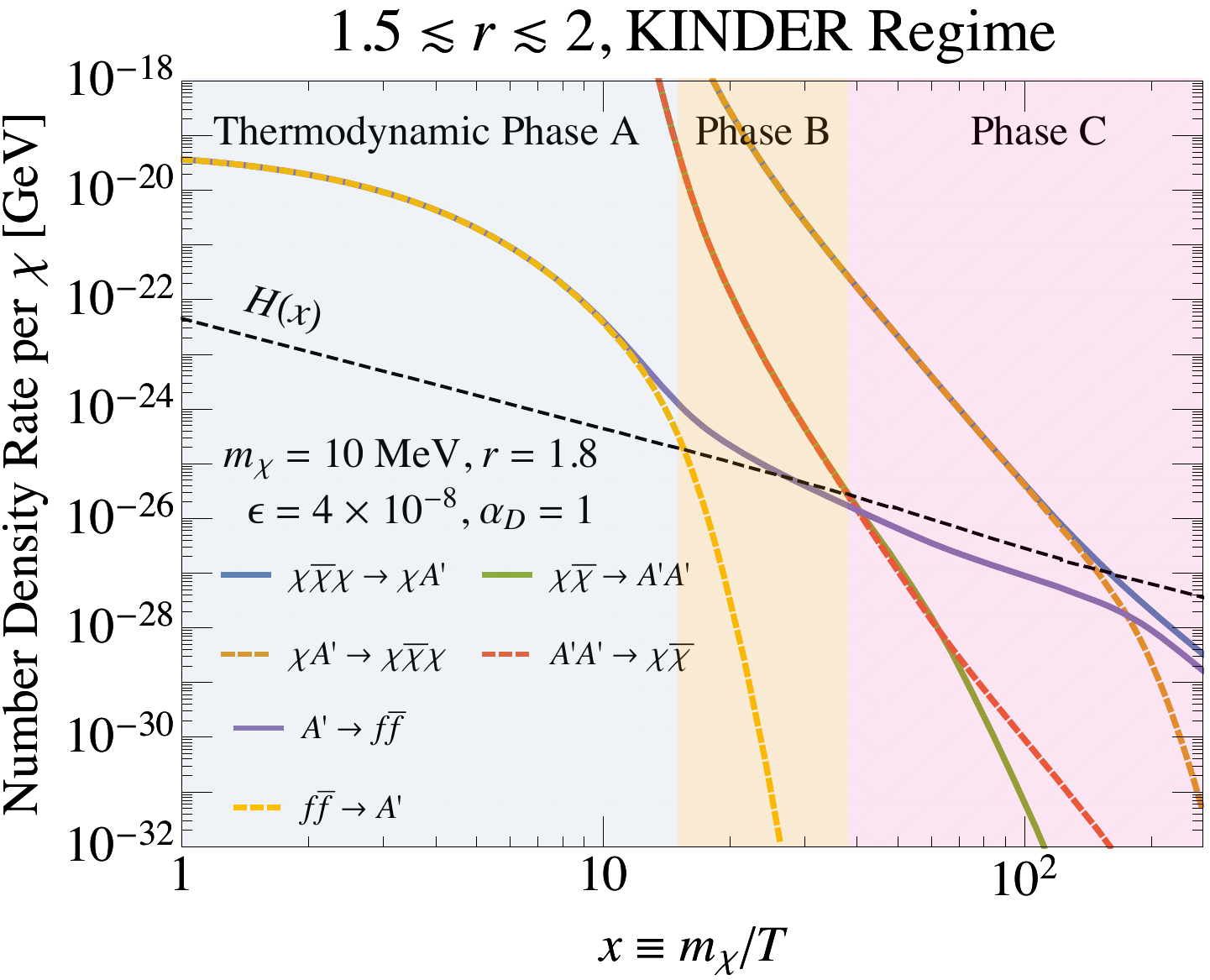}
    \qquad
    \includegraphics[width=0.47\textwidth]{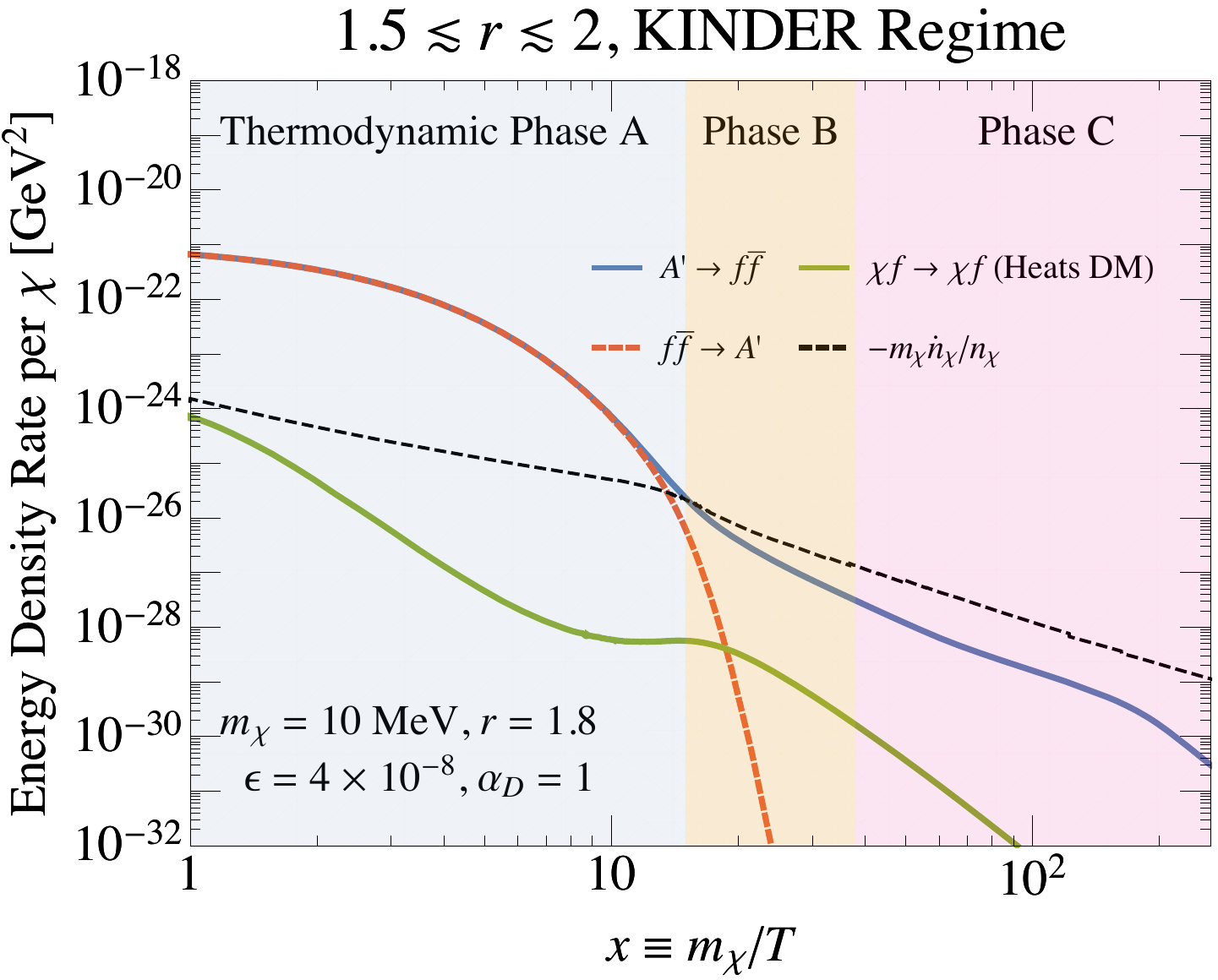}
    \caption{Rates of change in number density and energy density of the dark sector in the $1.5 \lesssim r \lesssim 2$ KINDER regime, with $m_\chi = \SI{10}{\mega\eV}$, $r = 1.4$, $\epsilon = 4 \times 10^{-8}$ and $\alpha_D = 1$ In both plots, thermodynamic phases A, B and C as defined in Sec.~\ref{subsec:fast_reactions_and_freezeout} are shown in light blue, yellow and pink respectively. (\textit{Left}) Number density rates for $\chi \overline{\chi} \chi \to \chi A'$ (blue line), $\chi A' \to \chi \overline{\chi} \chi$ (dark orange dashed line), $\chi \overline{\chi} \to A'A'$ (green line), $A'A' \to \chi \overline{\chi}$ (red dashed line) are shown, with solid lines indicating processes that net deplete $\chi$'s, and dashed lines indicating processes that net produce it instead. Also shown are the rates for $A' \to f \overline{f}$ (purple line) and $f \overline{f} \to A'$ (dashed yellow line). The Hubble parameter is shown in the black dashed line. (\textit{Right}) Energy density rates for $A' \to f \overline{f}$ (blue line), $f \overline{f} \to A'$ (red dashed line) and $\chi f \to \chi f$ (green line), which has the net effect of heating the dark sector. The rate at which the energy density of dark matter is changing $-m_\chi \dot{n}_\chi / n_\chi$ (black dashed line) is shown for reference.} 
    \label{fig:rates_NFDM}
\end{figure*}

As $\epsilon$ decreases further, processes that exchange energy between the dark and SM sectors become gradually less efficient; eventually, thermal equilibrium between the two sectors is lost even prior to $2 \leftrightarrow 2$ freezeout. This scenario, which we call the kinetic decoupling relic (KINDER) regime, is starkly different from the ``classic NFDM'' regime explained above. Notably, the abundance of DM after freezeout is governed primarily by when kinetic decoupling occurs, and therefore depends on both $\epsilon$ and $\alpha_D$. With thermal equilibrium between the two sectors lost prior to the freezeout of dark sector processes, the dark sector now goes through the different thermodynamic phases described in Sec.~\ref{subsec:fast_reactions_and_freezeout}. 

\subsubsection{General Features}
\label{subsubsec:KINDER_general_features}

In Fig.~\ref{fig:abundances_NFDM}, we show the abundances of $\chi$ and $A'$, as well as the dark sector temperature $T'$ as a function of $x$ for our benchmark parameter values in the KINDER regime: $m_\chi = \SI{10}{\mega\eV}$, $\alpha_D = 1$, $\epsilon = 4 \times 10^{-8}$, and $r = 1.8$. For ease of presentation, we plot the abundance as $m_\chi Y_\chi$ and $m_\chi Y_{A'}$, where $Y_i$ is defined in Eq.~\eqref{eq:relic_abundance_requirement}. In Fig.~\ref{fig:rates_NFDM}, we show the number density and energy density rates for the relevant dark sector processes; explicitly, these are the terms for each process that appear on the right-hand side of Eqs.~\eqref{eq:Boltz_chi} and~\eqref{eq:Boltz_Ap} divided by $n_\chi$ for number density rates, and the right-hand side of Eq.~\eqref{eq:Boltz_rho} divided by $n_\chi$ for energy density rates. At this parameter point (which is representative of the KINDER regime), the dark sector freezeout  proceeds through the following stages: 

\begin{enumerate}
    \item \textit{Kinetic decoupling, transition from thermodynamic phase A to B}. While either $\chi f \to \chi f$ and $A' \to f \overline{f}$ occur at rates larger than or comparable to the kinetic energy production rate of $\chi$ (i.e.\ the left-hand side of Eqs.~\eqref{eq:Ap_decay_kinetic_decoupling} or~\eqref{eq:elastic_scatter_kinetic_decoupling} are large compared to the RHS, kinetic equilibrium between the dark sector and SM particles is maintained at a common temperature $T=T'$. Once this is no longer true, i.e.\ after both $\chi f \to \chi f$ and $A' \to f \overline{f}$ become slow, kinetic decoupling occurs, and $T'$ begins to diverge from $T$. For our benchmark parameter values, kinetic decoupling occurs when $A' \to f \overline{f}$ becomes slow, as shown in Fig.~\ref{fig:rates_NFDM}.

    \item \textit{Cannibalization in thermodynamic phase B}. After this point, both $2 \leftrightarrow 2$ and $3 \leftrightarrow 2$ processes remain fast, and the dark sector enters thermodynamic phase B, where $T' \neq T$ and $\mu_\chi \approx \mu_{A'} \approx 0$, since both processes are fast. The net effect of the dark sector processes is to convert mass to kinetic energy in the dark sector so as to deplete $\chi$, and because this happens after the dark sector has kinetically decoupled from the SM, the dark sector heats up. This shares many similarities with dark matter models with a cannibal phase~\cite{Carlson:1992fn,Kuflik:2015isi,Pappadopulo:2016pkp,Kuflik:2017iqs,Farina:2016llk}, but with two different species involved in the $3 \leftrightarrow 2$ process sustaining cannibalization
    %cannibalization
    instead of one. Like other cannibal DM models, the dark sector particles have zero chemical potential, and $x'$ evolves in an approximately logarithmic manner with respect to $x$, with $Y_\chi$ evolving slowly. Unlike previous models, however, the entropy of the dark sector is not quite conserved, with $A' \to f \overline{f}$ decays remaining relatively efficient at depositing heat from the dark sector to the SM, but not fast enough to ensure equal temperatures; we will discuss this point in more detail below.

    \item \textit{Freezeout of $2 \leftrightarrow 2$ process, continued cannibalization}. After this point, the dark sector enters thermodynamic phase C with $2\mu_\chi \approx \mu_{A'}$, since the $3 \leftrightarrow 2$ process continues to be fast. Both $\chi$ and $A'$ develop a nonzero chemical potential in thermodynamic phase C, and the logarithmic evolution of $x'$ and slow evolution of $Y_\chi$ with respect to $x$ continues until the $3 \leftrightarrow 2$ process freezes out. This is an extension of the conventional cannibal dark matter scenario that we will investigate in greater detail below. 

    \item \textit{Freezeout of $3 \leftrightarrow 2$ process}. Finally, the $3 \to 2$ rate falls below the Hubble rate. With no other active number changing processes, the dark matter number density $n_\chi$ evolves proportionally to $a^{-3}$. 
\end{enumerate}

Because the slow evolution of $Y_\chi$ takes place from the time of kinetic decoupling until the freezeout of the $3 \leftrightarrow 2$ process, the DM thermal relic density is governed mainly by the kinetic decoupling process. In this regime, the vector-portal DM model therefore shares many similarities with elastically decoupling (ELDER) dark matter~\cite{Kuflik:2015isi}, with the main differences being the existence of thermodynamic phase C mentioned in the last paragraph, and the fact that kinetic decoupling in vector-portal DM is frequently governed by $A' \leftrightarrow f \overline{f}$, instead of elastic scattering processes, i.e.\ $\chi f \leftrightarrow \chi f$. The dark sector entropy is also not fully conserved due to the existence of $A' \leftrightarrow f \overline{f}$. 

Similarly to the boundary between the WIMP and ``classic NFDM'' regimes, we can estimate the value of $\epsilon$ at which we transition from the KINDER regime to the ``classic NFDM'' regime, by finding the value of $\epsilon$ for which kinetic decoupling and $3 \leftrightarrow 2$ freezeout occur at roughly the same time. We find that $A' \leftrightarrow f \overline{f}$ is often the process that governs kinetic decoupling, and so the boundary between these regimes occurs at the value of $\epsilon = \epsilon_{\text{K/N}}$ where both Eqs.~\eqref{eq:3_to_2_freezeout_condition} and~\eqref{eq:Ap_decay_kinetic_decoupling} are satisfied at the same SM temperature $T$. Analytically, we find
\begin{multline}
    \epsilon_\text{K/N} \sim 10^{-7} e^{9.9(r-1.6)} \left(\frac{\alpha_D}{1.0}\right)^{\frac{3(r-1)}{4}}  \\
    \times \left(\frac{1.6}{r}\right)^{9/4} \left(\frac{x_f}{20}\right)^{-\frac{r+3}{4}} \left( \frac{g_*(x_f)}{10.75} \right)^{- \frac{r-3}{8}}  \\
    \times \left(\frac{\SI{}{\giga\eV}}{m_\chi}\right)^{\frac{r-3}{4}} \left(\frac{f(r)}{105.7}\right)^{\frac{r-1}{4}}
    \label{eq:eps_condition_NFDM_KINDER}
\end{multline}
as the boundary between the `classic NFDM' regime and the KINDER regime, with $x_f$ denoting the dimensionless inverse temperature at the freezeout of the $3 \leftrightarrow 2$ process. To obtain an expression analogous to Eq.~\eqref{eq:eps_condition_relic_abundance_NFDM} under the additional assumption that the correct relic abundance is obtained, i.e.\ that Eq.~\eqref{eq:relic_abundance_requirement} is satisfied, we need to understand how the freezeout abundance of DM scales with the model parameters analytically in the KINDER regime. In the next few sections, we will review each thermodynamic phase of the KINDER regime, providing where possible an analytic understanding of the KINDER freezeout process. 

\subsubsection{Kinetic Decoupling and Cannibalization}
\label{subsubsec:kinetic_decoupling_cannibalization}

As we discussed in Sec.~\ref{subsec:fast_reactions_and_freezeout}, kinetic decoupling occurs at the point when both Eqs.~\eqref{eq:Ap_decay_kinetic_decoupling} and~\eqref{eq:elastic_scatter_kinetic_decoupling} have just been satisfied. We find that kinetic decoupling is usually controlled by $A' \leftrightarrow f \overline{f}$, i.e.\ the condition Eq.~\eqref{eq:Ap_decay_kinetic_decoupling} is fulfilled after Eq.~\eqref{eq:elastic_scatter_kinetic_decoupling}. Therefore, for the purpose of analytic estimates, we will assume that this is always true; our numerical results show that elastic scattering can become the process controlling kinetic decoupling at $m_\chi \sim \mathcal{O}(\SI{}{\giga\eV})$ and large $\alpha_D$.

Let us first obtain an analytic estimate of $x_d$, the dimensionless inverse temperature at kinetic decoupling, to see how it depends on the parameters of our model. Using the expression in Eq.~\eqref{eq:n_chi_dot} with $T' = T$ and $\mu_\chi = 0$, Eq.~\eqref{eq:Ap_decay_kinetic_decoupling} reads
\begin{alignat}{1}
    x_d^2 e^{(1-r)x_d} \approx 2.2 \frac{g_*^{1/2}(x_d)}{r^{7/2}} \frac{m_\chi^2}{M_\text{pl}\Gamma} \quad (\text{kinetic decoupling})\,.
    \label{eq:xd_dependence_Ap_decay}
\end{alignat}
For our benchmark parameters in this regime, the value of $x$ where this condition is met is shown in the right panel of Fig.~\ref{fig:rates_NFDM} at the transition between thermodynamic phases A and B.

After kinetic decoupling the dark sector temperature $T^{\prime}$ deviates from the SM temperature $T$, as indicated in Fig.~\ref{fig:abundances_NFDM}, while the $2 \rightarrow 2$ and $3 \rightarrow 2$ processes continue to proceed at rates larger than the Hubble expansion rate. The dark sector enters thermodynamic phase B, with both the $2 \leftrightarrow 2$ and $3 \leftrightarrow 2$ processes maintaining chemical equilibrium in the dark sector and forcing the chemical potentials to zero, as discussed in Sec.~\ref{subsec:fast_reactions_and_freezeout}. During this phase, the dark sector is cannibalistic, undergoing a net conversion of mass to kinetic energy in the dark sector, which then causes the dark sector to heat up. 

In the limit where no energy is transferred to the SM, the dark sector entropy $s_D a^3$ is conserved. The dark sector entropy density can be approximated as
\begin{alignat}{1}
    s_D &= \frac{ \rho_\chi + \rho_{A'} + P_\chi + P_{A'} - \mu_\chi n_\chi - \mu_{A'} n_{A'}}{T^{\prime}} \nonumber \\
    &\simeq \frac{m_\chi n_\chi - \mu_\chi n_\chi - \mu_{A'} n_{A'}}{T'} \,,
\label{eq:entropy}
\end{alignat}
where in the second line we can neglect $\rho_{A'}$ due to its relatively large Boltzmann suppression compared to $\rho_\chi$, and we used the fact that $P_{A'} \ll P_\chi = n_{\chi} T' \ll m_\chi n_\chi$ for $x' \gg 1$. Conservation of entropy enforces $ d \left( s_D a^{3} \right)/dt = 0$, with no processes active between the dark sector and the SM. In this limit, we have $\mu_\chi n_\chi \gg \mu_{A'} n_{A'}$ since $\mu_\chi$ and $\mu_{A'}$ are of the same order, and $\mu_\chi \dot{n}_\chi + \mu_{A'} \dot{n}_{A'} \simeq 0$, since the fast dark sector processes are responsible for both setting the chemical potentials and the number density evolution of the dark sector particles. Making use of Eq.~\eqref{eq:entropy} and the expression of $\dot{n}_\chi$ in Eq.~\eqref{eq:n_chi_dot}, entropy conservation in the dark sector implies the following relation between $T'$ and $T$:
\begin{multline}
    \frac{3}{T} \left(1 - \frac{\mu_\chi}{m_\chi}\right) \frac{dT}{dT'} \simeq \frac{1}{2T'} + \frac{m_\chi}{T'^2} + \left(1 - \frac{T'}{m_\chi}\right) \frac{d}{dT'} \left(\frac{\mu_\chi}{T'}\right) \\\qquad (s_Da^3\text{ conserved}).
    \label{eq:temperature_evolution_conserved_entropy}
\end{multline}
In thermodynamic phase B, we have $\mu_{\chi} \approx \mu_{A^{\prime}} \approx 0$ and $m_\chi \gg T'$, giving:
\begin{alignat}{1}
    \frac{m_\chi}{T'^2} \frac{dT'}{dT} \simeq \frac{3}{T}  \qquad (s_Da^3\text{ conserved}),
\end{alignat}
which we can integrate from $T_d' = T_d \equiv m_\chi / x_d$ up to some dark sector temperature $T'$ to get

\begin{alignat}{1}
    x' \simeq x_d + 3 \log \left(\frac{x}{x_d}\right) \qquad (s_Da^3\text{ conserved}).
    \label{eq:xp_NFDM_KINDER_with_entropy_conservation_Phase_B}
\end{alignat}
We see that the dark sector temperature $T^{\prime}$ is approximately fixed by the temperature of kinetic decoupling $T_{d}$, with $x'$ evolving slowly (logarithmically) with $x$ thereafter. If entropy were perfectly conserved, then the corresponding evolution in $n_\chi$ would be approximately
\begin{alignat}{1}
    n_\chi \approx n_{\chi,0}(T_d) \times \frac{T^3}{T_d^3} \qquad (s_Da^3\text{ conserved}). 
\end{alignat}
which would indicate an approximately constant $n_\chi a^3$ and $Y_\chi$ in phase B. 

While entropy conservation arguments are sufficient to get a crude approximation of the behavior of the dark sector in this phase, the true picture is significantly more complicated; for example, in Fig.~\ref{fig:abundances_NFDM}, while $Y_\chi$ stops exponentially decreasing in phase B, it is clearly not constant. In Fig.~\ref{fig:rates_NFDM}, we see that the dark sector enters thermodynamic phase B after kinetic decoupling occurs at around $x_d \sim 15$ for our benchmark parameters. After kinetic decoupling, the $A' \leftrightarrow f \overline{f}$ is no longer fast enough to keep the dark sector and SM in thermal equilibrium. As a result, the dark sector begins to heat, as shown in the bottom panel of Fig.~\ref{fig:abundances_NFDM}. With the increase in $T'$, however, comes an increase in $n_{A'}$, which also increases the rate at which energy density is transferred by $A' \to f \overline{f}$ to the SM. As a result, the energy density transfer from the dark sector to the SM remains relatively large even after kinetic decoupling; this can be seen in Fig.~\ref{fig:rates_NFDM}, which shows that this rate stays close to the rate of change of the dark sector energy density per $\chi$ particle, given approximately by $m_\chi \dot{n}_\chi / n_\chi$. Dark sector entropy is thus not quite conserved. 

\begin{figure}
    \centering
    \includegraphics[width=0.45\textwidth]{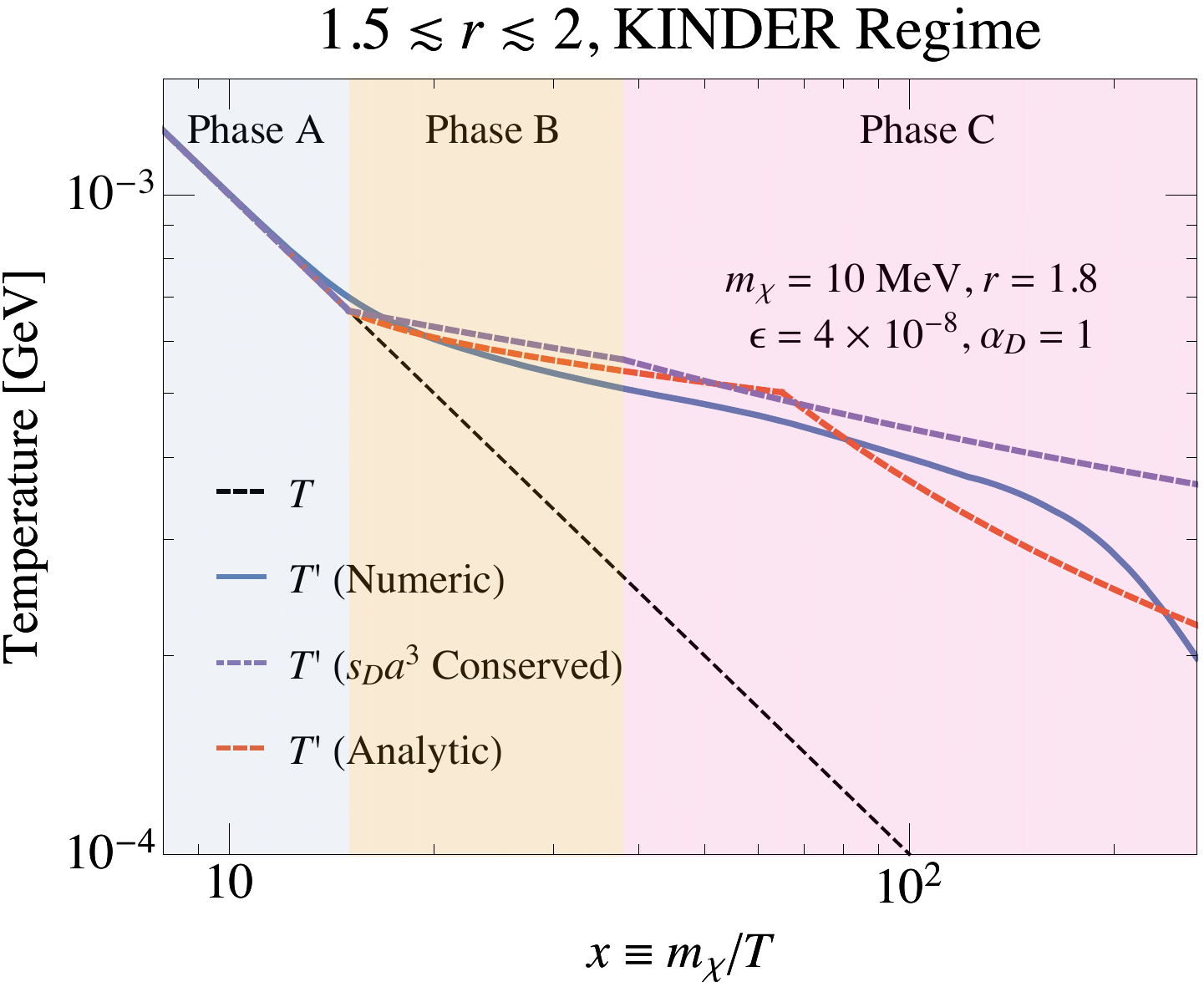}
    \caption{$1.5 \lesssim r \lesssim 2$, KINDER regime comparison between the improved analytic estimate of $T'$ (red dashed line) using Eq.~\eqref{eq:Tp_KINDER_thermo_phase_B} in thermodynamic phase B and Eq.~\eqref{eq:Tp_KINDER_thermo_phase_C_approx} in phase C and the full numeric calculation from the Boltzmann equations (blue line). We also show the predicted temperature assuming dark sector entropy conservation, using Eq.~\eqref{eq:xp_NFDM_KINDER_with_entropy_conservation_Phase_B} in phase B and Eq.~\eqref{eq:Tp_KINDER_thermo_phase_C_approx} in phase C is given by the purpled dashed line. The SM temperature $T$ is shown for reference (black dashed line), with the thermodynamic phases A, B and C marked in light blue, yellow and pink.}
    \label{fig:Tp_analytic_KINDER_NFDM}
\end{figure}

A better analytic understanding for the dark sector evolution thermodynamic phase B can be obtained from the argument above: since $T'$ always evolves in such a way as to keep $A' \to f \overline{f}$ relatively efficient at transferring energy from the dark sector to the SM, we find that
\begin{alignat}{1}
    m_{A'} \Gamma \frac{n_{A'}}{n_\chi} \approx \frac{m_\chi \dot{n}_\chi}{n_\chi} \,.
\end{alignat}
Thermodynamic phase B is characterized by zero chemical potentials for both species, i.e.\ $n_\chi \approx n_{\chi,0}(T')$, and likewise for $n_{A'}$. Given the expression for $\dot{n}_\chi$ in Eq.~\eqref{eq:n_chi_dot}, this approximation gives
\begin{alignat}{1}
    \frac{3}{4} \Gamma r^{5/2} e^{(1-r)x'} \approx H \left(\frac{3x}{2x'} + x \right) \frac{dx'}{dx} \,.
\end{alignat}
Taking $3x/2x' \ll x$ and $x^2 H(x) \approx x_d^2 H(x_d)$, this differential equation is easily integrated to get 
\begin{multline}
    x' \approx x_d \\
    + \frac{1}{r-1} \log \left[1 + \frac{3}{8} \frac{\Gamma r^{5/2} (r-1)e^{(1-r)x_d}}{x_d^2 H(x_d)}(x^2 - x_d^2)\right] \,.
    \label{eq:Tp_KINDER_thermo_phase_B}
\end{multline}
Fig.~\ref{fig:Tp_analytic_KINDER_NFDM} shows the comparison between this analytic temperature evolution and the numerical evolution computed directly from the Boltzmann equations. We see that the analytic result assuming entropy conservation overestimates the temperature somewhat, since it neglects the transfer of energy to the SM, and our modified analytic estimate is in better agreement with the phase B numerical results. 

As we indicated earlier, a very similar logarithmic evolution of $x'$ in a kinetically decoupled dark sector with zero chemical potential has already been found in other dark sector models~\cite{Carlson:1992fn,Kuflik:2015isi,Pappadopulo:2016pkp,Kuflik:2017iqs,Farina:2016llk}. However, as discussed above in Sec.~\ref{subsubsec:KINDER_general_features}, in the dark photon model parameter space we are studying, a second stage of cannibalization begins when the universe expands and cools to the point where the $2 \leftrightarrow 2$ process freezes out. 

\subsubsection{\texorpdfstring{$2 \leftrightarrow 2$ Freezeout and Continued Cannibalization}{2 <-> 2 Freezeout and Continued Cannibalization}}
\label{subsubsec:NFDM_KINDER_thermo_phase_C}

The $2 \leftrightarrow 2$ process freezes out when the $\chi \overline{\chi} \to A'A'$ rate falls below the Hubble expansion rate, triggering a nonzero chemical potential in the dark sector; this is indicated on the left panel of Fig.~\ref{fig:rates_NFDM} by the transition from phase B to C. We will label the temperatures of the SM and dark sector at which $2 \leftrightarrow 2$ freezeout occurs as $T_2$ and $T_2'$ respectively, and correspondingly $x_2$ and $x_2'$.

To understand the behavior of the dark sector in this phase analytically, we rely on Eq.~\eqref{eq:n_chi_evolution} and drop the contribution from elastic scattering, which is unimportant by the time the dark sector is in thermodynamic phase C. This gives
\begin{alignat}{1}
    \dot{n}_\chi + 3 H n_\chi \simeq \frac{r}{8(1-r)} \langle \sigma v^2 \rangle \left[n_\chi^3 - \frac{n_{\chi,0}^2}{n_{A',0}} n_\chi n_{A'} \right] 
    \label{eq:n_chi_evolution_no_elastic_scattering}
\end{alignat}
for the $\chi$ number density evolution, and
\begin{multline}
    \dot{n}_{A'} + 3 H n_{A'} \simeq - \frac{1}{8(1-r)} \langle \sigma v^2 \rangle \left[n_\chi^3 - \frac{n_{\chi,0}^2}{n_{A',0}} n_\chi n_{A'} \right] \\
    - \Gamma \left[n_{A'} - n_{A',0}(T)\right] \,.
    \label{eq:n_Ap_evolution_no_elastic_scattering}
\end{multline}
for the $A'$ number density.

In general, provided that the dark-sector number densities $n_i$ ($i=\chi, A'$) are such that they would be in a steady state in the absence of the cosmic expansion, their time derivatives will be parametrically controlled by $H$ and can be approximated as being of order $H n_i$ (the prefactor, of course, being important to the details of the solution). During the two cannibalization stages, when the comoving number density evolution is slow, we furthermore expect the prefactor to be an $\mathcal{O}(1)$ number.  Therefore, Eq.~\eqref{eq:n_chi_evolution_no_elastic_scattering} shows that:
\begin{alignat}{1}
	H n_{\chi} \approx \frac{r}{8(r-1)} \langle \sigma v^{2} \rangle \left[ n_\chi^3 - \frac{ n_{\chi,0}^2 }{ n_{A',0} } n_\chi n_{A'} \right] \, .
	\label{eq:NFDM_KINDER_3_to_2_rate_is_Hubble}
\end{alignat}

The $3 \leftrightarrow 2$ term on the right-hand side of Eq.~\eqref{eq:NFDM_KINDER_3_to_2_rate_is_Hubble} also appears in the Boltzmann equation for $A'$ shown in Eq.~\eqref{eq:Boltz_Ap}; however, since in general $n_{A'} \ll n_\chi$, we see that
\begin{alignat}{1}
    \frac{1}{8(r-1)} \langle \sigma v^2 \rangle \left[ n_\chi^3 - \frac{n_{\chi0}^2}{n_{A',0}} n_\chi n_{A'} \right] \gg H n_{A'} \,.
    \label{eq:3_to_2_fast_for_Ap}
\end{alignat}
In the parameter space of interest for obtaining the correct relic abundance in thermodynamic phase C, we generally have $\Gamma \gg H$ by the time $T \sim m_\chi$, as well as $n_{A'} \gg n_{A',0}(T)$, i.e.\ 
\begin{alignat}{1}
    \Gamma (n_{A'} - n_{A',0}) \approx \Gamma n_{A'} \gg H n_{A'} \,.
    \label{eq:decay_fast}
\end{alignat}
As we argued above, we expect the right-hand side of Eq.~\eqref{eq:n_Ap_evolution_no_elastic_scattering} to be on the order of $H n_{A'}$; since both terms on the right-hand side are large compared to $H n_{A'}$, we expect these terms to be comparable in magnitude. Given that the $3 \leftrightarrow 2$ rate is on the order of $H n_\chi$ as shown in Eq.~\eqref{eq:NFDM_KINDER_3_to_2_rate_is_Hubble}, we therefore arrive at the following important approximate relation that is valid in phase C:
\begin{alignat}{1}
	\frac{r}{8(r-1)} \langle \sigma v^2 \rangle \left[n_\chi^3 - \frac{n_{\chi,0}^2}{n_{A',0}} n_\chi n_{A'} \right] \approx H n_\chi \approx r \Gamma n_{A'}  \,.
	\label{eq:Hn_chi_roughly_Gamma_n_Ap}
\end{alignat}
How well the last approximation in the equation above is satisfied determines the accuracy of our analytic results: in Fig.~\ref{fig:rates_NFDM}, we see that this approximation is satisfied up to a factor of 3 throughout phase C. 

In thermodynamic phase C with a fast $3 \leftrightarrow 2$ process, recall from Eq.~\eqref{eq:mu_relation_3_to_2} that the chemical potentials of $\chi$ and $A'$ are related by $\mu_{A'} \approx 2 \mu_\chi$. We can therefore rewrite Eq.~\eqref{eq:Hn_chi_roughly_Gamma_n_Ap} as
\begin{alignat}{1}
    \frac{4}{3} r^{-3/2} e^{(r-1)x'} e^{-\mu_\chi/T'} \approx \frac{r \Gamma}{H(T)}\,.
\end{alignat}
At the point of $2 \to 2$ freezeout, with the dark sector and SM temperatures being $T_2'$ and $T_2$ respectively, we still have $\mu_\chi(T_2') = 0$, and so we have
\begin{alignat}{1}
    \frac{4}{3} r^{-3/2} e^{(r-1)x_2'} \approx \frac{r \Gamma}{H(T_2)} \,,
\end{alignat}
from which we finally obtain the following approximate relation for $\mu_\chi$ as a function of $T$ and $T'$:
\begin{alignat}{1}
	\frac{ \mu_\chi}{T^{\prime} } \approx \left( r - 1 \right) m_{\chi} \left[ \frac{1}{T^{\prime}} - \frac{1}{T_{2}^{\prime} } \right] - \log \left[ \frac{ H (T_2) }{ H (T) } \right] \, .
	\label{eq:mu_KINDER_phase_C}
\end{alignat}

\begin{figure}
    \centering
    \includegraphics[width=0.45\textwidth]{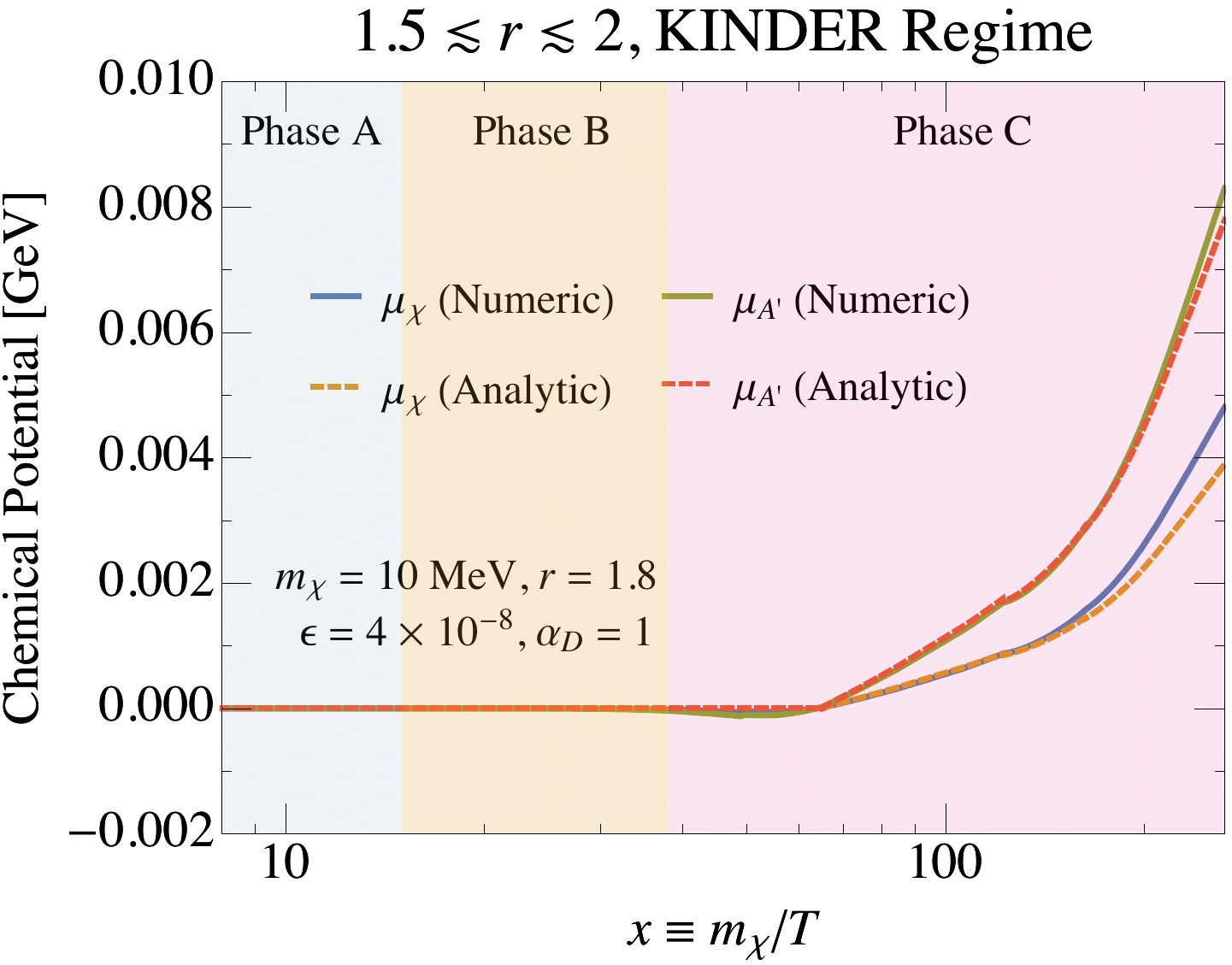}
    \caption{$1.5 \lesssim r \lesssim 2$, KINDER regime comparison between the analytic estimates of $\mu_\chi$ (orange dashed line) and $\mu_{A'}$ (red dashed line) based on Eq.~\eqref{eq:mu_NFDM_KINDER_with_T_mu} and $\mu_{A'} \approx 2 \mu_\chi$ with the numerical $\mu_\chi$ (blue line) and $\mu_{A'}$ (green line) based on integrating the full Boltzmann equations.}
    \label{fig:mu_KINDER_NFDM}
\end{figure}

To obtain a full, analytic understanding of the dark sector evolution, we now need to determine $T'$ as a function of $T$ after $2 \rightarrow 2$ freezeout. We can once again obtain a rough approximation by taking the dark sector entropy to be conserved, in which case Eq.~\eqref{eq:temperature_evolution_conserved_entropy} determines the evolution of $T'$ as a function of $T$. In order to get analytic control of the temperature evolution, we can make the approximations $T' \ll m_\chi$ and $\mu_\chi \ll m_\chi$; the latter condition is true early in phase C since the chemical potential starts at zero. Using the expression for $\mu_\chi/T'$ derived in Eq.~\eqref{eq:mu_KINDER_phase_C}, we find
\begin{alignat}{1}
    \frac{dT}{dT'} \approx \frac{T}{2T'} + (2-r)\frac{m_\chi T}{T'^2} \qquad (s_Da^3\text{ conserved}).
    \label{eq:temp_evol_intermediate_regime}
\end{alignat}
After $2 \to 2$ freezeout, for values of $r$ that are not too close to 2, we typically have $2(2-r)x' \gg 1$, and so we may drop the first term in the equation above to find that
\begin{alignat}{1}
    \frac{dT}{dT'} \approx (2-r) \frac{m_\chi T}{T'^2} \qquad (s_Da^3\text{ conserved}).
    \label{eq:approx_dT_dTp_phase_C}
\end{alignat}
We may integrate this approximate expression to obtain
\begin{alignat}{1}
    x' \approx x_2' + \frac{1}{2-r} \log \left(\frac{x}{x_2}\right) \qquad (s_Da^3\text{ conserved}).
    \label{eq:Tp_KINDER_thermo_phase_C_approx}
\end{alignat}
which shows that even during thermodynamic phase C with a nonzero chemical potential in the dark sector, the dark sector temperature $T'$ still evolves logarithmically with the SM temperature $T$. After the freezeout of the $2 \leftrightarrow 2$ process, the $3 \leftrightarrow 2$ process alone is sufficient to maintain cannibalization of the dark sector, even though a nonzero dark chemical potential $\mu$ has developed. This second stage of cannibalization which occurs in the KINDER scenario is an extension of the conventional cannibalization scenario. It is a critical part of the thermal history of KINDER, because it ensures that after $2 \leftrightarrow 2$ freezeout and before $3 \leftrightarrow 2$ freezeout, the dark sector temperature $T^{\prime}$ and comoving number density $\left( n_{\chi} a^{3} \right)$ continue to evolve slowly, as Fig.~\ref{fig:abundances_NFDM} shows, remaining mostly fixed by their values at kinetic decoupling. We will explore this slow evolution of $n_\chi$ in more detail in Sec.~\ref{subsubsec:3_to_2_freezeout_and_relic_abundance_NFDM_KINDER}.

As before, entropy conservation is not strictly obeyed due to the fact that $A' \to f \overline{f}$ remains quite efficient at transferring energy from the dark sector to the SM; a more sophisticated analytic understanding can once again be attained by examining the Boltzmann equations closely. First, with elastic scattering being unimportant, Eq.~\eqref{eq:2_to_2_and_3_to_2_relation} shows that there is an approximate relationship between the $2 \leftrightarrow 2$ and $3 \leftrightarrow 2$ rates that is applicable even after $2 \leftrightarrow 2$ freezeout:
\begin{multline}
    \langle \sigma v \rangle_{A'A' \to \chi \overline{\chi}} \left[n_{A'}^2 - \frac{n_{A',0}^2}{n_{\chi,0}^2} n_\chi^2\right] \\
    \simeq \frac{2-r}{8(1-r)} \langle \sigma v^2 \rangle \left[n_\chi^2 - \frac{n_{\chi,0}^2 }{n_{A',0}} n_\chi n_{A'} \right] \,.
\end{multline}
As we argued in Eq.~\eqref{eq:NFDM_KINDER_3_to_2_rate_is_Hubble}, the $3 \leftrightarrow 2$ rate is comparable to $H n_\chi$, which leads us to conclude that
\begin{alignat}{1}
 \langle \sigma v \rangle_{A'A' \to \chi \overline{\chi}} \left[n_{A'}^2 - \frac{n_{A',0}^2}{n_{\chi,0}^2} n_\chi^2\right] \approx \frac{2-r}{r} H n_\chi \,.
\end{alignat}
This expression demonstrates that just after the point of $2 \leftrightarrow 2$ freezeout, defined in Eq.~\eqref{eq:2_to_2_freezeout_condition}, the $\chi \overline{\chi} \to A'A'$ and $A'A' \to \chi \overline{\chi}$ rates remain close to each other, until 
\begin{alignat}{1}
    n_\chi \langle \sigma v \rangle_{\chi \overline{\chi} \to A'A'} \approx \frac{2 - r}{r} H \,.
    \label{eq:nonzero_chemical_potential_in_KINDER_phase_C}
\end{alignat}
The fact that these rates are close even after $2 \leftrightarrow 2$ freezeout can be seen in Fig.~\ref{fig:rates_NFDM}, immediately after the transition between phases B and C. 

Before the condition in Eq.~\eqref{eq:nonzero_chemical_potential_in_KINDER_phase_C} is satisfied, we must therefore have $\mu_\chi \approx \mu_{A'}$ as well, which together with the fast $3 \leftrightarrow 2$ requirement that $\mu_{A'} \approx 2 \mu_\chi$ maintains the chemical potential of the dark sector at approximately zero. Moreover, temperature evolution continues to obey the temperature evolution derived in phase B, shown in Eq.~\eqref{eq:Tp_KINDER_thermo_phase_B}. Eventually, $n_\chi$ decreases to a point where Eq.~\eqref{eq:nonzero_chemical_potential_in_KINDER_phase_C} becomes satisfied at some SM temperature $T_\mu$ and corresponding $x_\mu \equiv m_\chi / T_\mu$.

Above $x_\mu$, the previous argument used to obtain Eq.~\eqref{eq:mu_KINDER_phase_C} can be used to obtain a similar expression: 
\begin{alignat}{1}
    \frac{\mu_\chi}{T'} \approx (r-1) m_\chi \left[\frac{1}{T'} - \frac{1}{T_\mu'}\right] - \log \left[\frac{H(T_\mu)}{H(T)}\right] \,,
    \label{eq:mu_NFDM_KINDER_with_T_mu}
\end{alignat}
and the condition shown in Eq.~\eqref{eq:Hn_chi_roughly_Gamma_n_Ap} reduces the $\chi$ number density evolution to the following compact form:
 \begin{alignat}{1}
     \dot{n}_\chi + 3 H n_\chi \simeq -r \Gamma n_{A'} \,.
 \end{alignat}
Using the expression for $\dot{n}_\chi$ found in Eq.~\eqref{eq:n_chi_dot} as well as the expression for the chemical potential derived in Eq.~\eqref{eq:mu_KINDER_phase_C}, we obtain
\begin{multline}
    - \left[\frac{3}{2T'} + (2-r) \frac{m_\chi}{T'^2}\right] \frac{dT'}{dT} \\
    \simeq -\frac{1}{T} - \frac{r \Gamma}{H(T_\mu)T} \frac{3}{4} r^{3/2} e^{(1-r)m_\chi/T_\mu'} \,,
\end{multline}

If we make the approximation that $3/2 \ll (2-r) m_\chi/T'$, we can integrate this expression to obtain  
\begin{alignat}{1}
    x' \approx x_\mu' + \frac{1+C}{2-r} \log \left(\frac{x}{x_\mu}\right) \,,
\end{alignat}
where
\begin{alignat}{1}
    C \equiv \frac{r \Gamma}{H(T_\mu)} \frac{3}{4} r^{3/2} e^{(1-r)x_\mu'} \,.
\end{alignat}
Compared to the estimate for $x'$ in phase C obtained using entropy conservation in Eq.~\eqref{eq:Tp_KINDER_thermo_phase_C_approx}, we see that this more sophisticated analytic treatment \textit{(i)} correctly identifies the delay in the onset of a nonzero chemical potential, and \textit{(ii)} introduces a correction to the temperature evolution encapsulated by the factor $C$ ($C \simeq 3.4$ for our benchmark parameters). The result of our analytic estimate for the temperature is shown in Fig.~\ref{fig:Tp_analytic_KINDER_NFDM}, and shows reasonable agreement with the fully numerical solution, up till the complete freezeout of the dark sector at $x \sim 200$. The agreement between the analytic estimate and the numerical result deteriorates at larger $x$ as the approximation $H n_\chi \approx r \Gamma n_{A'} $ becomes poor (we should only expect them to be equal up to an $\mathcal{O}(1)$ factor). The result for our improved analytic estimate for the chemical potentials using Eq.~\eqref{eq:mu_NFDM_KINDER_with_T_mu} is shown in Fig.~\ref{fig:mu_KINDER_NFDM}, and shows good agreement with the numerical results.

\subsubsection{\texorpdfstring{$3 \rightarrow 2$ Freezeout and Relic Abundance}{3 -> 2 Freezeout and Relic Abundance}}
\label{subsubsec:3_to_2_freezeout_and_relic_abundance_NFDM_KINDER}

Cannibalization of the dark sector continues until the universe expands and cools to the point at which $3 \leftrightarrow 2$ annihilations freeze out at temperature $T_3$ (and corresponding $x_3$). This marks the freezeout of DM $\chi$, at $x_{3} \sim 200$ for our benchmark KINDER parameter point, as demonstrated in Figs.~\ref{fig:abundances_NFDM} and~\ref{fig:rates_NFDM}. After freezeout, the comoving DM abundance $Y_\chi$ settles to its constant relic value, and the dark sector temperature begins to evolve as $T^{\prime} \propto T^{2}$, as expected for a completely decoupled nonrelativistic fluid. 

Given the analytic estimates derived in the previous sections, we can now obtain an analytic estimate for the number density of DM particles at $3 \leftrightarrow 2$ freezeout, given by the condition shown in Eq.~\eqref{eq:3_to_2_freezeout_condition}. We use the assumption of dark sector entropy conservation for simplicity, although a similar conclusion can be reached by using the more accurate analytic results described previously. 

The number density of DM at freezeout can be written given the chemical potential in Eq.~\eqref{eq:mu_KINDER_phase_C}, giving
\begin{alignat}{1}
    n_\chi(T_3') \approx 4 \left(\frac{m_\chi^2}{2\pi x_3'}\right)^{3/2} \frac{x_2^2}{x_3^2} e^{(r-2)x_3'} e^{(1-r)x_2'} \,. 
\end{alignat}
However, the approximate expression for the temperature evolution in thermodynamic phase C found in Eq.~\eqref{eq:Tp_KINDER_thermo_phase_C_approx} allows us to rewrite this as
\begin{alignat}{1}
    n_\chi(T_3') \approx 4 \left(\frac{m_\chi T_3'}{2 \pi}\right)^{3/2} \frac{x_2^3}{x_3^3} e^{-x_2'} \,.
\end{alignat}
Finally, using the expression for the temperature evolution during thermodynamic phase B in Eq.~\eqref{eq:Tp_KINDER_thermo_phase_B}, we can rewrite $x_2'$ in terms of $x_d$, the dimensionless inverse temperature at which kinetic decoupling occurs, giving
\begin{alignat}{1}
    n_{\chi}(T_3') \simeq 4 \left(\frac{m_\chi^2}{2 \pi x_3'}\right)^{3/2} e^{-x_d} \frac{x_d^3}{x_3^3} \,.
    \label{eq:n_chi_at_freezeout_KINDER_prelim_approx}
\end{alignat}
This remarkable expression shows explicitly that the freezeout abundance is mostly controlled by kinetic decoupling, being exponentially sensitive to $x_d$, up to small power law corrections. 

Since the temperature of the dark sector evolves logarithmically after kinetic decoupling, we can make the approximation $x_3' \approx x_d$ in Eq.~\eqref{eq:n_chi_at_freezeout_KINDER_prelim_approx}. Substituting the resulting expression into Eq.~\eqref{eq:3_to_2_freezeout_condition}, we obtain the following estimate for $n_\chi$ at freezeout: 
\begin{alignat}{1}
    n_\chi (T_3') \simeq 8.2 e^{x_d/2} \left(\frac{g_*^{1/2}(x_3)}{x_d M_\text{pl} \langle \sigma v^2 \rangle}\right)^{3/4} \,.
    \label{eq:n_chi_at_freezeout_KINDER_approx}
\end{alignat}

We are now ready to obtain an analytic estimate for $\epsilon_{K/N}$ as shown in Eq.~\eqref{eq:eps_condition_NFDM_KINDER}, when the ``classic NFDM'' regime transitions into the KINDER regime in the $\alpha_D$--$\epsilon$ plane, but now with the requirement that the correct relic abundance is achieved by choosing $m_\chi$ appropriately at each point in this parameter space. At the regime boundary, kinetic decoupling and $3 \to 2$ freezeout occur at roughly the same time, i.e. $x_d \approx x_3$. Combining the requirement shown in Eq.~\eqref{eq:relic_abundance_requirement} for the correct relic abundance of $\chi$ with Eq.~\eqref{eq:n_chi_at_freezeout_KINDER_approx} gives
\begin{multline}
    \left(\frac{m_\chi}{\SI{}{\giga\eV}}\right) \sim 0.1 \left(\frac{20}{x_d}\right)^{4/3} \left(\frac{g_{*,s}(x_d)}{10.75}\right)^{2/3}  \\
    \times \left(\frac{g_*(x_d)}{10.75}\right)^{-1/6} \left(\frac{\alpha_D}{1.0}\right) \left(\frac{f(r)}{105.7}\right)^{1/3} \,.
\end{multline}
Note that typical values of $x_d$ are $x_d \simeq 18.5$ for $m_\chi \simeq \SI{}{\mega\eV}$ and $x_d \simeq 23.4$ for $m_\chi \simeq \SI{}{\giga\eV}$. Substituting this expression into Eq.~\eqref{eq:eps_condition_NFDM_KINDER} leads to
\begin{multline}
    \epsilon_\text{K/N} \sim 4\times 10^{-8} e^{10.5(r - 1.6)} \left(\frac{1.6}{r}\right)^{9/4}  \\
    \times \left(\frac{g_{*,s}(x_d)}{10.75} \right)^\frac{3-r}{6} \left(\frac{g_*(x_d)}{10.75}\right)^\frac{3-r}{12} \\
    \times \left(\frac{x_d}{20}\right)^\frac{r-21}{12} \left(\frac{\alpha_D}{1.0}\right)^{r/2} \left(\frac{f(r)}{105.7}\right)^{r/6} \,.
    \label{eq:eps_condition_NFDM_KINDER_relic_abundance}
\end{multline}

\subsubsection{\texorpdfstring{Summary of regimes and boundaries for $1.5 < r < 2$}{Summary of regimes and boundaries for 1.5 ~< r ~< 2}}

\begin{figure*}
    \centering
    \includegraphics[width=0.47\textwidth]{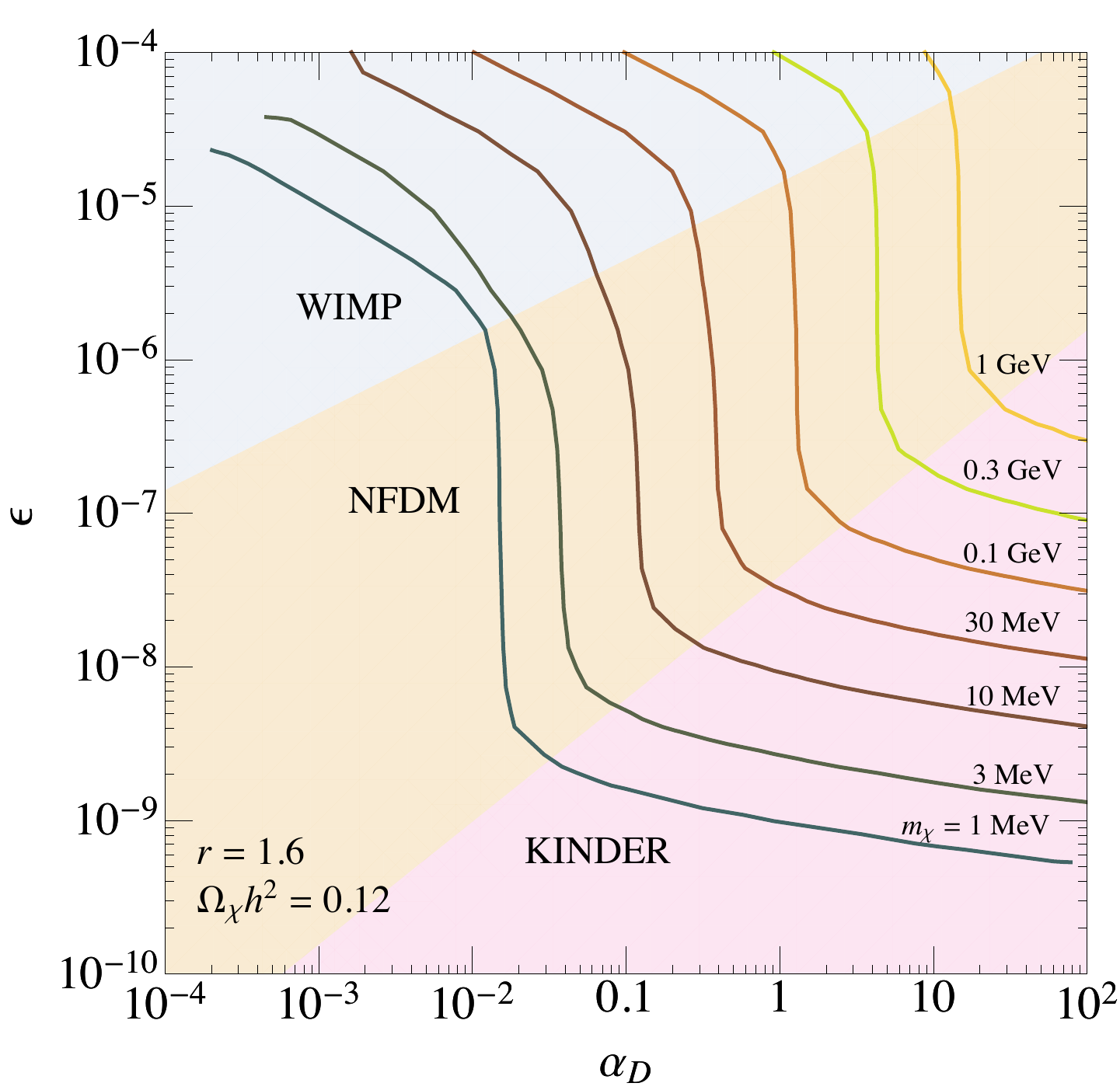} \qquad
    \includegraphics[width=0.47\textwidth]{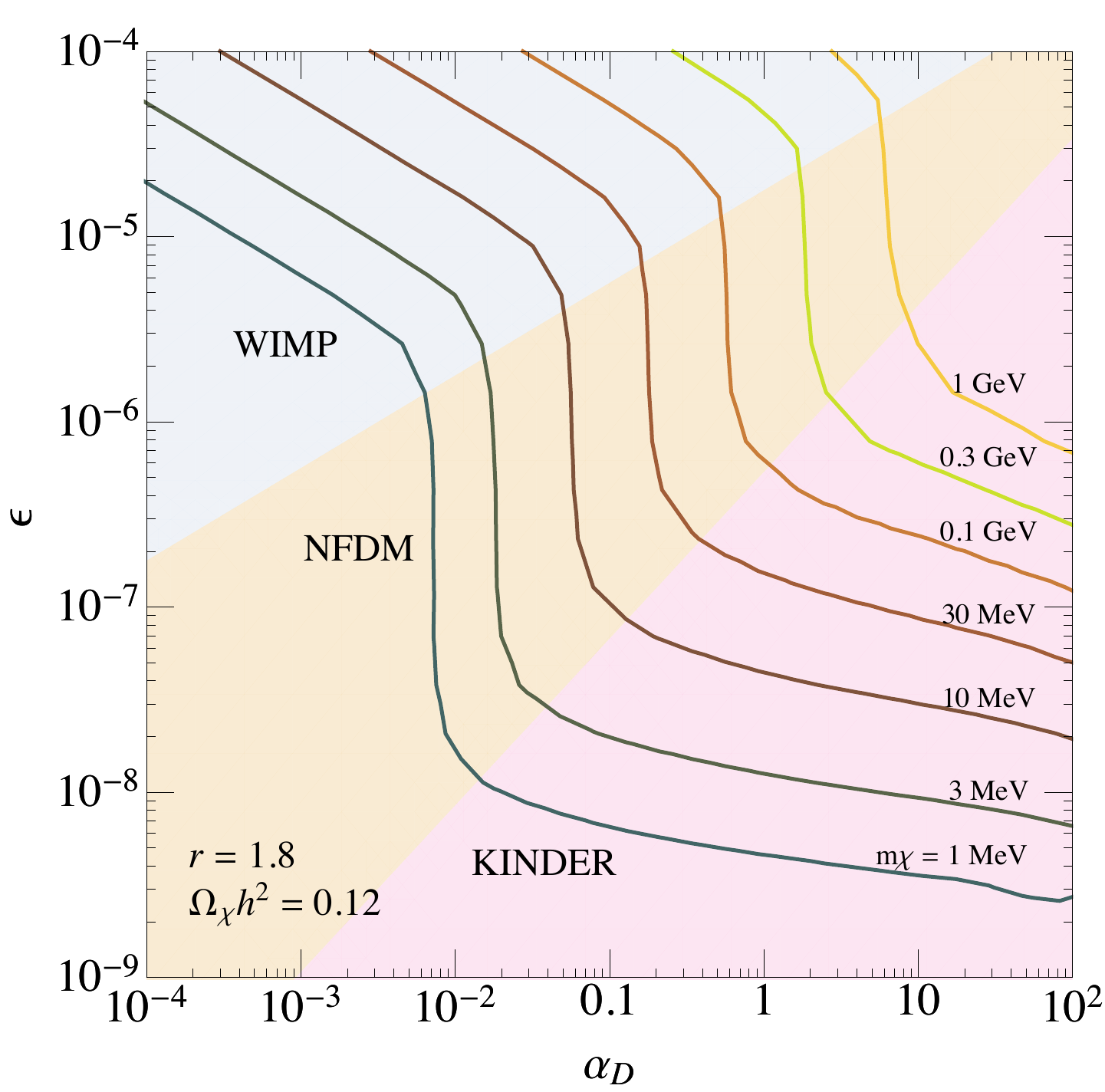}
    \caption{Contours of the observed relic abundance ($\Omega_\chi h^2 = 0.12$) in the $\alpha_D$--$\epsilon$ plane for $1.5 \lesssim r \lesssim 2$, for \textit{(left)} $r = 1.6$ and \textit{(right)} $r = 1.8$, for various values of $m_\chi$. The KINDER regime (pink), ``classic NFDM'' regime (orange) and WIMP regime (blue) are indicated, with the boundaries obtained using Eq.~\eqref{eq:eps_condition_WIMP_NFDM} for the WIMP/``classic NFDM'' boundary and Eq.~\eqref{eq:eps_condition_NFDM_KINDER} for the ``classic NFDM''/KINDER boundary.} 
    \label{fig:Eps_Alpha_NFDM_Plots}
\end{figure*}

Fig.~\ref{fig:Eps_Alpha_NFDM_Plots} shows the $\epsilon$--$\alpha_D$ parameter space of this model with $1.5 \lesssim r \lesssim 2$, with contours at fixed values of $m_\chi$ indicating the values of $\epsilon$ and $\alpha_D$ for each $m_\chi$ at which the observed relic abundance of $\Omega_\chi h^2 = 0.12$ is obtained. We show $r = 1.6$ and $r = 1.8$ as two examples for this range of $r$ values. The three different regimes that we have discussed in this section --- the WIMP, ``classic NFDM'' and KINDER regimes --- are shown in this parameter space, with the boundaries between the regimes given by $\epsilon_\text{N/W}$ defined in Eq.~\eqref{eq:eps_condition_WIMP_NFDM} between the WIMP and ``classic NFDM'' regimes, and by $\epsilon_\text{K/N}$ defined in Eq.~\eqref{eq:eps_condition_NFDM_KINDER_relic_abundance} between the ``classic NFDM'' and KINDER regimes.

For large $\epsilon$ values above $\epsilon_\text{N/W}$, the contours follow a constant value of $\epsilon^2 \alpha_D$, the parameter combination that appears in the expression for $\langle \sigma v \rangle_{\chi \overline{\chi} \to f \overline{f}}$; this corresponds to the WIMP regime. 

Below $\epsilon_\text{N/W}$, the freezeout of the dark sector transitions into the `classic NFDM' regime, with the dark sector remaining in thermal contact up till the point of freezeout, and with the abundance controlled solely by when the $3 \leftrightarrow 2$ process freezes out.  Consequently --- as previously discussed in Sec.~\ref{subsec:classic_NFDM} --- the correct relic abundance does not depend on $\epsilon$ and is only determined by the value of $\alpha_D$, leading to vertical contours. 

For yet smaller values of $\epsilon$, we eventually encounter the NFDM-KINDER boundary $\epsilon_\text{K/N}$. Within the KINDER regime, the dark matter abundance is determined by the kinetic decoupling process; over much of the parameter space this process is controlled by $A' \leftrightarrow f \overline{f}$, which \emph{only} depends on $\epsilon$, leading to roughly horizontal contours of approximately constant $\epsilon$. At larger values of $m_\chi$, the elastic scattering process (which depends on $\alpha_D$) becomes more important, and starts to play a bigger role in determining when kinetic decoupling occurs.

\section{\texorpdfstring{$1 \lesssim r \lesssim 1.5$}{1 ~< r ~< 1.5}}
\label{sec:FDM_region}

We will now focus on the behavior of the dark sector when $1 \lesssim r \lesssim 1.5$. For these values of $r$, the $2 \leftrightarrow 2$ process freezes out after the $3 \leftrightarrow 2$ process, leading to qualitatively different behavior in the dark sector. Solving the full Boltzmann equations given in Eqs.~\eqref{eq:Boltz_chi}--\eqref{eq:Boltz_rho} reveals a rich and complicated picture, with both the freezeout of DM and the temperature of the dark sector showing drastically different behavior depending on the parameter values. 

For $\epsilon \gtrsim 10^{-4}$, the dark sector is once again in the WIMP regime, and freezes out via $\chi \overline{\chi} \to f \overline{f}$. For smaller values of $\epsilon$, we find four different regimes when $1 \lesssim r \lesssim 1.5$:

\begin{enumerate}

    \item \textit{Regime I: the ``classic forbidden'' scenario.} $\epsilon$ is large enough that $A' \leftrightarrow f \overline{f}$ is fast, so that $n_{A'} \simeq n_{A',0}(T)$; furthermore, $\chi f \to \chi f$ elastic scattering is sufficiently fast to ensure that the dark sector temperature is nearly equal to the SM temperature throughout the freezeout. The dark sector stays in thermodynamic phase A until the $2 \leftrightarrow 2$ process freezes out, and no dark sector number-changing processes remain. This regime is precisely the limit studied in Ref.~\cite{PhysRevLett.115.061301}.

    \item \textit{Regime II: $n_{A'} = n_{A',0}(T)$, slight cooling.} At slightly smaller values of $\epsilon$, the process $A' \leftrightarrow f \overline{f}$ is still fast enough to maintain $n_{A'} \simeq n_{A',0}(T)$. However, this condition is insufficient to keep the dark sector in thermal contact with the SM, which cools due to the net conversion of kinetic energy in $\chi$ particles into rest mass of the heavier $A'$ particles through $\chi \overline{\chi} \to A'A'$. In regime II, $\epsilon$ is large enough for the elastic scattering process, $\chi f \leftrightarrow \chi f$, to transfer some heat from the SM to the dark sector, slowing the cooling.

    \item \textit{Regime III: $n_{A'} = n_{A',0}(T)$, rapid cooling.} Going to still smaller values of $\epsilon$, the $A' \leftrightarrow f \overline{f}$ process is still fast enough to lock the number density of $A'$ to $n_{A',0}(T)$, but $\chi f \to \chi f$ is too inefficient to transfer any heat from the SM to the dark sector at any point after $3 \leftrightarrow 2$ freezeout. In this limit, the rate of cooling is independent of $\epsilon$, and the dark sector cools in a manner that only depends on $\alpha_D$.

    \item \textit{Regime IV: KINDER.} For the smallest values of $\epsilon$ that we consider, kinetic decoupling of the dark sector from the SM occurs while both the $3 \to 2$ and $2 \to 2$ processes have rates that are much faster than Hubble. This shares many of the features of the KINDER regime discussed for $1.5 \lesssim r \lesssim 2$: the dark sector first enters thermodynamic phase B with zero chemical potential and logarithmic evolution of $T'$ with respect to $T$, and then transitions to thermodynamic phase C after $3 \leftrightarrow 2$ freezeout.

\end{enumerate}

We will first discuss the broad features of how the dark sector temperature evolves for $1 \lesssim r \lesssim 1.5$, before examining each of these regimes in turn, focusing on getting some analytic intuition for them. All of our results are once again obtained by solving the Boltzmann equations, Eqs.~\eqref{eq:Boltz_chi}--~\eqref{eq:Boltz_rho}, numerically.

\subsection{Dark Sector Temperature Evolution}

In Regime I, the ``classic forbidden'' DM regime, the temperature evolution of the dark sector is trivially given by $T' = T$. For the other regimes, the $3 \leftrightarrow 2$ freezeout divides the dark sector temperature evolution into two important phases.

\subsubsection{\texorpdfstring{Temperature Evolution Before $3 \leftrightarrow 2$ Freezeout}{Temperature Evolution Before 3<->2 Freezeout}}

In Regimes II and III, while both the $2 \leftrightarrow 2$ and $3 \leftrightarrow 2$ processes are fast, the simultaneous conditions imposed on the chemical potentials shown in Eqs.~\eqref{eq:mu_relation_3_to_2} and~\eqref{eq:mu_relation_2_to_2} are satisfied only if $\mu_\chi \approx \mu_{A'} \approx 0$. At the same time, $\epsilon$ is large enough such that $n_{A'} = n_{A',0}(T)$; therefore, we must have $n_\chi = n_{\chi,0}(T)$ as well, i.e.\ $T' = T$. Prior to $3 \leftrightarrow 2$ freezeout, Regimes II and III thus stay in thermodynamic phase A. 

For the KINDER-like Regime IV during this phase, the temperature evolution is identical to the KINDER regime with $1.5 \lesssim r \lesssim 2$, with $T' = T$ prior to kinetic decoupling, and the dark sector entering thermodynamic phase B once decoupling occurs. While in thermodynamic phase B, the dark sector particles have zero chemical potential, and the temperature evolves as in Eq.~\eqref{eq:Tp_KINDER_thermo_phase_B}. 

\subsubsection{\texorpdfstring{Temperature Evolution After $3 \leftrightarrow 2$ Freezeout}{Temperature Evolution After 3<->2 Freezeout}}

Once the $3 \leftrightarrow 2$ process freezes out, the only process which depletes $\chi$ particles is  $\chi \overline{\chi} \to A'A'$. This process converts lighter $\chi$ particles into heavier $A'$ particles, removing kinetic energy from the dark sector, resulting in a cooling of the dark sector. The $2 \leftrightarrow 2$ process enforces $\mu_\chi \approx \mu_{A'}$, which start to take on nonzero values. 

As we derived in Sec.~\ref{subsec:boltzmann_equations}, the Boltzmann equations enforce certain relations between the rates of the $3 \leftrightarrow 2$ process, the $2 \leftrightarrow 2$ process, $A' \leftrightarrow f \overline{f}$ and elastic scattering in the nonrelativistic limit. As shown in Eq.~\eqref{eq:n_chi_evolution}, we can approximately express the number density evolution of $\chi$ particles purely in terms of the elastic scattering rate and the $3 \leftrightarrow 2$ rate. In Regime III, the number density evolution between $3 \leftrightarrow 2$ and $2 \leftrightarrow 2$ freezeout is dominated solely by the $3 \leftrightarrow 2$ rate, with the elastic scattering term being negligible. Since the $3 \to 2$ rate has dropped below the Hubble rate in this phase, Regime III is characterized by $n_\chi a^3$ being approximately constant, with the dark sector temperature being dependent only on the $3 \to 2$ rate. In Regime II, the number density evolution is instead dominated by the elastic scattering rate before $2 \leftrightarrow 2$ freezeout, leading to more rapid evolution of $n_\chi$, and less deviation of $T'$ from the SM temperature. In the limit of large elastic scattering, $n_\chi \to n_{\chi,0}(T)$ with $T' \to T$, which is the condition found in Regime I. 

To understand the behavior of Regimes II and III more quantitatively, we can expand $\dot{n}_\chi$ in Eq.~\eqref{eq:n_chi_evolution} using Eq.~\eqref{eq:n_chi_dot} to obtain
\begin{multline}
    \left[\frac{3}{2T'} + \frac{m_\chi}{T'^2} + \frac{d}{dT'} \left(\frac{\mu_\chi}{T'}\right)\right] \frac{dT'}{dT}  \\
    \simeq \frac{3}{T} - \frac{r}{8(1-r) HT} \langle \sigma v^2 \rangle \left[n_\chi^2 - \frac{n_{\chi,0}^2}{n_{A',0}} n_{A'} \right] \\ 
    + \frac{n_f}{(1-r)m_\chi H T} \langle \sigma v \delta E \rangle_{\chi f \to \chi f} \,.
    \label{eq:fdm_temperature_evolution}
\end{multline}

In Regimes II and III, approximations for $\mu_\chi/T'$ after $3 \to 2$ freezeout can be found. In these regimes, the value of $\epsilon$ is large enough such that 
\begin{alignat}{1}
    n_{A'} \approx n_{A',0}(T) = 3 \left(\frac{rm_\chi T}{2 \pi}\right)^{3/2} e^{-r m_\chi/T} \,.
\end{alignat}
We emphasize, however, that the dark sector temperature $T'$ is \textit{not} equal to $T$; rather, the chemical potential $\mu_{A'}$ evolves in such a way as to maintain the relation above. The $\chi \overline{\chi} \to A' A'$ process removes kinetic energy from the dark sector, and the exact evolution of $T'$ depends on the efficiency of the heat exchange processes between the dark sector and the SM. Writing out the full expression for $n_{A'}$ in Eq.~\eqref{eq:n_Ap} and making use of the fact that while the $2 \to 2$ process is the only process that is fast, Eq.~\eqref{eq:mu_relation_2_to_2} must hold i.e.\ $\mu_\chi \approx \mu_{A'}$, we find that the chemical potential must satisfy the following relation:
\begin{alignat}{1}
    e^{\mu_\chi/T'} \approx e^{\mu_{A'}/T'} \approx \left(\frac{x'}{x}\right)^{3/2} e^{-r(x - x')} \,.
    \label{eq:phase_II_chemical_potential}
\end{alignat}
Furthermore, the ratio of $n_\chi$ and $n_{A'}$ is completely specified by $x'$ since the chemical potentials cancel out: 
\begin{alignat}{1}
    \frac{n_\chi}{n_{A'}} \approx \frac{4}{3} r^{-3/2} e^{(r-1)x'} \,.
    \label{eq:phase_II_n_chi_over_n_Ap}
\end{alignat}
Eqs.~\eqref{eq:phase_II_chemical_potential} and~\eqref{eq:phase_II_n_chi_over_n_Ap} show that given $T'$ as a function of $T$, we will be able to obtain $n_\chi$ and $n_{A'}$ as a function of the SM temperature in Regimes II and III. Eq.~\eqref{eq:phase_II_chemical_potential} provides an expression for $\mu_\chi/T'$, which combined with Eq.~\eqref{eq:fdm_temperature_evolution} gives an expression for $T'$ as a function of $T$ after the freezeout of the $3 \leftrightarrow 2$ process, with $n_{A'}(T') \approx n_{A',0}(T)$: 
\begin{multline}
    (1-r) \frac{m_\chi}{T'^2}  \frac{dT'}{dT} \approx - \frac{r m_\chi}{T^2} + \frac{3}{2T}\\
    - \frac{r}{8(1-r)HT} \langle \sigma v^2 \rangle \left[ n_\chi^2  - \frac{n_{\chi,0}^2}{n_{A',0}} n_{A'}\right]\\
    + \frac{n_f}{(1-r)m_\chi H T} \langle \sigma v \delta E \rangle_{\chi f \to \chi f} \,.
\end{multline}
If we make the further approximation that $m_\chi \gg T,T'$, this equation takes a particularly simple form, 
\begin{multline}
    \frac{dT'}{dT} \approx \frac{r}{r-1} \frac{T'^2}{T^2} - \frac{3 T'^2}{2(r-1) m_\chi T} \\
    - \frac{r T'^2}{8(r-1)^2 H m_\chi T} \langle \sigma v^2 \rangle \left[n_\chi^2 - \frac{n_{\chi,0}^2}{n_{A',0}} n_{A'} \right] \\
    + \frac{n_f T'^2}{(r-1)^2 H T m_\chi^2} \langle \sigma v \delta E \rangle_{\chi f \to \chi f} \,. 
    \label{eq:fdm_phaseII_dTprime_dT}
\end{multline}
We note that the second term on the right-hand side is typically smaller than the term before it since $T \ll m_\chi$, but has been included to improve the accuracy of this analytic result. In terms of $x'$ and $x$, we have
\begin{multline}
    \frac{dx'}{dx} \approx \frac{r}{r-1} - \frac{3}{2(r-1)x} \\
    - \frac{r}{8(r-1)^2 H x} \langle \sigma v^2 \rangle \left[n_\chi^2 - \frac{n_{\chi,0}^2}{n_{A',0}} n_{A'} \right] \\
    + \frac{n_f}{(r-1)^2 m_\chi H x} \langle \sigma v \delta E \rangle_{\chi f \to \chi f} \,.
    \label{eq:fdm_phaseII_dxprime_dx}
\end{multline}

The relative importance of each term on the right-hand side of Eq.~\eqref{eq:fdm_phaseII_dTprime_dT}, which governs the temperature evolution after $3 \leftrightarrow 2$ freezeout, separates Regimes I--III. Since the $3 \leftrightarrow 2$ term is typically less than $\mathcal{O}(1)$, the different regimes are distinguished by how large the elastic scattering term is compared to $r/(r-1)$. In Regime I, throughout the period between $3 \leftrightarrow 2$ freezeout and $2 \leftrightarrow 2$ freezeout, we have
    \begin{alignat}{1}
        n_f \langle \sigma v \delta E \rangle_{\chi f \to \chi f} \gg r(r-1) \frac{H m_\chi^2}{T^2} (T' - T) \quad \text{(Regime I)}\,,
        \label{eq:regime_I_elastic_scattering_condition}
    \end{alignat}
keeping in mind that $n_f \langle \sigma v \delta E \rangle_{\chi f \to \chi f} \propto (T' - T)$ (see Eq.~\eqref{eq:elastic_scattering_rate_full} for an expression for $\langle \sigma v \delta E \rangle_{\chi f \to \chi f}$). The fast elastic scattering enforces $T' \simeq T$, the assumption of the ``classic forbidden'' regime. In Regime II, we have instead
\begin{alignat}{1}
    n_f \langle \sigma v \delta E \rangle_{\chi f \to \chi f} \sim r(r-1) \frac{H m_\chi^2}{T^2} (T' - T)\quad \text{(Regime II)}\,
    \label{eq:regime_II_elastic_scattering_condition}
\end{alignat}
at some point between the two dark sector freezeout events. In this regime, since $n_f \langle \sigma v \delta E \rangle_{\chi f \to \chi f} \propto (T' - T)$, the dark sector begins to cool immediately after $3 \to 2$ freezeout, but once $T'$ starts differing significantly from $T$, the elastic scattering term becomes large enough to slow the cooling process. 

Finally, in Regime III, between the $3 \leftrightarrow 2$ and $2 \leftrightarrow 2$ freezeout events, we always have
\begin{alignat}{1}
    n_f \langle \sigma v \delta E \rangle_{\chi f \to \chi f}\ll r(r-1) \frac{H m_\chi^2}{T^2} (T' - T) \,\, \text{(Regime III)}\,,
    \label{eq:regime_III_elastic_scattering_condition}
\end{alignat}
This is the limit where the elastic scattering process is too inefficient to transfer heat between the two sectors, and therefore the dark sector cooling is rapid and becomes independent of $\epsilon$.

\subsection{Regime Boundaries and Characteristics}
\label{subsec:FDM_region_regimes}

\begin{figure}
    \includegraphics[width=0.47\textwidth]{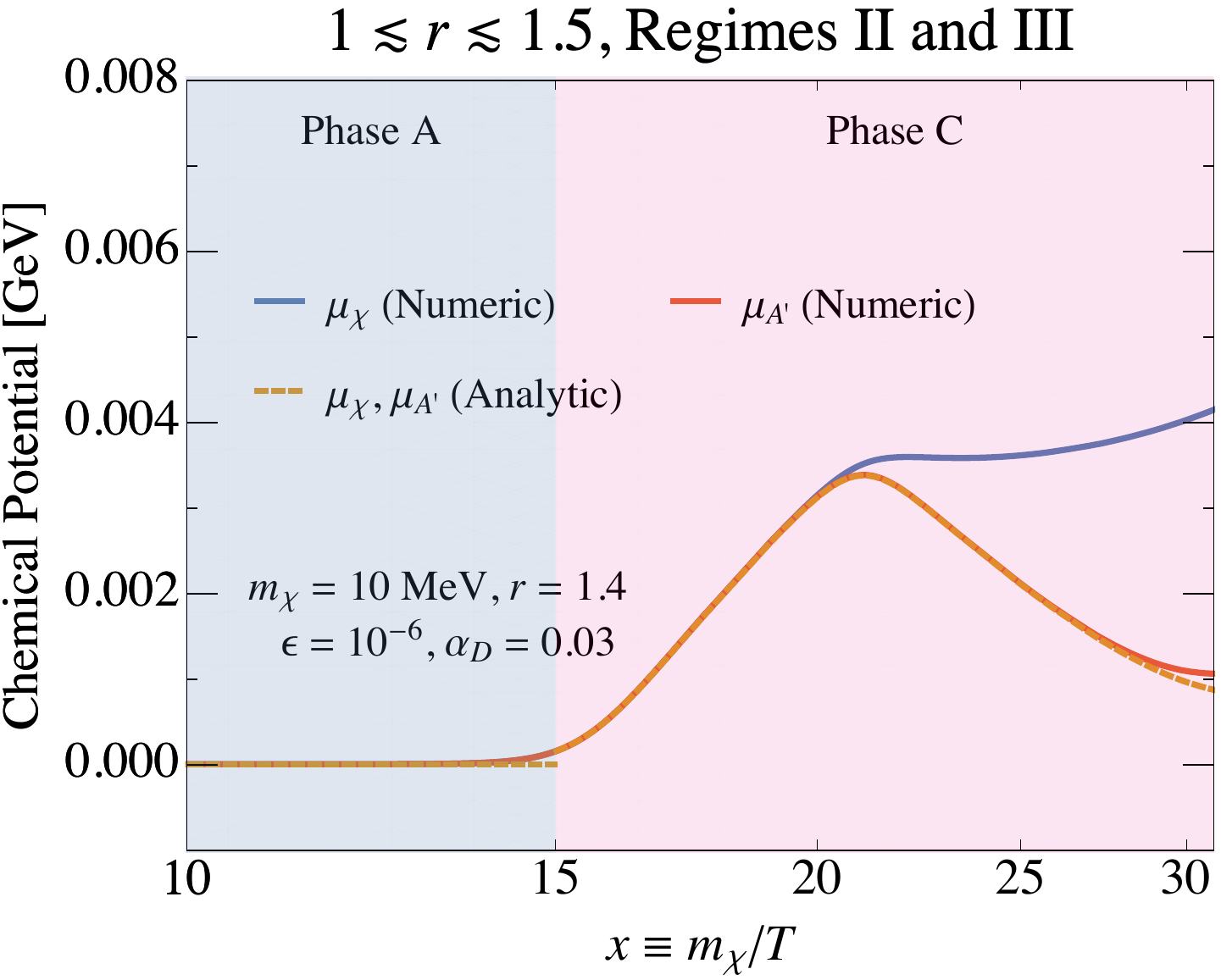}
    \caption{Chemical potential in the dark sector in Regimes II and III. Here, we choose a set of benchmark parameters ($m_\chi = \SI{10}{\mega\eV}$, $r = 1.4$, $\epsilon = 10^{-6}$, $\alpha_D = 0.03$) within Regime II, but a similar result is obtained in Regime III as well. The numerically computed chemical potentials of $\chi$ (blue line) and $A'$ (red line) are shown, together with the analytic result from Eq.~\eqref{eq:phase_II_chemical_potential} (orange dashed line).}
    \label{fig:mu_FDM_Regime_II}
\end{figure}

We will now describe some general characteristics of each regime, providing where we can an analytic description of the dark sector freezeout process. We also explain how to numerically estimate the value of $\epsilon$ on the $\alpha_D$--$\epsilon$ plane at which the boundary between the regimes is located. 

\subsubsection{Regime I}

For $\epsilon \gtrsim 10^{-4}$, freezeout of the dark sector is controlled by $\chi \overline{\chi} \leftrightarrow f \overline{f}$, corresponding to the conventional WIMP regime. For values of $\epsilon$ smaller than this, we enter regime I, the ``classic forbidden'' regime, with $T' \approx T$ until the final freezeout of the dark sector. This regime was studied in Ref.~\cite{PhysRevLett.115.061301}, where they showed that the dark sector freezeout is determined entirely by when the $2 \leftrightarrow 2$ freezeout occurs, a purely dark sector process which is independent of $\epsilon$. 

The ``classic forbidden''-WIMP boundary occurs when the $2 \leftrightarrow 2$ dark sector process freezes out and approximately the same time as $\chi \overline{\chi} \to f \overline{f}$, i.e.\
\begin{alignat}{1}
    n_\chi \langle \sigma v \rangle_{\chi \overline{\chi} \to A'A'} \approx H \approx n_\chi \langle \sigma v \rangle_{\chi \overline{\chi} \to f \overline{f}} \quad \text{(WIMP/I)} \,.
\end{alignat}
If we further require the freezeout to produce the observed relic abundance and fulfil Eq.~\eqref{eq:relic_abundance_requirement}, we obtain the following analytic estimate for $\epsilon_{\text{WIMP/I}}$, the value of $\epsilon$ at the WIMP/Regime I boundary, and specializing to $r = 1.4$ for illustration:
\begin{multline}
    \epsilon_{\text{WIMP/I}} \sim 4 \times 10^{-4} \left( \frac{\alpha_D}{1.0} \right)^{3/14}  \\
    \times \left( \frac{g_*(x_f)}{10.75} \right)^{1/14} \left( \frac{20}{x_f} \right)^{3/7} \left( \frac{g_{*,s}(x_f)}{10.75} \right)^{1/7} \,,
\end{multline}
where $x_f \sim 20$ gives the temperature of freezeout of both the $2 \leftrightarrow 2$ and the $\chi \overline{\chi} \leftrightarrow f \overline{f}$ processes.

At the low-$\epsilon$ end of Regime I, the elastic scattering energy transfer rate becomes gradually small enough such that Eq.~\eqref{eq:regime_I_elastic_scattering_condition} is no longer satisfied at all points between $3 \leftrightarrow 2$ freezeout and $2 \leftrightarrow 2$ freezeout, and the dark sector transitions into Regime II. The boundary between Regimes I and II is therefore marked by when the elastic scattering condition for Regime II, Eq.~\eqref{eq:regime_II_elastic_scattering_condition}, becomes fulfilled just as $2 \to 2$ freezeout occurs, i.e.\ 
\begin{alignat}{1}
    n_f \langle \sigma v \delta E \rangle_{\chi f \to \chi f} &\approx r(r-1) \frac{H(T_2) m_\chi^2}{T_2^2} (T_2' - T_2) \,, \nonumber \\ 
    n_\chi \langle \sigma v \rangle_{\chi \overline{\chi} \to A' A'} &\approx H(T_2) \qquad\qquad\qquad\qquad\quad \text{(I/II)} \,,
    \label{eq:fdm_regime_I_II_boundary}
\end{alignat}
where $T_2$ and $T_2'$ are the SM and dark sector temperatures at $2 \to 2$ freezeout. Note that both $\langle \sigma v \delta E \rangle_{\chi f \to \chi f}$ and $\langle \sigma v \rangle_{\chi \overline{\chi} \to A'A'}$ depend on $T$ and $T'$. Together with Eq.~\eqref{eq:relic_abundance_requirement} for the relic abundance, we can obtain a numerical estimate for $\epsilon_\text{I/II}$, the value of $\epsilon$ as a function of $\alpha_D$ at the boundary between Regimes I and II. 

\subsubsection{Regime II}

\begin{figure*}
    \centering
    \includegraphics[width=0.47\textwidth]{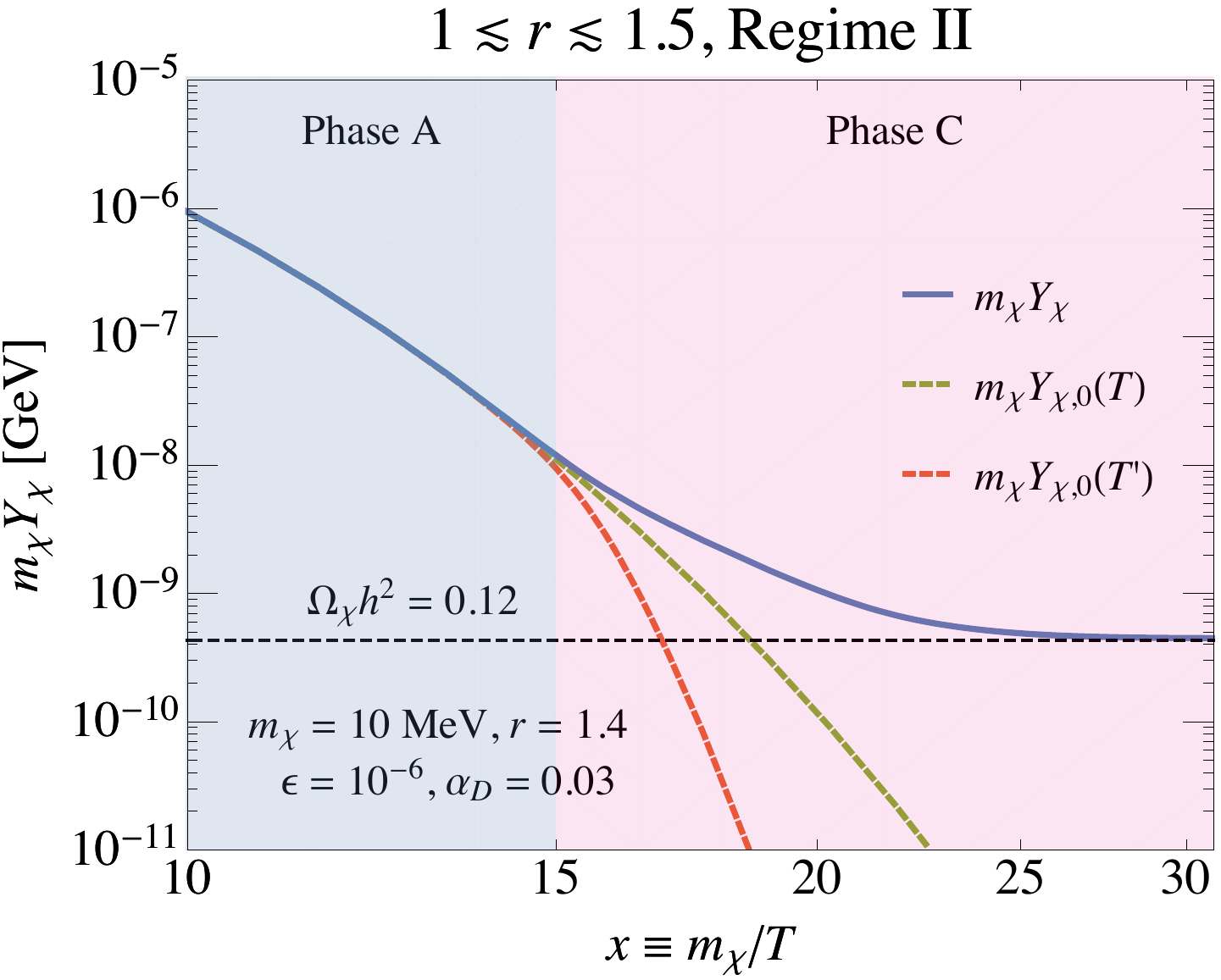}
    \qquad
    \includegraphics[width=0.47\textwidth]{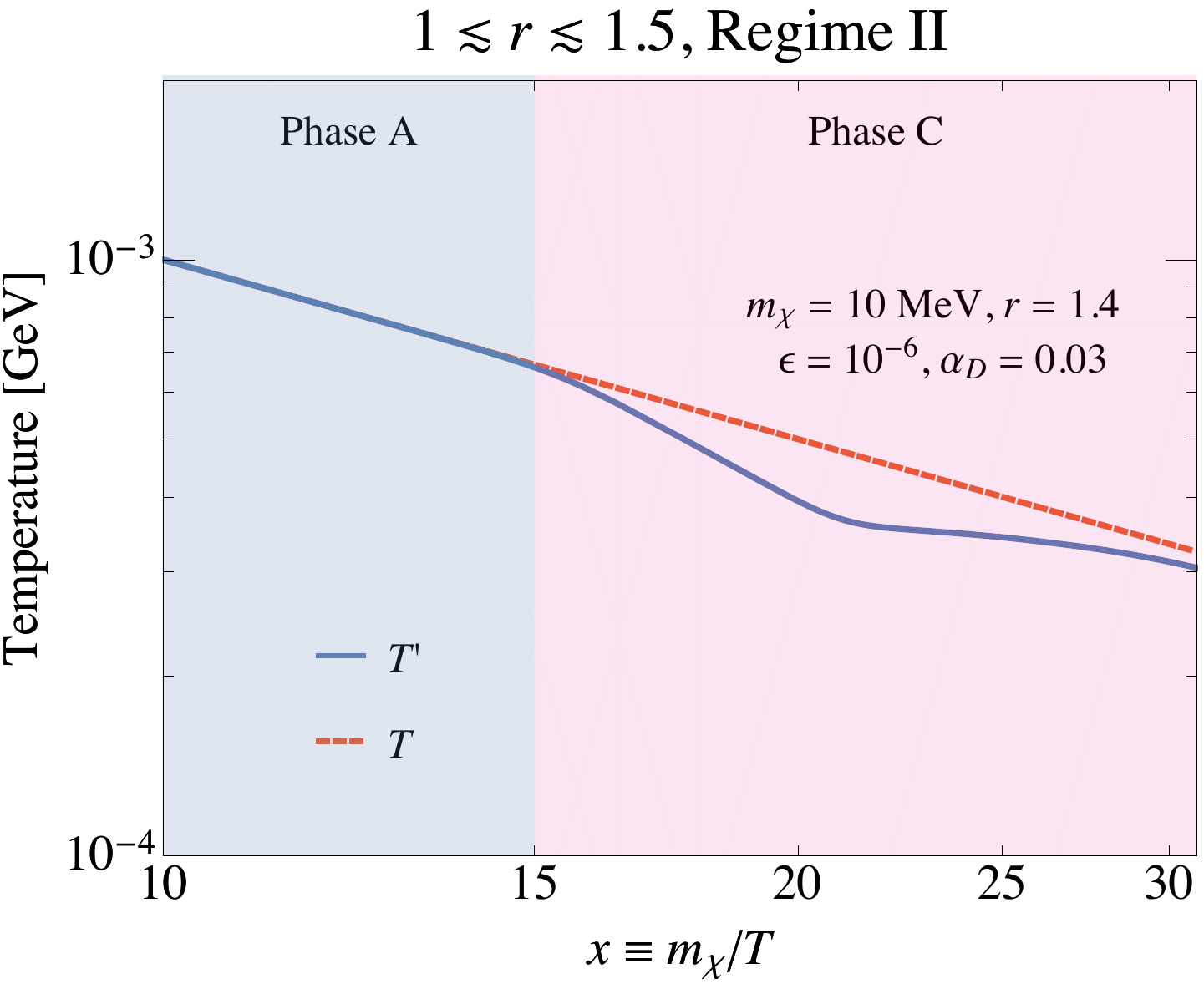}
    \caption{Dark sector evolution in Regime II for $1 \lesssim r \lesssim 1.5$, with parameters $m_\chi = \SI{10}{\mega\eV}$, $r = 1.4$, $\epsilon = 10^{-6}$ and $\alpha_D = 0.03$. In both plots, thermodynamic phases A and C as defined in Sec.~\ref{subsec:fast_reactions_and_freezeout} are shown in light blue and pink respectively. (\textit{Left}) $\chi$ abundance (given as $m_\chi Y_\chi$) as a function of $x$ (blue line), with the zero chemical potential abundance at the SM temperature $m_\chi Y_{\chi,0}(T)$ (green dashed line) and the dark sector temperature $m_\chi Y_{\chi,0}(T')$ (red dashed line) shown for reference. The observed DM abundance is indicated by the horizontal black dashed line, as defined in Eq.~\eqref{eq:relic_abundance_requirement}. (\textit{Right}) The dark sector temperature $T'$ (blue line), as a function of the SM temperature (red dashed line). The $A'$ abundance evolves trivially as $n_{A'} = n_{A',0}(T)$ in this regime.} 
    \label{fig:abundances_FDM_Regime_II}
\end{figure*}

\begin{figure*}
    \centering
    \includegraphics[width=0.47\textwidth]{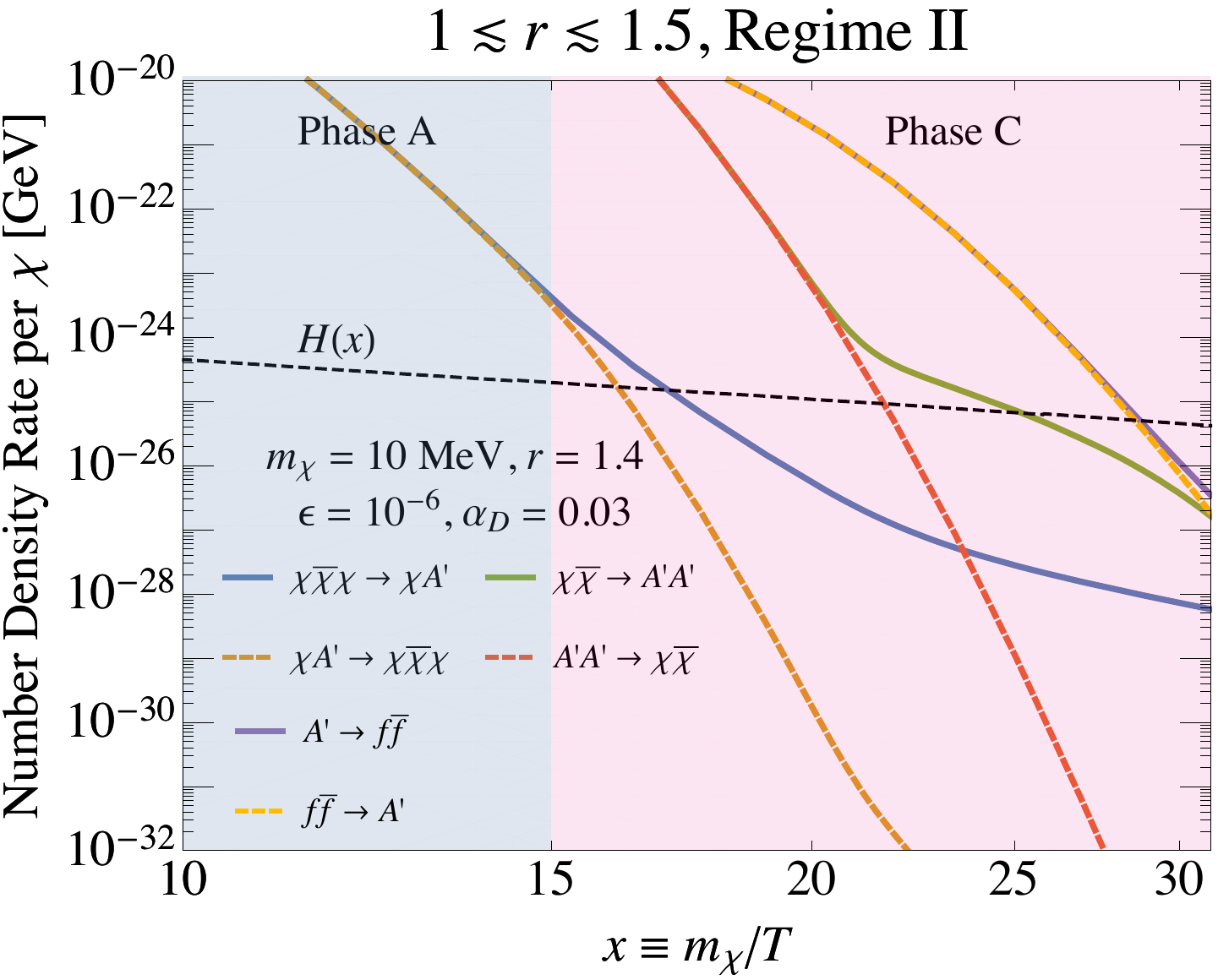}
    \qquad
    \includegraphics[width=0.47\textwidth]{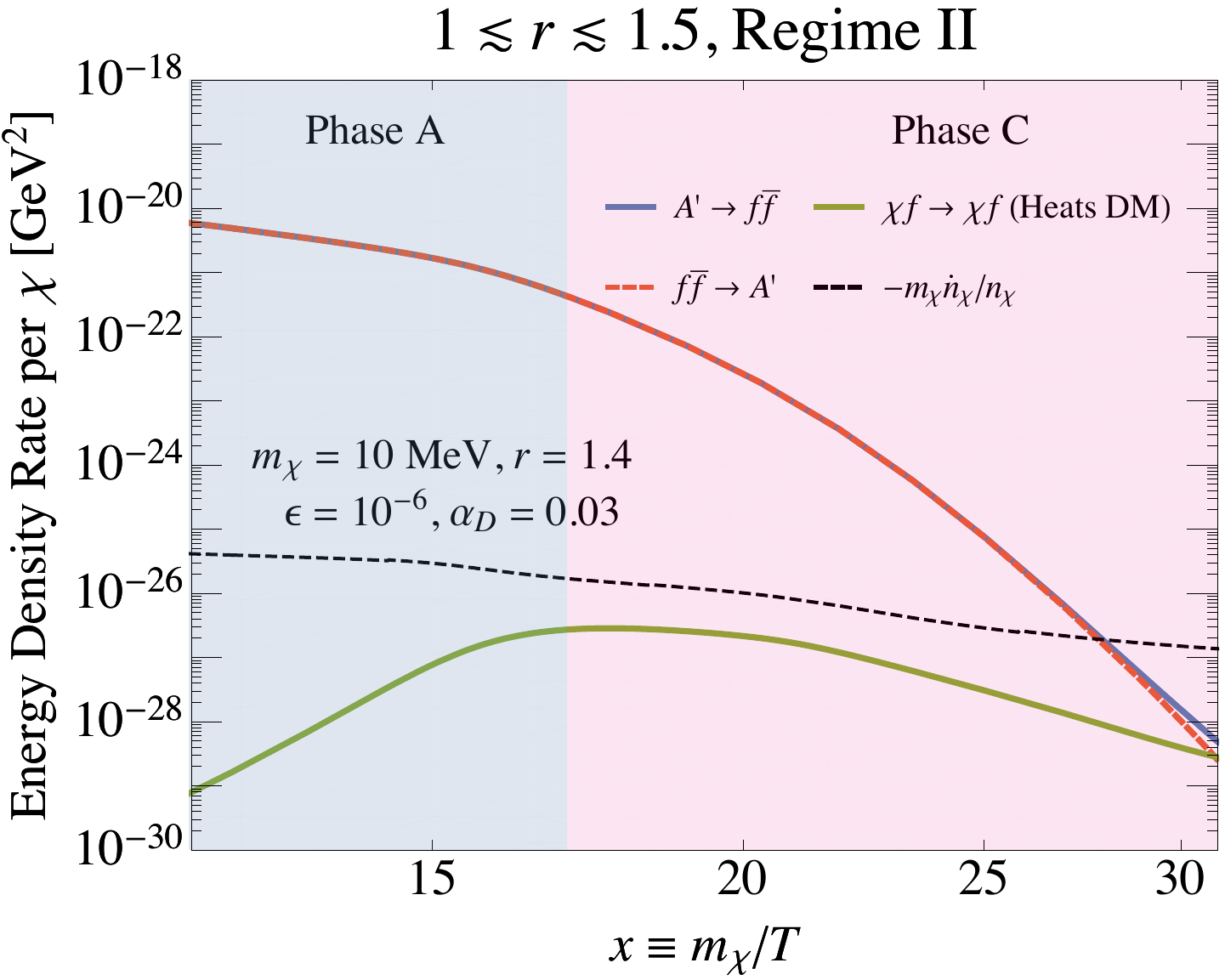}
    \caption{Rates of change in number density and energy density of the dark sector in Regime II for $1 \lesssim r \lesssim 1.5$, with parameters $m_\chi = \SI{10}{\mega\eV}$, $r = 1.4$, $\epsilon = 10^{-6}$ and $\alpha_D = 0.03$. In both plots, thermodynamic phases A and C as defined in Sec.~\ref{subsec:fast_reactions_and_freezeout} are shown in light blue and pink respectively. (\textit{Left}) $\chi$ Number density rates for $\chi \overline{\chi} \chi \to \chi A'$ (blue line), $\chi A' \to \chi \overline{\chi} \chi$ (dark orange dashed line), $\chi \overline{\chi} \to A'A'$ (green line), $A' A' \to \chi \overline{\chi}$ (red dashed line), $A' \to f \overline{f}$ (purple line) and $f \overline{f} \to A'$ (yellow dashed line) are shown. The Hubble rate is shown as a black dashed line. (\textit{Right}) Energy density rates for $A' \to f \overline{f}$ (blue line), $f \overline{f} \to A'$ (red dashed line) and $\chi f \leftrightarrow \chi f$ (green line), which has the net effect of heating the dark sector. The rate at which the energy density of DM is changing $- m_\chi \dot{n}_\chi / n_\chi$ (black dashed line) is shown for reference.} 
    \label{fig:rates_FDM_Regime_II}
\end{figure*}

Regime II is characterized by Eq.~\eqref{eq:regime_II_elastic_scattering_condition} between $3 \leftrightarrow 2$ and $2 \leftrightarrow 2$ freezeout, which ensures that $T' < T$ due to the $\chi \overline{\chi} \to A'A'$ process, but with some heat being transferred from the SM to the dark sector to impede the cooling of the dark sector due to $\chi f \to \chi f$. At the same time, the decay rate $\Gamma$ is large enough such that $n_{A'} \approx n_{A',0}(T)$ throughout the freezeout of the dark sector. This condition immediately determines the chemical potentials $\mu_\chi$, given analytically by the expression Eq.~\eqref{eq:phase_II_chemical_potential}, as well as $\mu_{A'} \approx \mu_\chi$. In Fig.~\ref{fig:mu_FDM_Regime_II}, we show this analytic result in comparison with the numeric calculation of the chemical potential, for our Regime II benchmark point of $m_\chi = \SI{10}{\mega\eV}$, $r = 1.4$, $\epsilon = 10^{-6}$ and $\alpha_D = 0.03$. Note that the agreement deteriorates rapidly once $2 \leftrightarrow 2$ freezeout occurs at $x \sim 21$, after which the DM particle has completely frozen out, and the assumption that $\mu_\chi \approx \mu_{A'}$ breaks. A similar result is obtained in Regime III as well, where $n_{A'} \approx n_{A',0}(T)$ also holds.

Fig.~\ref{fig:abundances_FDM_Regime_II} shows the evolution of the $\chi$ number density and $T'$ at the same benchmark parameters. $n_{A'}$ evolves trivially as $n_{A',0}(T)$, and therefore need not be separately plotted. Since a chemical potential develops immediately after the dark sector kinetically decouples from the SM at the point of $3 \leftrightarrow 2$ freezeout at $x \sim 15$, the dark sector passes from thermodynamic phase A to C directly. The characteristic cooling of the dark sector is apparent in the right panel of Fig.~\ref{fig:abundances_FDM_Regime_II}, and is governed by Eq.~\eqref{eq:fdm_phaseII_dxprime_dx}. In this regime, this differential equation does not appear to be analytically integrable; we show only the numerical result, obtained directly from the full Boltzmann equations.

In Fig.~\ref{fig:rates_FDM_Regime_II}, we show the number density and energy density rates of all relevant dark sector processes. The transition between phases A and C occurs at roughly $x \sim 15$, when the backward and forward $3 \leftrightarrow 2$ rates cease to be approximately equal. This occurs when the $3 \to 2$ rate is still much larger than the Hubble rate, due to the relation between the rates of the $3 \leftrightarrow 2$ and $2 \leftrightarrow 2$ processes enforced by Eq.~\eqref{eq:2_to_2_and_3_to_2_relation}, where the $2 \leftrightarrow 2$ rate being of order $H n_\chi$ allows the $3 \leftrightarrow 2$ total rate to be much larger than the Hubble rate. Once the dark sector transitions into phase C, we see that the elastic scattering energy density rate per $\chi$ particle becomes just a factor of a few smaller than $- m_\chi \dot{n}_\chi / n_\chi$, meeting the Regime II criterion laid out in Eq.~\eqref{eq:regime_II_elastic_scattering_condition}. This shows that a significant amount of heat is transferred from the SM to the dark sector, slowing the cooling rate compared to what happens in Regime III, which we will discuss next. 

Within Regime II, as $\epsilon$ decreases still further, Eq.~\eqref{eq:regime_II_elastic_scattering_condition} is met increasingly earlier, leading to a colder dark sector due to the diminishing ability of $\chi f \to \chi f$ to heat the dark sector. Eventually, the condition Eq.~\eqref{eq:regime_II_elastic_scattering_condition} is only met at the point of $3 \to 2$ freezeout, and no significant amount of heat is transferred to the dark sector after that. This marks the boundary between Regime II and III, i.e.\
\begin{alignat}{1}
    n_f \langle \sigma v \delta E \rangle_{\chi f \to \chi f} &\approx r(r-1) \frac{H(T_3) m_\chi^2}{T_3^2}(T_3' - T_3) \,, \nonumber \\
    \frac{1}{4} n_\chi^2 \langle \sigma v^2 \rangle &\approx H(T_3) \qquad\qquad\qquad\qquad\quad \text{(II/III)} \,,
    \label{eq:fdm_boundary_regimes_II_III} 
\end{alignat}
where $T_3$ and $T_3'$ are the SM and dark sector temperatures at $3 \leftrightarrow 2$ freezeout respectively. An analytic estimate for $\epsilon_{\text{II/III}}$, the value of $\epsilon$ when these two conditions are satisfied, is
\begin{multline}
    \epsilon_{\text{II/III}} \sim 8 \times 10^{-7} \left(\frac{1.0}{\alpha_D}\right)  \left(\frac{g_*(x_f)}{10.75}\right)^{1/4} \\
    \times \left(\frac{r}{1.4}\right)^{5/2} \left(\frac{r-1}{0.4}\right)^{1/2} \left(\frac{m_\chi}{\SI{}{\giga\eV}}\right)^{1/2}  \left(\frac{x_f}{20}\right)^3 \,.
\end{multline}
Once again, combining the boundary conditions shown above with the observed relic abundance in Eq.~\eqref{eq:relic_abundance_requirement} allows us to eliminate $m_\chi$ and $x_f$ from the expression above numerically. This numerical expression for $\epsilon_\text{II/III}$ forms the boundary between Regimes II and III.

\subsubsection{Regime III}
\label{subsubsec:FDM_regime_III}

\begin{figure*}
    \centering
    \includegraphics[width=0.47\textwidth]{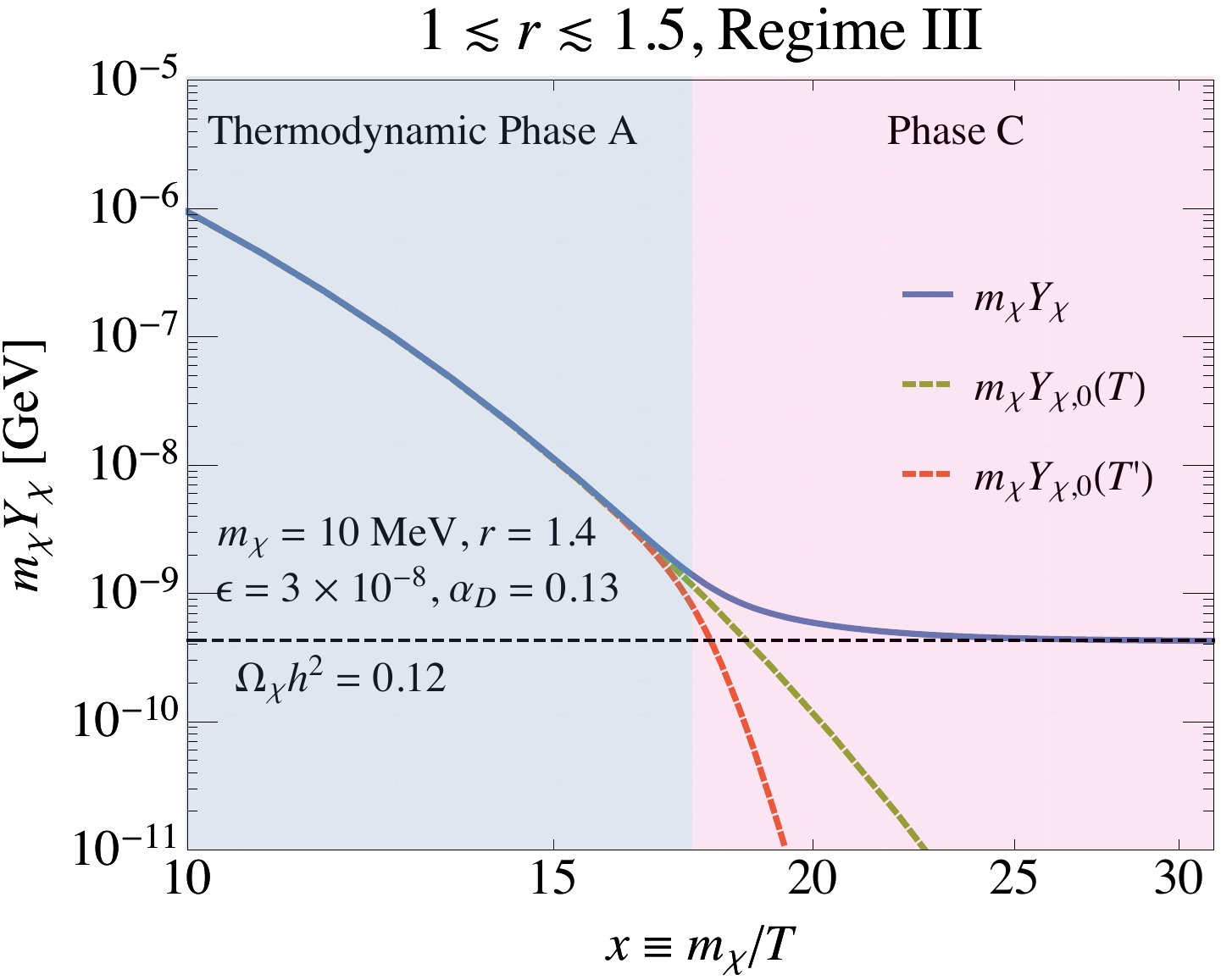}
    \qquad
    \includegraphics[width=0.47\textwidth]{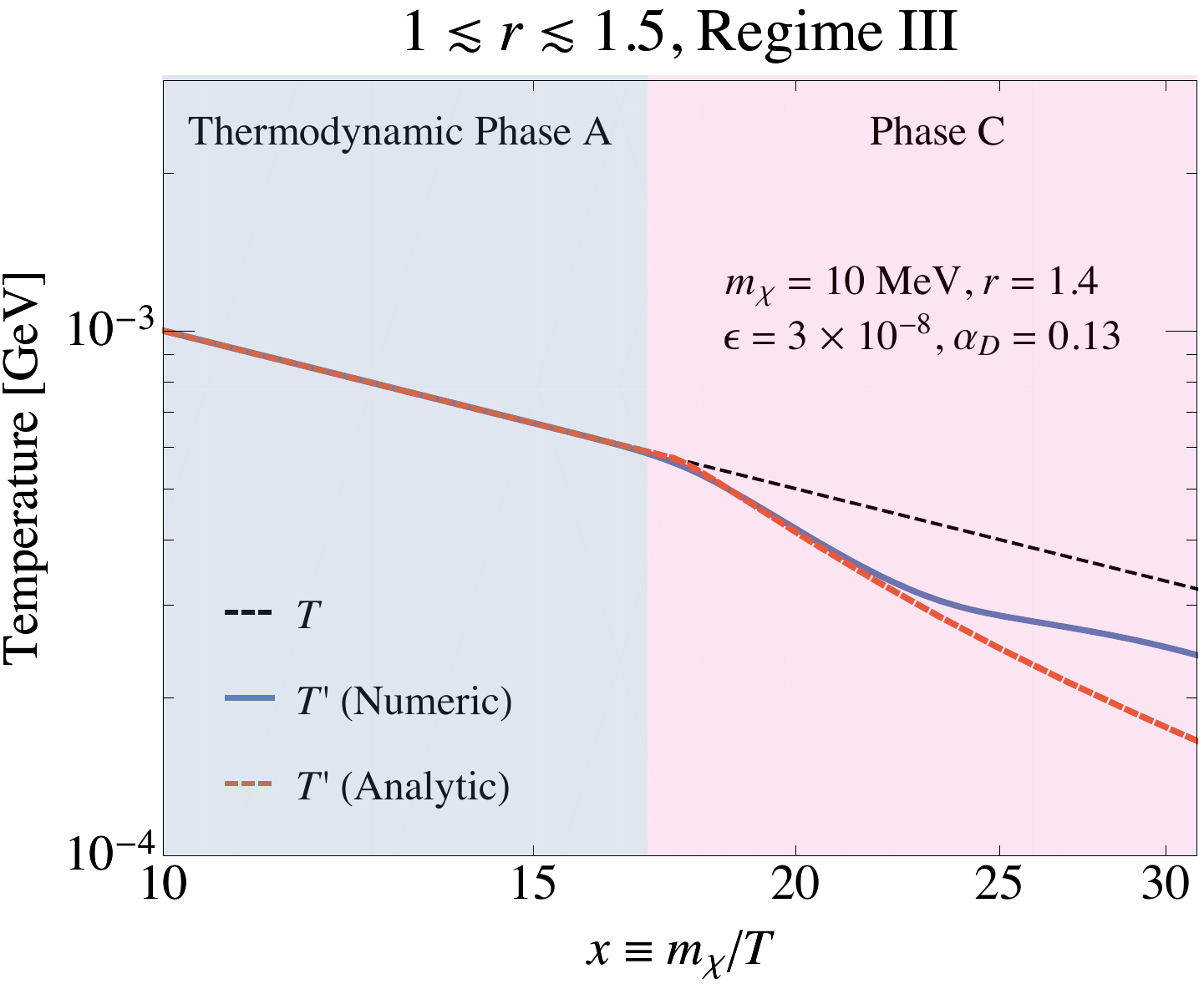}
    \caption{Dark sector evolution in Regime III for $1 \lesssim r \lesssim 1.5$, with parameters $m_\chi = \SI{10}{\mega\eV}$, $r = 1.4$, $\epsilon = 3 \times 10^{-8}$ and $\alpha_D = 0.13$. In both plots, thermodynamic phases A and C as defined in Sec.~\ref{subsec:fast_reactions_and_freezeout} are shown in light blue and pink respectively. (\textit{Left}) $\chi$ abundance (given as $m_\chi Y_\chi$) as a function of $x$ (blue line), with the zero chemical potential abundance at the SM temperature $m_\chi Y_{\chi,0}(T)$ (green dashed line) and the dark sector temperature $m_\chi Y_{\chi,0}(T')$ (red dashed line) shown for reference. The observed DM abundance ($\Omega_\chi h^2 = 0.12$) is indicated by the horizontal black dashed line, as defined in Eq.~\eqref{eq:relic_abundance_requirement}. (\textit{Right}) The dark sector temperature $T'$ (blue line), as a function of the SM temperature (black dashed line). An analytic estimate for $T'$, given in Eq.~\eqref{eq:Tp_analytic_FDM_regime_III}, is shown by the red dashed line. The $A'$ abundance evolves trivially as $n_{A'} = n_{A',0}(T)$ in this regime.} 
    \label{fig:abundance_and_Tprime_FDM_Regime_III}
\end{figure*}

\begin{figure*}
    \centering
    \includegraphics[width=0.47\textwidth]{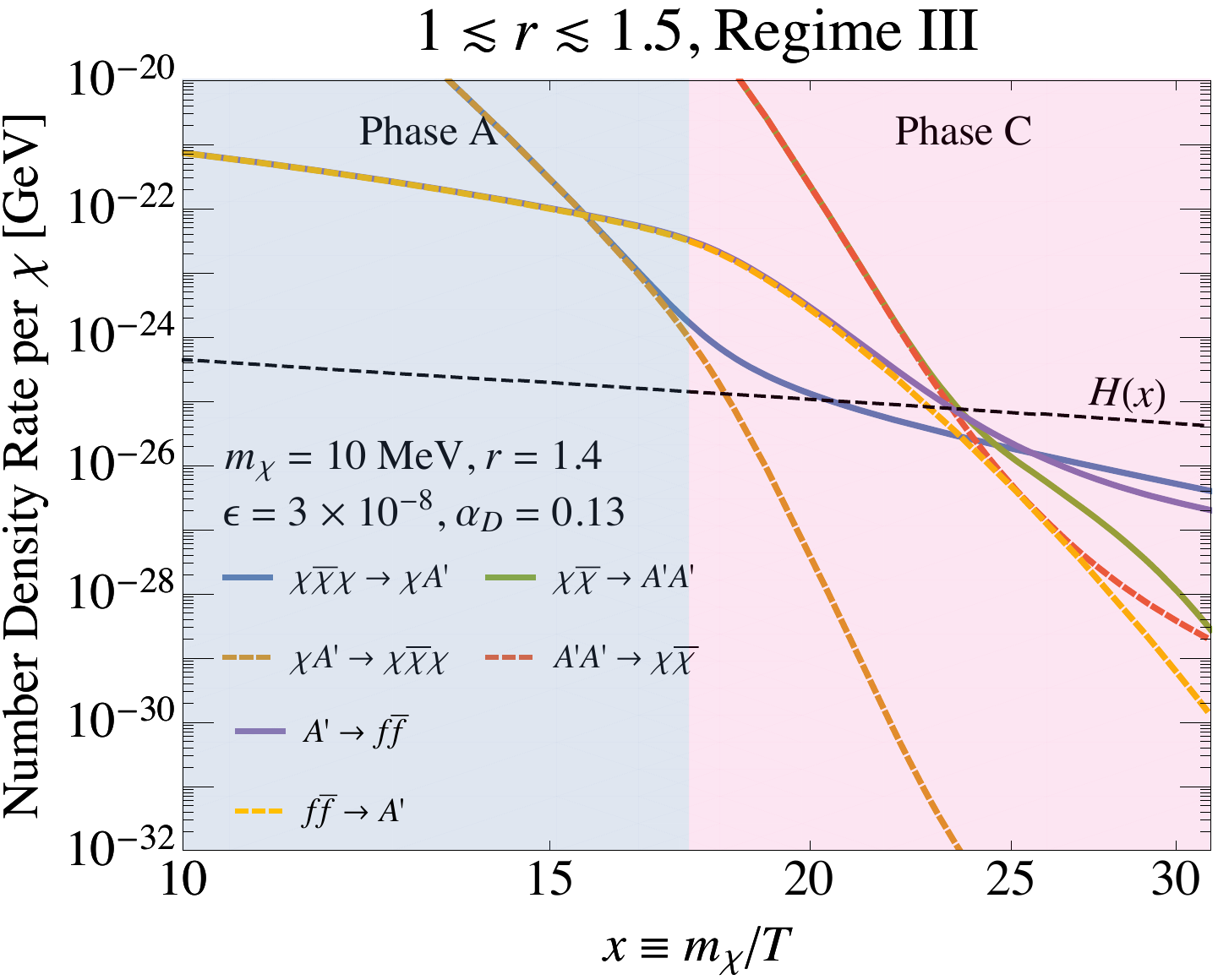}
    \qquad
    \includegraphics[width=0.47\textwidth]{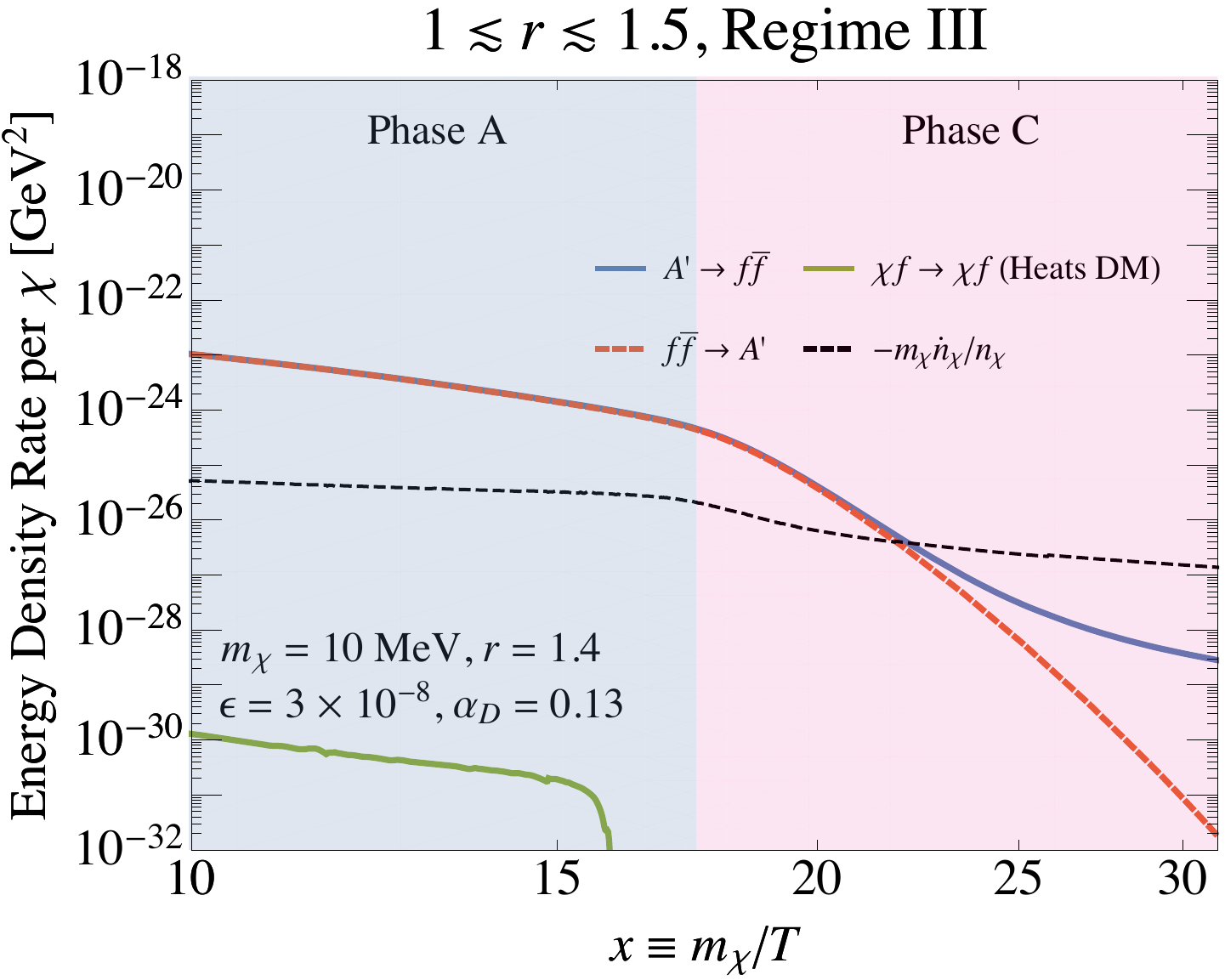}
    \caption{Rates of change in number density and energy density of the dark sector in Regime III for $1 \lesssim r \lesssim 1.5$, with parameters $m_\chi = \SI{10}{\mega\eV}$, $r = 1.4$, $\epsilon = 3 \times 10^{-8}$ and $\alpha_D = 0.13$. In both plots, thermodynamic phases A and C as defined in Sec.~\ref{subsec:fast_reactions_and_freezeout} are shown in light blue and pink respectively. (\textit{Left}) $\chi$ Number density rates for $\chi \overline{\chi} \chi \to \chi A'$ (blue line), $\chi A' \to \chi \overline{\chi} \chi$ (dark orange dashed line), $\chi \overline{\chi} \to A'A'$ (green line), $A' A' \to \chi \overline{\chi}$ (red dashed line), $A' \to f \overline{f}$ (purple line) and $f \overline{f} \to A'$ (yellow dashed line) are shown. The Hubble rate is shown as a black dashed line. (\textit{Right}) Energy density rates for $A' \to f \overline{f}$ (blue line), $f \overline{f} \to A'$ (red dashed line) and $\chi f \to \chi f$ (green line), which has the net effect of heating the dark sector. The rate at which the energy density of DM is changing $- m_\chi \dot{n}_\chi / n_\chi$ (black dashed line) is shown for reference.}
    \label{fig:rates_FDM_Regime_III}
\end{figure*}

Fig.~\ref{fig:abundance_and_Tprime_FDM_Regime_III} shows the evolution of the $\chi$-abundance and the dark sector temperature in Regime III, for our benchmark parameters in this regime, $m_\chi = \SI{10}{\mega\eV}$, $r = 1.4$, $\epsilon = 3 \times 10^{-8}$ and $\alpha_D = 0.13$. In Fig.~\ref{fig:rates_FDM_Regime_III}, we show the number density and energy density rates per $\chi$ particle through the dark sector freezeout. In this regime, the dark sector temperature once again cools rapidly after $3 \to 2$ freezeout and enters thermodynamic phase C; unlike Regime II, however, elastic scattering plays no significant role in influencing this evolution between $3 \leftrightarrow 2$ freezeout and $2 \leftrightarrow 2$ freezeout, as can be seen in the right panel of Fig.~\ref{fig:rates_FDM_Regime_III}. The dark sector temperature evolution after $3 \leftrightarrow 2$ freezeout can be obtained by setting $\langle \sigma v \delta E \rangle \to 0$ in Eq.~\eqref{eq:fdm_phaseII_dxprime_dx} and neglecting the $2 \to 3$ rate (which is much smaller than the forward rate after $3 \to 2$ freezeout), i.e.\
\begin{alignat}{1}
    \frac{dx'}{dx} \approx \frac{r}{r-1} - \frac{3}{2(r-1)x} - \frac{r}{8(r-1)^2 Hx} \langle \sigma v^2 \rangle n_\chi^2 \,.
\end{alignat}
Given the approximation for the chemical potential $\mu_\chi$ in Eq.~\eqref{eq:phase_II_chemical_potential}, this differential equation can be integrated exactly, starting from $x_3 = x_3'$, to give
\begin{multline}
    x' \approx x_3 + \frac{r}{r-1}(x - x_3) - \frac{3}{2(r-1)} \log \left(\frac{x}{x_3}\right) \\
    - \frac{1}{2(r-1)} \log \left[1 + \frac{r m_\chi^6 \langle \sigma v^2 \rangle e^{-2x_3}}{8 \pi^3(r-1) H(T_3) x_3^3} \left(1 - \frac{x_3^4}{x^4} \right) \right] \,.
    \label{eq:Tp_analytic_FDM_regime_III}
\end{multline}
In the right panel of Fig.~\ref{fig:abundance_and_Tprime_FDM_Regime_III}, we show this analytic result in comparison with the numeric result obtained from the full Boltzmann equation, and find excellent agreement between them, up to $2\leftrightarrow 2$ freezeout at $x \sim 23$. 

Throughout Regime III, $n_{A'} \approx n_{A',0}(T)$ due to the highly efficient $A' \leftrightarrow f \overline{f}$ process; as $\epsilon$ decreases, however, $A' \leftrightarrow f \overline{f}$ becomes less and less rapid, and eventually this process becomes too inefficient to keep the dark sector in thermal equilibrium at the point of $3 \leftrightarrow 2$ freezeout. Below this point, kinetic decoupling between the two sectors occurs before either of the dark sector processes freezes out, leading to the KINDER-like Regime IV. We can estimate the boundary between Regimes III and IV by requiring the $3 \to 2$ freezeout and kinetic decoupling to occur at the same time, i.e.\  
\begin{alignat}{1}
    \frac{r m_\chi \Gamma}{n_{\chi,0}(T_3)} n_{A',0}(T_3) &\approx \frac{m_\chi \dot{n}_{\chi,0}(T_3)}{n_{\chi,0}(T_3)} \,, \nonumber \\
    \frac{1}{4} n_\chi^2 \langle \sigma v^2 \rangle &\approx H(T_3) \qquad\qquad\qquad\quad \text{(III/IV)} \,.
    \label{eq:fdm_boundary_regimes_III_IV}
\end{alignat}
These conditions are however identical to the conditions used for estimating the boundary between the KINDER and the NFDM regime for $1.5 \lesssim r \lesssim 2$ in Eq.~\eqref{eq:eps_condition_NFDM_KINDER}. This equation can be restated as
\begin{multline}
    \epsilon_{\text{III/IV}} \sim 10^{-8} e^{9.6(r-1.4)} \left(\frac{\alpha_D}{1.0}\right)^{\frac{3(r-1)}{4}}  \\
    \times \left(\frac{1.4}{r}\right)^{9/4} \left(\frac{x_f}{20}\right)^{-\frac{r+3}{4}} \left(\frac{g_*(x_f)}{10.75}\right)^{-\frac{r-3}{8}} \\
    \times \left(\frac{\SI{}{\giga\eV}}{m_\chi}\right)^{\frac{r-3}{4}} \left(\frac{f(r)}{23.7}\right)^\frac{r-1}{4} \,,
    \label{eq:eps_III_IV}
\end{multline}
where $\epsilon_{\text{III/IV}}$ is the value of $\epsilon$ between Regimes III and IV as a function of various model parameters. Finally, we may once again combine Eq.~\eqref{eq:eps_III_IV} with the condition for the observed relic abundance in Eq.~\eqref{eq:relic_abundance_requirement} to numerically derive the boundary between these regimes.

\subsubsection{Regime IV}

\begin{figure*}[t!]
    \centering
    \includegraphics[width=0.47\textwidth]{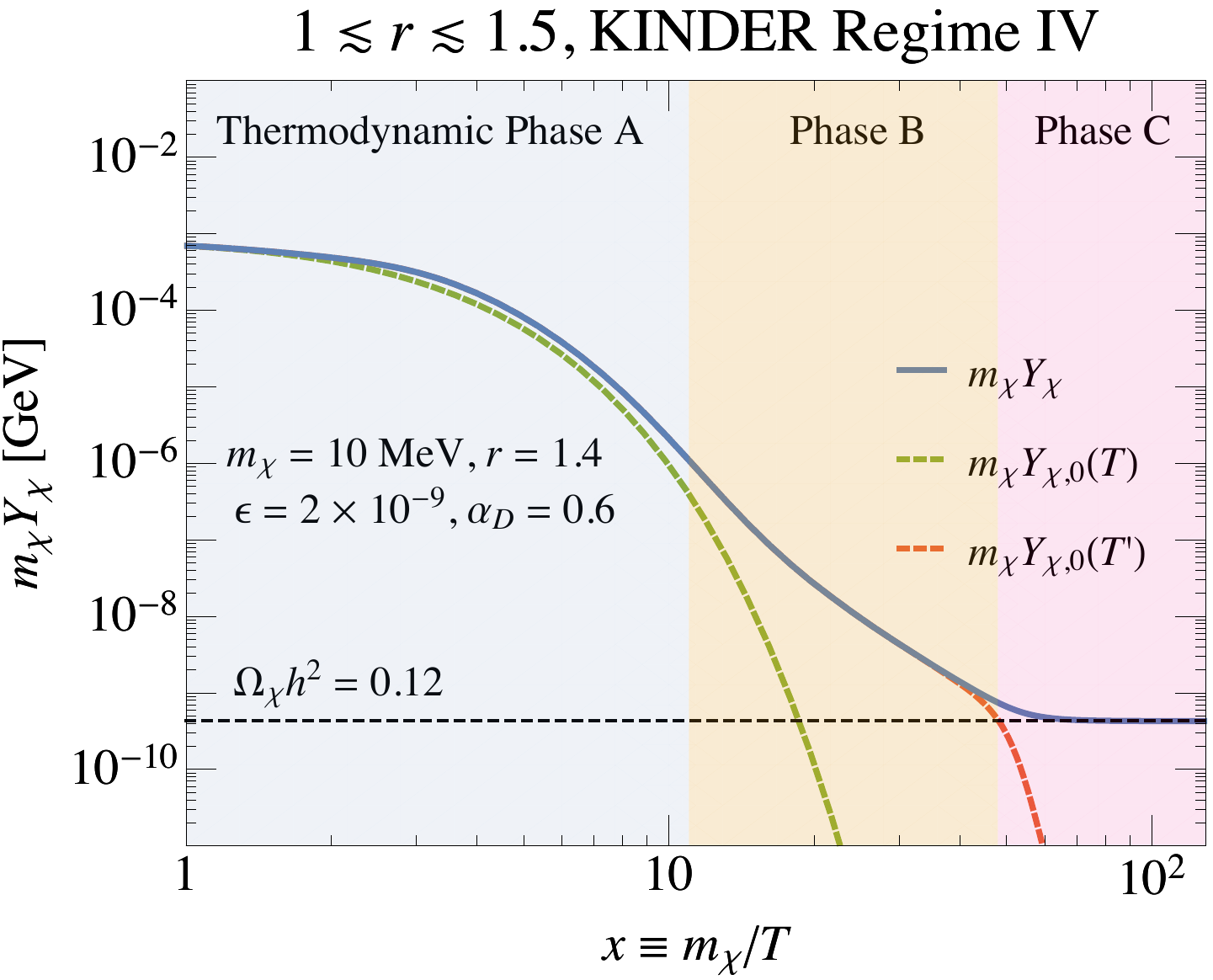}
    \qquad
    \includegraphics[width=0.47\textwidth]{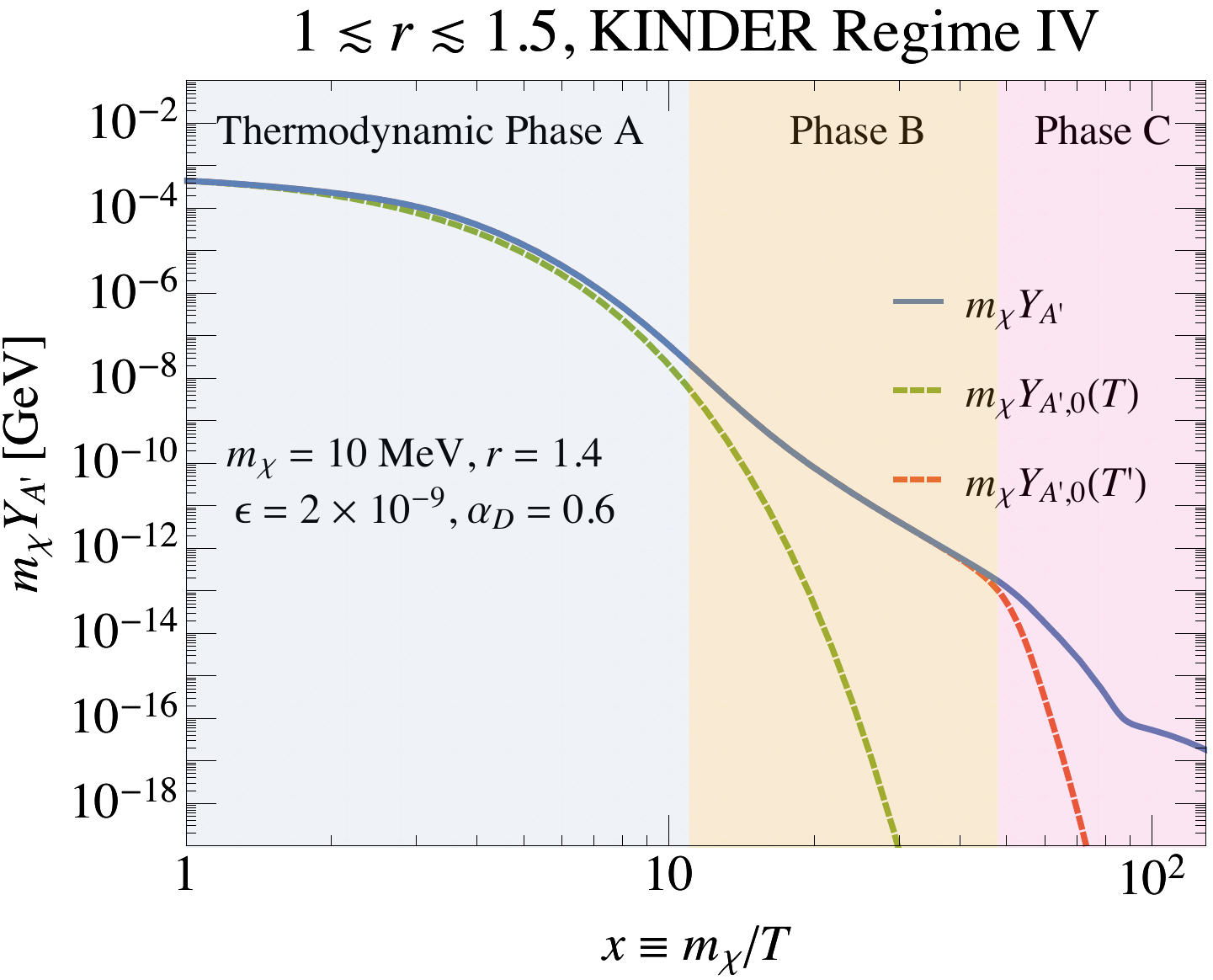}
    \includegraphics[width=0.47\textwidth]{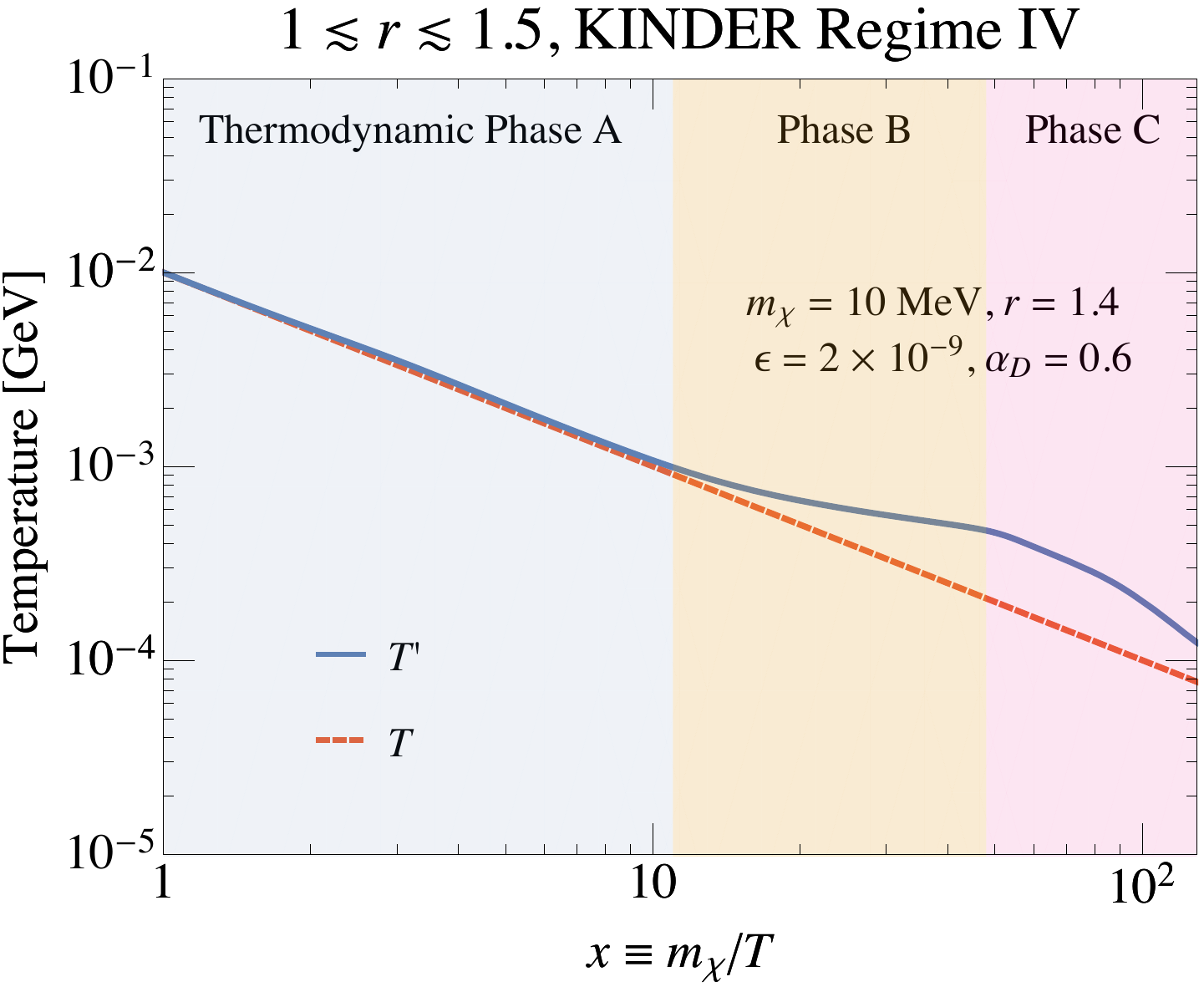}
    \caption{Dark sector evolution in regime IV for $1 \lesssim r \lesssim 1.5$, with parameters $m_\chi = \SI{10}{\mega\eV}$, $r = 1.4$, $\epsilon = 2 \times 10^{-9}$ and $\alpha_D = 0.6$. In all three plots, thermodynamic phases A, B and C as defined in Sec.~\ref{subsec:fast_reactions_and_freezeout} are shown in light blue, yellow and pink respectively. (\textit{Top left}) $\chi$ abundance (given as $m_\chi Y_\chi$) as a function of $x$ (blue line), with the zero chemical potential abundance at the SM temperature $m_\chi Y_{\chi,0}(T)$ (green dashed line) and the dark sector temperature $m_\chi Y_{\chi,0}(T')$ (red dashed line) shown for reference. The observed DM abundance is indicated by the horizontal black dashed line, as defined in Eq.~\eqref{eq:relic_abundance_requirement}. (\textit{Top right}) $A'$ abundance (given as $m_\chi Y_{A'}$) as a function of $x$ (blue line), with $Y_{A',0}(T)$ (green dashed line) and $Y_{A',0}(T')$ (red dashed line) once again given for reference. (\textit{Bottom}) The dark sector temperature $T'$ (blue line), as a function of the SM temperature (red dashed line).} 
    \label{fig:abundances_FDM_KINDER}
\end{figure*}
\begin{figure*}[t!]
    \centering
    \includegraphics[width=0.47\textwidth]{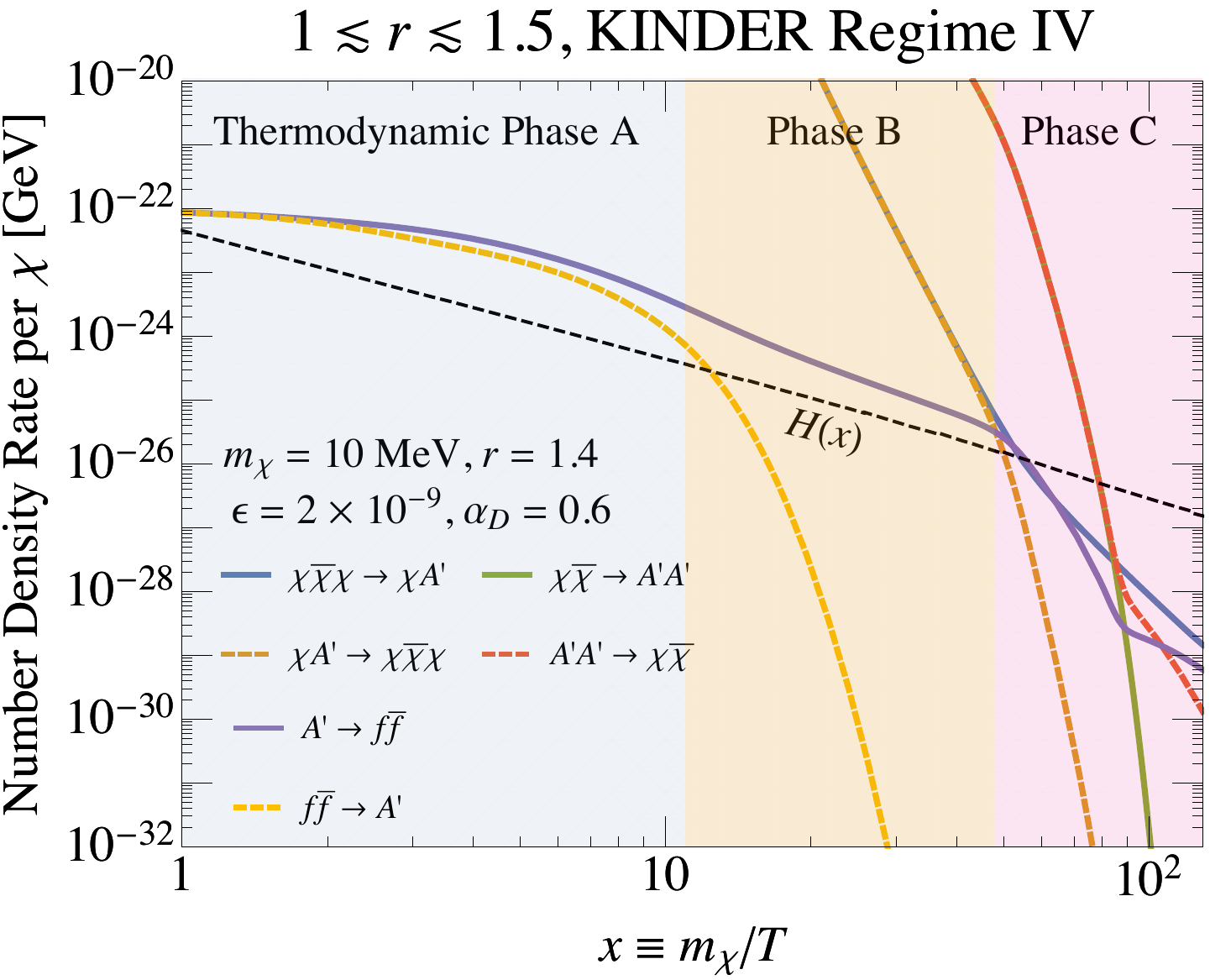}
    \qquad
    \includegraphics[width=0.47\textwidth]{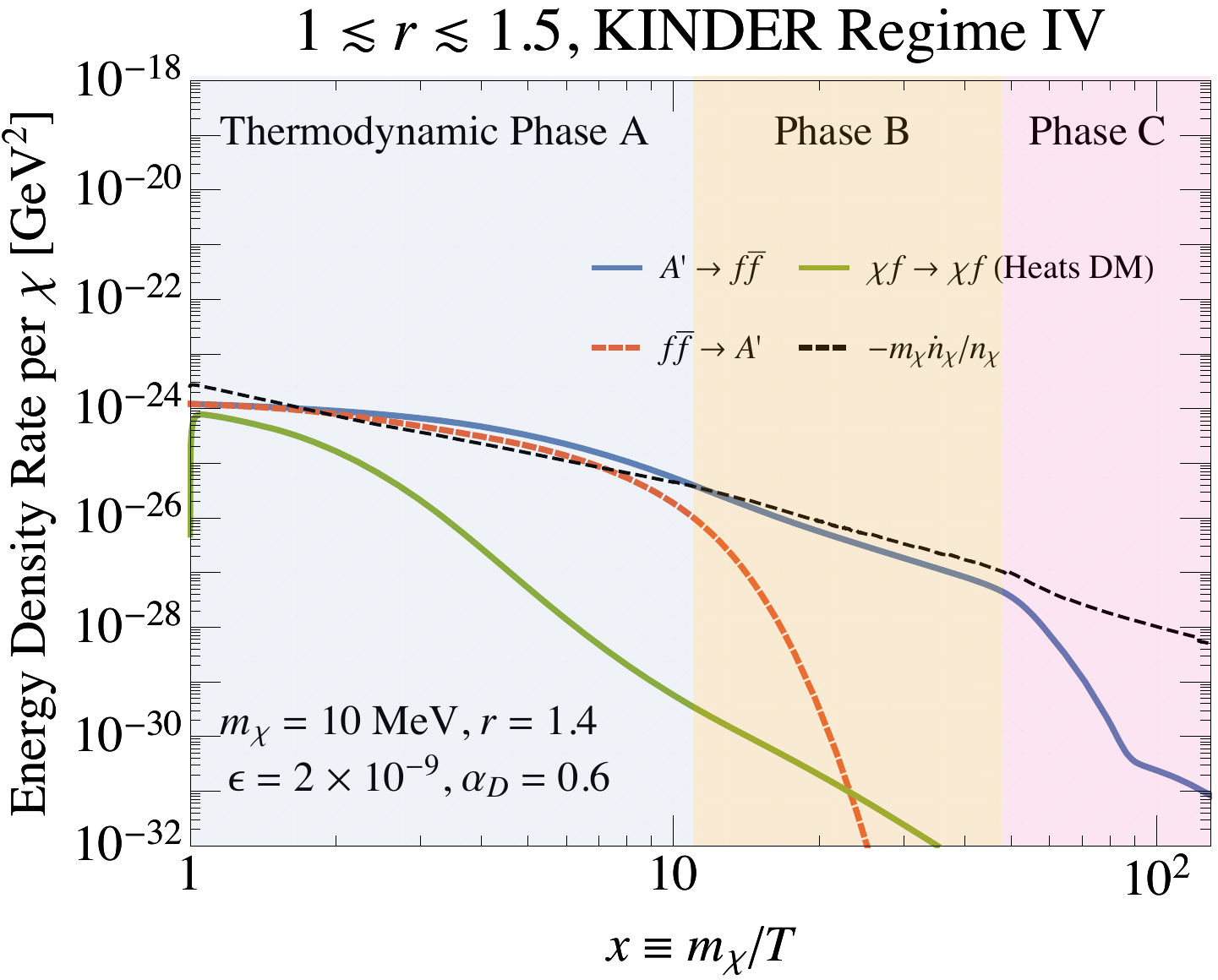}
    \caption{Rates of change in number density and energy density per $\chi$ particle of the dark sector in regime IV for $1 \lesssim r \lesssim 1.5$; the model parameters are $m_\chi = \SI{10}{\mega\eV}$, $r = 1.4$, $\epsilon = 2 \times 10^{-9}$ and $\alpha_D = 0.6$. In both plots, thermodynamic phases A, B and C as defined in Sec.~\ref{subsec:fast_reactions_and_freezeout} are shown in light blue, yellow and pink respectively. (\textit{Left}) Number density rates for $\chi \overline{\chi} \chi \to \chi A'$ (blue line), $\chi A' \to \chi \overline{\chi} \chi$ (dark orange dashed line), $\chi \overline{\chi} \to A'A'$ (green line), $A'A' \to \chi \overline{\chi}$ (red dashed line), $A' \to f \overline{f}$ (purple line) and $f \overline{f} \to A'$ (yellow dashed line) are shown. Also shown are the rates for $A' \to f \overline{f}$ (purple line) and $f \overline{f} \to A'$ (dashed yellow line). The Hubble parameter is shown in the black dashed line. (\textit{Right}) Energy density rates for $A' \to f \overline{f}$ (blue line), $f \overline{f} \to A'$ (red dashed line) and $\chi f \leftrightarrow \chi f$ (green line), which has the net effect of heating the dark sector. The rate at which the energy density of dark matter is changing $-m_\chi \dot{n}_\chi / n_\chi$ (black dashed line) is shown for reference.} 
    \label{fig:rates_FDM_KINDER}
\end{figure*}

Fig.~\ref{fig:abundances_FDM_KINDER} shows the evolution of the $\chi$-abundance and the dark sector temperature in Regime IV, for our benchmark parameters in this regime, $m_\chi = \SI{10}{\mega\eV}$, $r = 1.4$, $\epsilon = 2 \times 10^{-9}$ and $\alpha_D = 0.6$. In Fig.~\ref{fig:rates_FDM_KINDER}, we show the number density and energy density rates per $\chi$ particle throughout dark sector freezeout. In Regime IV, kinetic decoupling occurs before either of the $2 \leftrightarrow 2$ or $3 \leftrightarrow 2$ processes become slow. This regime is similar to the KINDER regime with $1.5 \lesssim r \lesssim 2$, exhibiting heating in the dark sector, with the key difference being that the $3 \to 2$ process is now slower than the $2 \to 2$ process. In thermodynamic phase A and B, the physics in this regime is identical to that of the KINDER regime with $1.5 \lesssim r \lesssim 2$, as discussed in Sec.~\ref{subsubsec:kinetic_decoupling_cannibalization}. Kinetic decoupling occurs first at a temperature given approximately by Eq.~\eqref{eq:xd_dependence_Ap_decay}, after which the dark sector enters phase B. An approximation for the evolution of $T'$ can be obtained by assuming dark sector entropy conservation, leading to
\begin{alignat}{1}
    x' \simeq x_d + 3 \log \left(\frac{x}{x_d}\right) \,,
\end{alignat}
while a more detailed examination of the Boltzmann equations leads to the improved approximation in Eq.~\eqref{eq:Tp_KINDER_thermo_phase_B}, i.e.\ 
\begin{multline}
    x' \approx x_d \\
    + \frac{1}{r-1} \log \left[1 + \frac{3}{8} \frac{\Gamma r^{5/2}(r-1)e^{(1-r)x_d}}{x_d^2 H(x_d)} (x^2 - x_d^2) \right]  \,.
\end{multline}

Once the $3 \leftrightarrow 2$ process freezes out, the dark sector enters thermodynamic phase C. As before, the $\chi$ number density evolution is given by Eq.~\eqref{eq:n_chi_evolution}, i.e.\
\begin{alignat}{1}
    \dot{n}_\chi + 3 H n_\chi \simeq \frac{r}{8(1-r)}\langle \sigma v^2 \rangle \left[n_\chi^3 - \frac{n_{\chi,0}^2}{n_{A',0}} n_\chi n_{A'} \right] \,.
    \label{eq:restated_n_chi_evolution}
\end{alignat}
For $1 \lesssim r \lesssim 1.5$, however, the $3 \to 2$ process is slow in phase C, meaning that 
\begin{alignat}{1}
    \dot{n}_\chi + 3 H n_\chi \simeq 0 \,,
\end{alignat}
i.e.\ $n_\chi \propto a^{-3}$ in phase C, with $\chi$ frozen out.  

\begin{figure}
    \centering
    \includegraphics[width=0.45\textwidth]{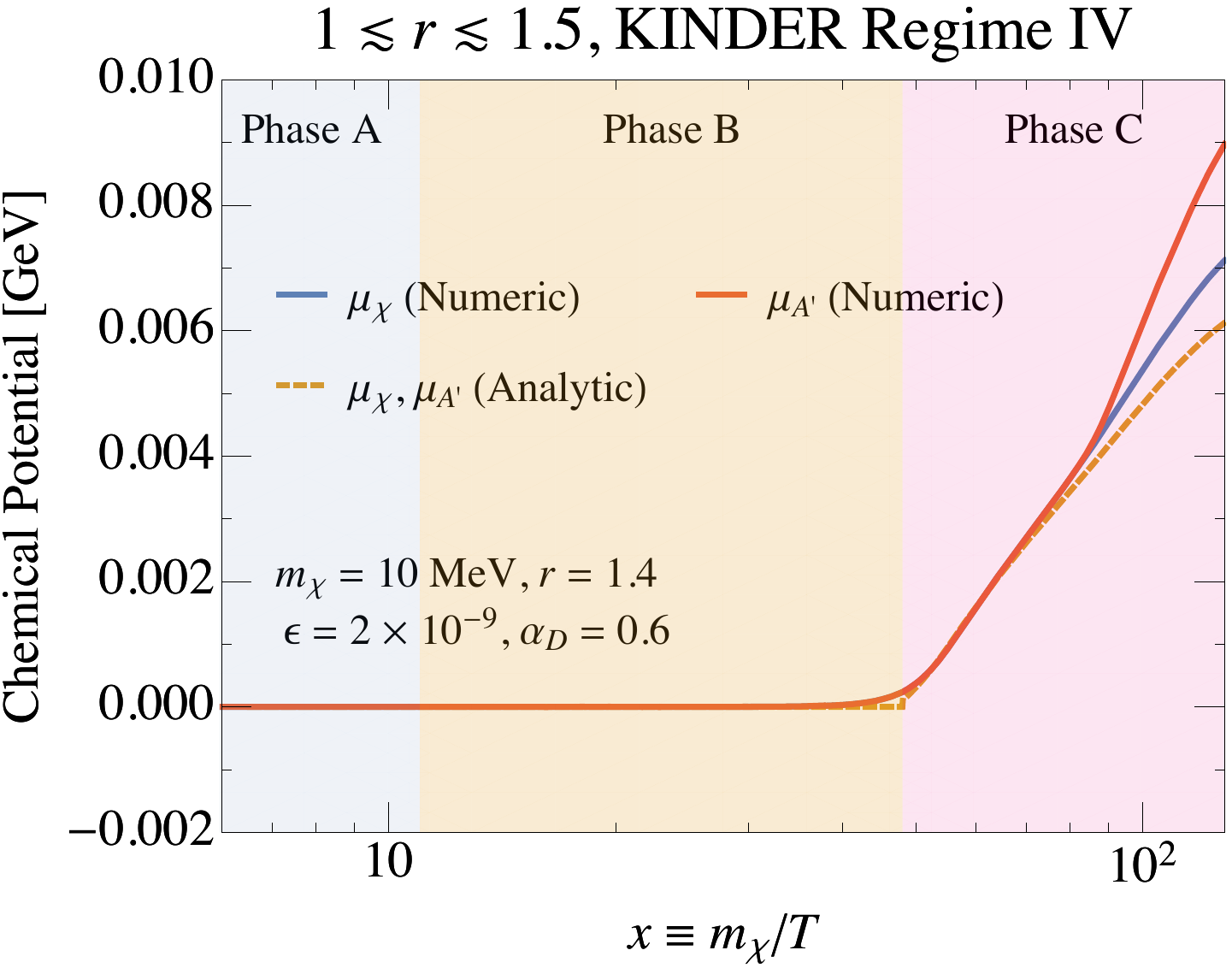}
    \caption{$1 \lesssim r \lesssim 1.5$, KINDER Regime IV (benchmark values $m_\chi = \SI{10}{\mega\eV}$, $r = 1.4$, $\epsilon = 2 \times 10^{-9}$ and $\alpha_D = 0.6$) comparison between the analytic estimate of $\mu_\chi \approx \mu_{A'}$ (orange dashed line) given in Eq.~\eqref{eq:chemical_potential_FDM_KINDER}, and the numerical computation of $\mu_\chi$ (blue line) and $\mu_{A'}$ (red line) based on integrating the full Boltzmann equations.}
    \label{fig:mu_KINDER_FDM}
\end{figure}

More accurately, Eqs.~\eqref{eq:3_to_2_fast_for_Ap} and~\eqref{eq:decay_fast} are still true in this regime, since $n_{A'} \ll n_\chi$; we therefore still have the following approximate relation after $3 \leftrightarrow 2$ freezeout occurs:
\begin{alignat}{1}
    \frac{r}{8(r-1)} \langle \sigma v^2 \rangle n_\chi^3 \approx r \Gamma n_{A'} \,,
\end{alignat}
where we have neglected the $2 \to 3$ rate since the $3 \leftrightarrow 2$ freezeout has occurred. This approximate relation gives us an expression for $\mu_\chi \simeq \mu_{A'}$:
\begin{multline}
    \frac{\mu_\chi}{T'} \approx \frac{3-r}{2}x' + \frac{3}{2} \log x' \\
    + \frac{1}{2} \log \left[ \frac{8(r-1) \Gamma}{\langle \sigma v^2 \rangle} \frac{3(2\pi)^3 r^{3/2}}{64 m_\chi^6}\right] \,.
    \label{eq:chemical_potential_FDM_KINDER}
\end{multline}
A comparison between this analytic approximation and the numerical result in phase C is shown in Fig.~\ref{fig:mu_KINDER_FDM}, demonstrating good agreement up till the point of $2 \leftrightarrow 2$ freezeout. 

\begin{figure}
    \centering
    \includegraphics[width=0.45\textwidth]{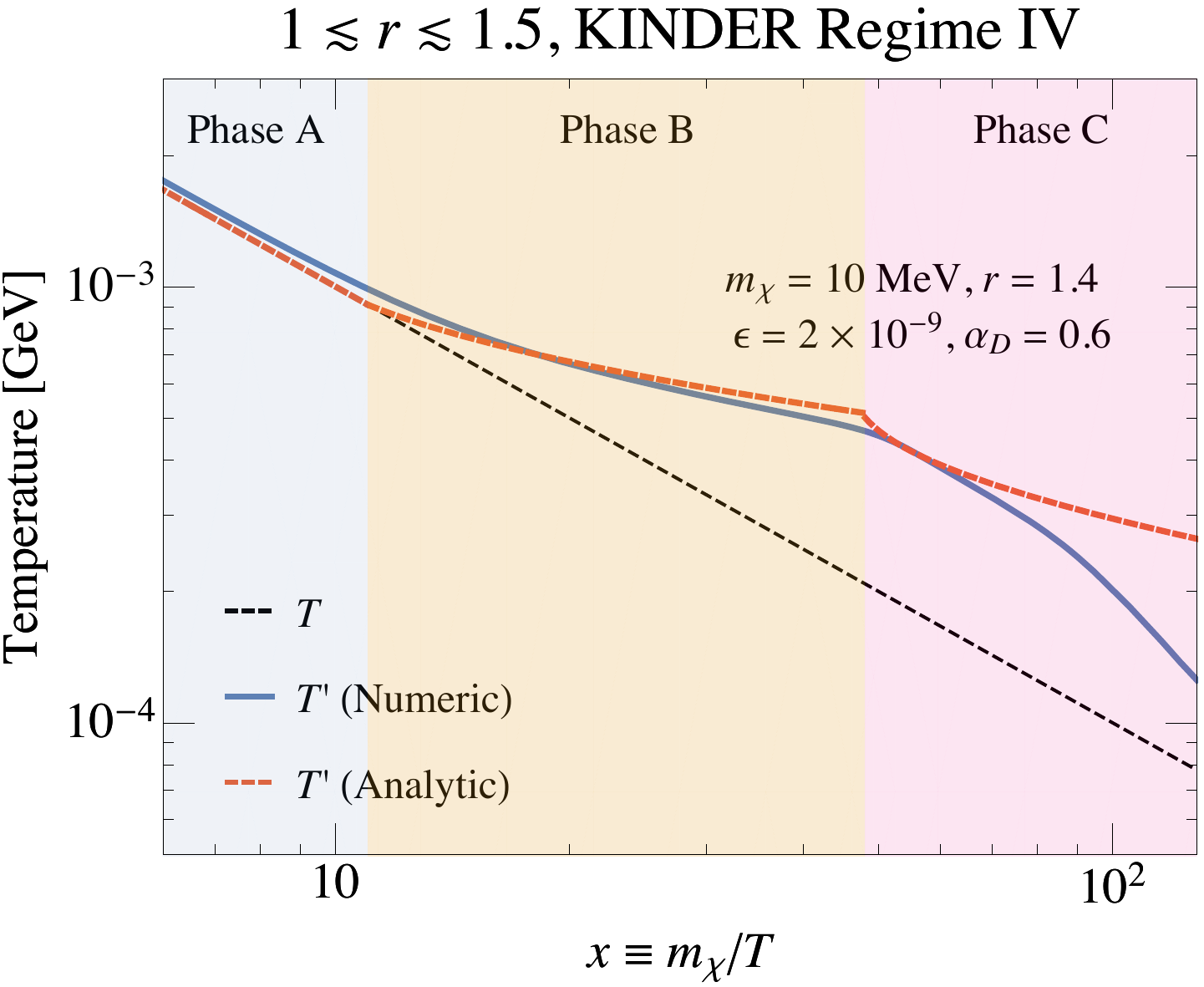}
    \caption{$1 \lesssim r \lesssim 1.5$, KINDER Regime IV (benchmark values $m_\chi = \SI{10}{\mega\eV}$, $r = 1.4$, $\epsilon = 2 \times 10^{-9}$ and $\alpha_D = 0.6$) comparison between the analytic estimate of $T'$ (red dashed line) given in Eq.~\eqref{eq:xp_FDM_Regime_IV_analytic}, and the numerical computation of $T'$ (blue line) based on integrating the full Boltzmann equations. The SM temperature is shown for reference (black dashed line).}
    \label{fig:Tp_analytic_KINDER_FDM}
\end{figure}

We can substitute our analytic expression for $\mu_\chi/T'$ into Eq.~\eqref{eq:restated_n_chi_evolution} using the expression for $\dot{n}_\chi$ in Eq.~\eqref{eq:n_chi_dot}, giving
\begin{multline}
    \frac{1-r}{2} \frac{m_\chi}{T'^2} \frac{dT'}{dT} \approx -\frac{3}{T} - \frac{r \Gamma}{H(T_3)} \frac{T_3^2}{T^3} \frac{3}{4} r^{3/2} e^{(1-r)m_\chi/T'} \,,
\end{multline}
where $T_3$ is the temperature at $3 \leftrightarrow 2$ freezeout. This expression can be integrated exactly to give
\begin{multline}
    x' \approx x_3' + \frac{6}{r-1} \log \left(\frac{x}{x_3}\right) \\
    + \frac{1}{r-1} \log \left[1 + \frac{\Gamma}{H(T_3)} \frac{3}{8} r^{5/2} \left(1 - \frac{x_3^4}{x^4}\right) e^{(1-r)x_3'} \right] \,.
    \label{eq:xp_FDM_Regime_IV_analytic}
\end{multline}
This analytic prediction in comparison with the numerical temperature evolution is shown in Fig.~\ref{fig:Tp_analytic_KINDER_FDM}, showing good agreement until near the $2 \leftrightarrow 2$ freezeout, when $\mu_\chi$ and $\mu_{A'}$ begin to diverge. 

\subsubsection{\texorpdfstring{Summary of regimes and boundaries for $1 \lesssim r \lesssim 1.5$}{Summary of regimes and boundaries for 1 ~< r ~< 1.5}}

\begin{figure*}
    \centering
    \includegraphics[width=0.47\textwidth]{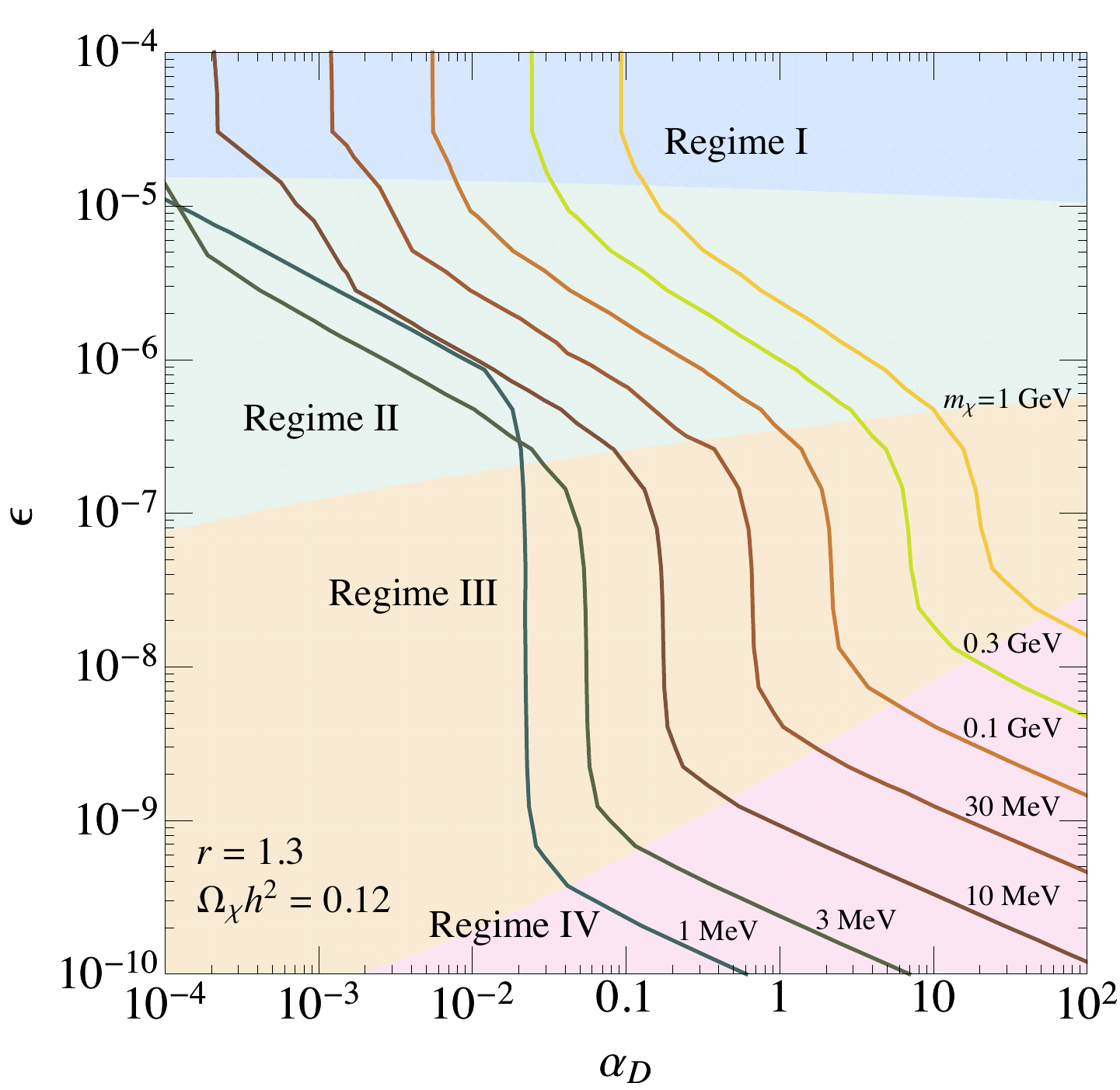} \qquad
    \includegraphics[width=0.47\textwidth]{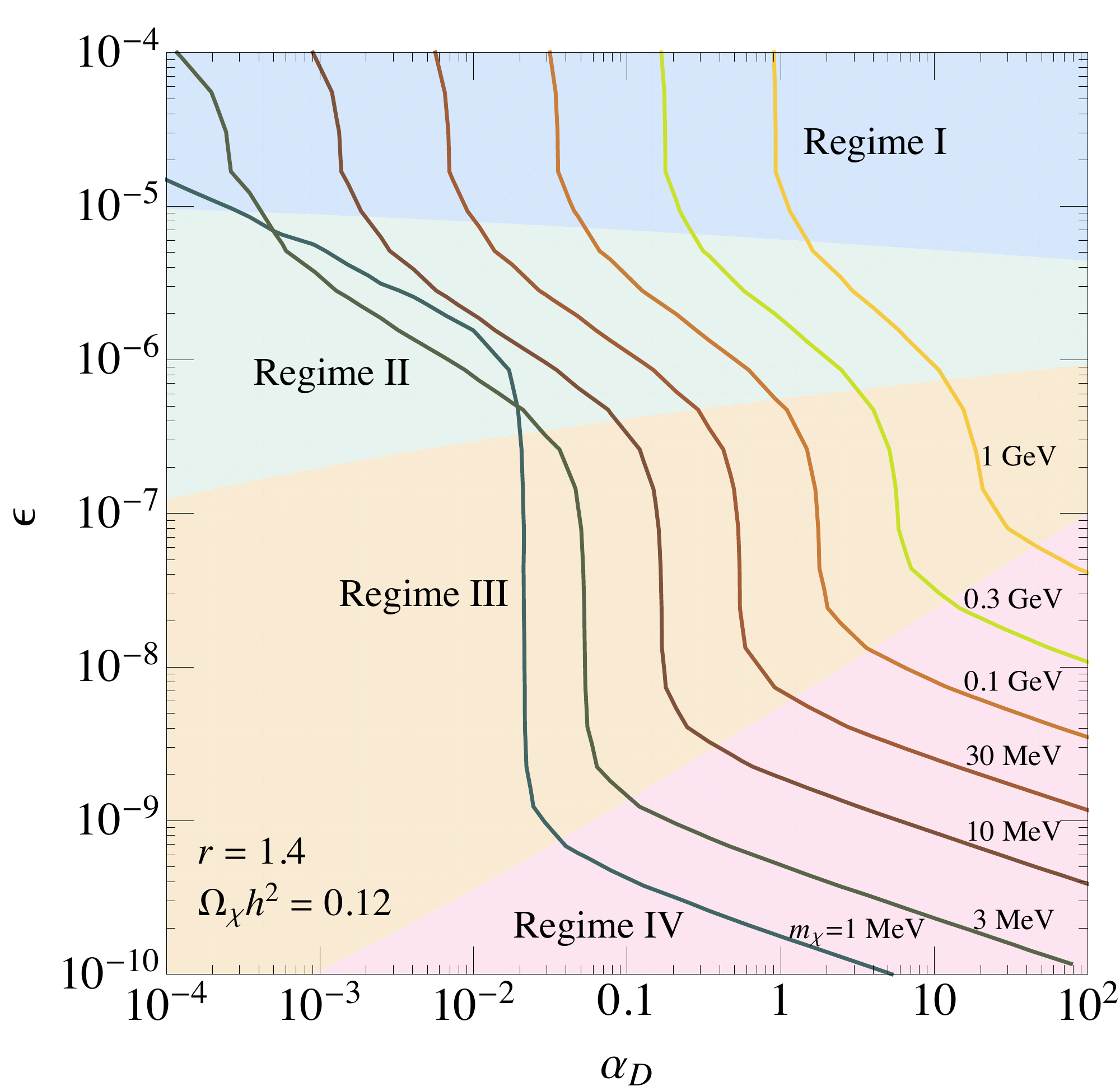}
    \caption{Contours of the observed relic abundance ($\Omega_\chi h^2 = 0.12$) in the $\alpha_D$--$\epsilon$ plane for $1 \lesssim r \lesssim 1.5$, for \textit{(left)} $r = 1.3$ and \textit{(right)} $r = 1.4$, for various values of $m_\chi$. The ``classical forbidden'' Regime I (blue), Regime II (green), Regime III (orange) and KINDER Regime IV (pink) are indicated, with the boundaries obtained using Eq.~\eqref{eq:fdm_regime_I_II_boundary} between I/II, Eq.~\eqref{eq:fdm_boundary_regimes_II_III} between II/III and Eq.~\eqref{eq:fdm_boundary_regimes_III_IV} between III/IV, all in conjunction with the relic abundance condition given in Eq.~\eqref{eq:relic_abundance_requirement}.} 
    \label{fig:Eps_Alpha_FDM_Plots}
\end{figure*}

Fig.~\ref{fig:Eps_Alpha_FDM_Plots} shows contours for fixed values of $m_\chi$ in the $\alpha_D$--$\epsilon$ parameter space for which the observed relic abundance of $\Omega_\chi h^2 = 0.12$ is attained. We show the same set of contours for $r = 1.3$ and $r = 1.4$ as two representative values of $r$ in the case of $1 \lesssim r \lesssim 1.5$. The four regimes can be made out by changes in behavior of the contour lines. Note that the boundary between the WIMP regime and Regime I occurs at $\epsilon$ values above the maximum $\epsilon$ shown in Fig.~\ref{fig:Eps_Alpha_FDM_Plots}. 

In Regime I, the relic abundance is controlled entirely by the $2 \leftrightarrow 2$ freezeout, which only depends on $\alpha_D$, leading to vertical contours in the $\alpha_D$--$\epsilon$ plane. Decreasing $\epsilon$ into Regime II, the relic abundance is controlled by when the freezeout of $3 \leftrightarrow 2$ and of $2 \leftrightarrow 2$ occur, as well as how efficiently $\chi f \leftrightarrow \chi f$ heats the dark sector and impedes the cooling due to $\chi \overline{\chi} \to A'A'$, leading to some nontrivial dependence on $\epsilon$ and $\alpha_D$. Once we arrive at Regime III however, elastic scattering becomes extremely inefficient, and the rate of dark sector cooling after $3 \leftrightarrow 2$ freezeout depends only on the $3 \to 2$ rate itself. Since all of the physically important processes are purely dark sector processes, the contours are once again independent of $\epsilon$. Finally, in Regime IV, the relic abundance is determined by when kinetic decoupling occurs, but also by the long power-law decrease in $n_\chi$ in phase B, which is dictated by dark-sector-only processes. This once again leads to contours that depend on both $\alpha_D$ and $\epsilon$.

We note that the contour of $\Omega_\chi h^2 = 0.12$ for $m_\chi \lesssim \SI{5}{\mega\eV}$ shows an abrupt change in behavior in Regime II compared to higher DM masses. This occurs due to the fact that in Regime II thermodynamic phase C, DM particles with masses below $\sim \SI{5}{\mega\eV}$ undergo elastic scattering with nonrelativistic, rather than relativistic, electrons throughout most of the freezeout process. The Boltzmann suppression of nonrelativistic electrons leads to a sharp decrease in $\langle \sigma v \delta E \rangle$, which controls the cooling rate of the dark sector in this phase, and hence the relic abundance of DM. The correct relic abundance is thus achieved at a higher value of $\epsilon$ than expected, in order for the stronger mixing to compensate for the decrease in electron number density. We refer the reader to App.~\ref{sec:Rate} for more details on how $\langle \sigma v \delta E \rangle_{\chi f \to \chi f}$ is computed. 

\section{Experimental Probes and Constraints}
\label{sec:constraints}

\begin{figure*}
    \centering
    \includegraphics[width=0.47\textwidth]{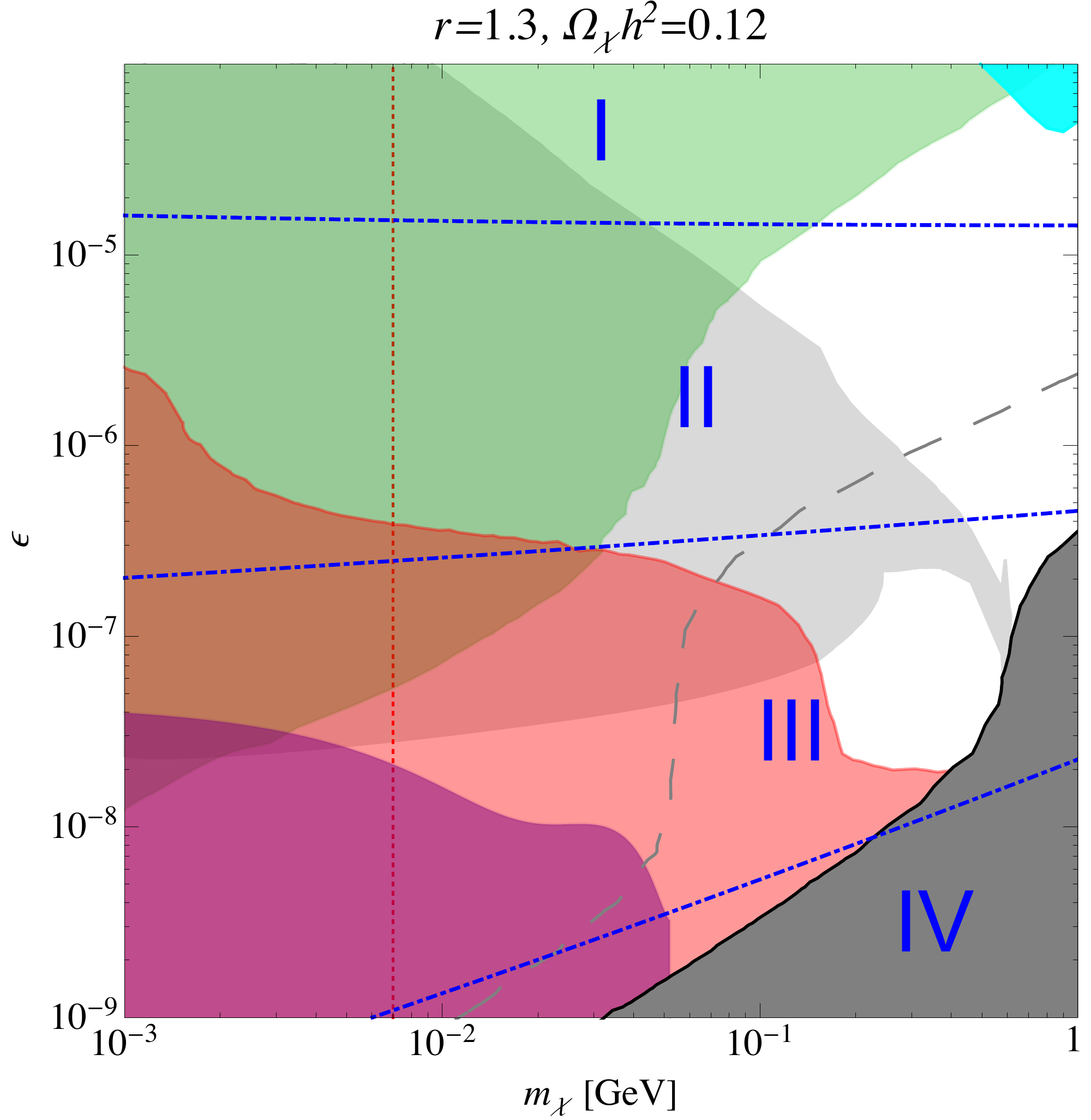}
    \includegraphics[width=0.47\textwidth]{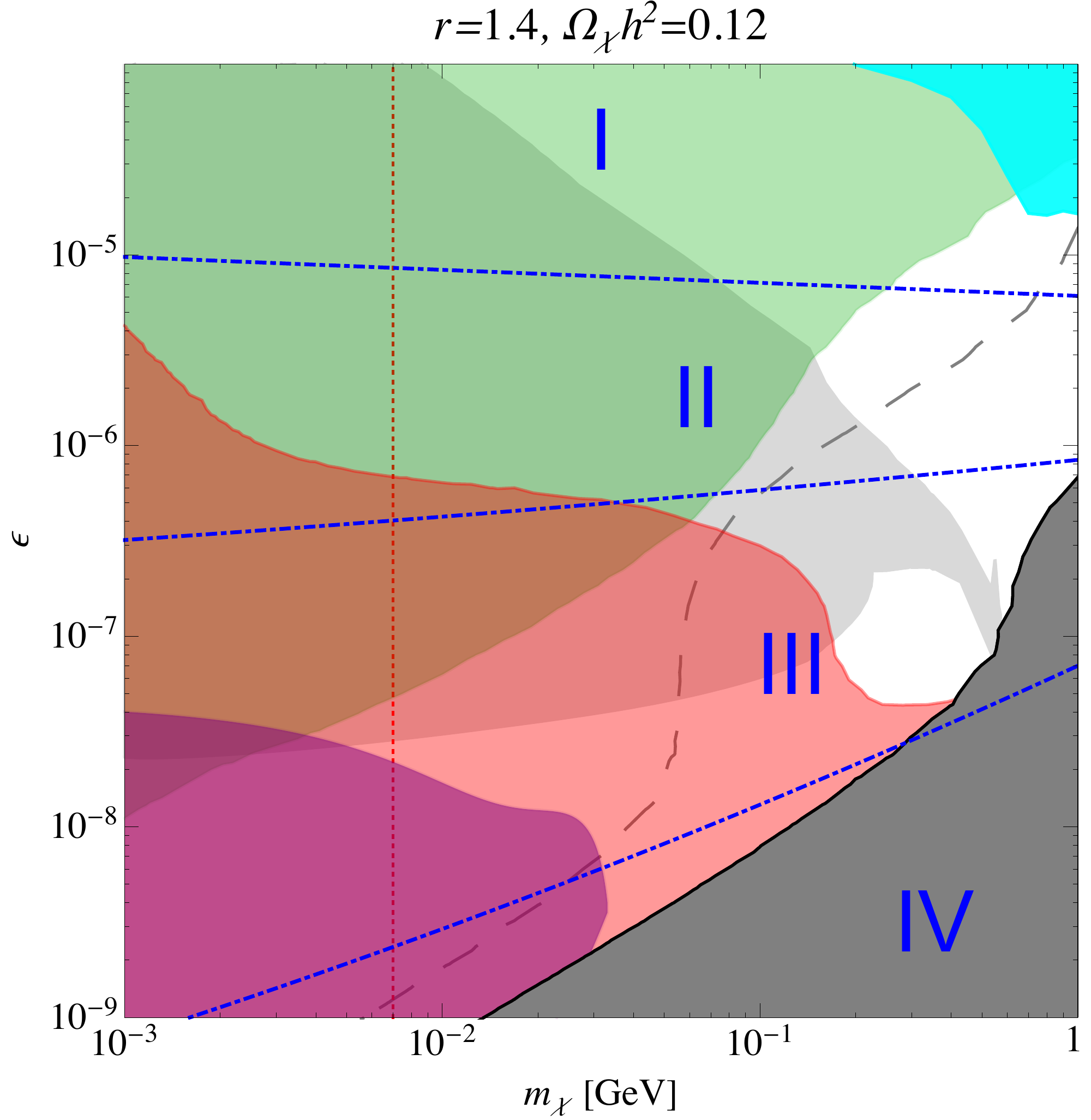}
    \qquad
    \includegraphics[width=0.47\textwidth]{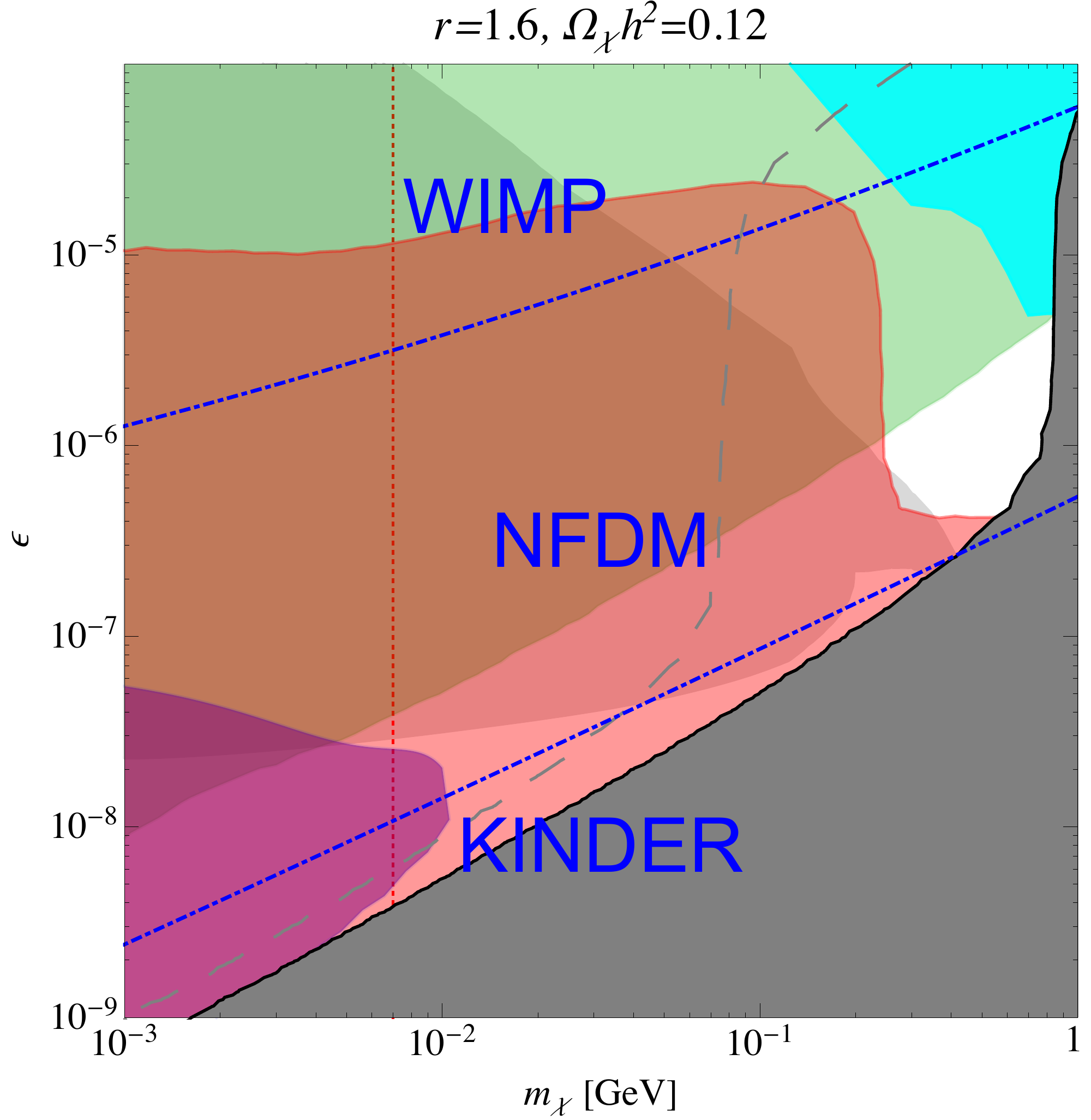}
    \includegraphics[width=0.47\textwidth]{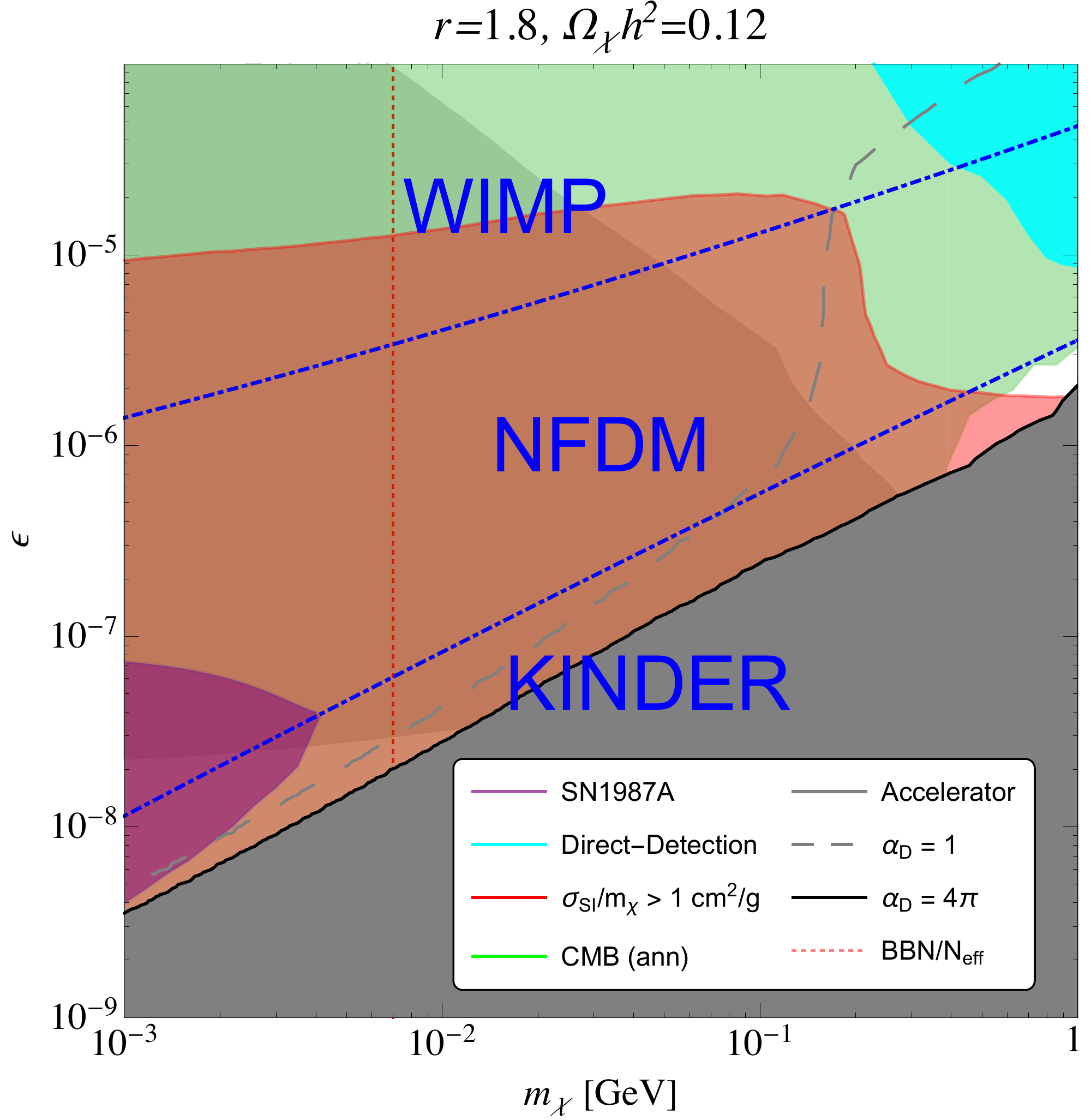}
    \caption{Constraints on our dark matter model for \textit{(upper left)} $r = 1.3$, \textit{(upper right)} $r = 1.4$, \textit{(lower left)} $r = 1.6$, and \textit{(lower right)} $r = 1.8$. The purple shaded regions are the constraints from SN1987A. The cyan regions on the upper-right corners of the plots are from direct detection experiments. We plot the self-interaction constraints as red shaded regions, and CMB $s$-wave annihilation limits with green. We also show the accelerator constraints as light gray. The constraint from BBN is shown as a red dotted line. Finally, we add the $\alpha_D = 1$ contour (gray dashed curve) and shade out the region where $\alpha_D>4\pi$ with dark gray.
    For $r = 1.3$ and 1.4, the boundaries (blue dot-dashed lines) between Regimes I, II, III, and IV (see~\cref{subsec:FDM_region_regimes}) are shown. For $r = 1.6$ and 1.8, the WIMP, NFDM (Section~\ref{subsec:classic_NFDM}), and KINDER (Section~\ref{subsec:KINDER}) regimes are separated by blue dot-dashed lines.} 
    \label{fig:const}
\end{figure*}

There are significant constraints on dark photons from both terrestrial experiments and supernova observations.
There are also cosmological constraints on the DM itself, from DM annihilation to electrons and positrons affecting the anisotropies of the CMB, and from modifications to the number of effective degrees of freedom during Big Bang nucleosynthesis (BBN) and the CMB epoch. DM self-interactions mediated by the dark photon exchange can be large, and can be probed by observations of galactic structure. Finally, a sufficiently warm dark sector can be constrained by measurements of the matter power spectrum.
We will discuss these constraints in this section, and plot the results in Fig.~\ref{fig:const}.

\subsection{Accelerator and Direct-Detection Experiments}

For $1 \lesssim r \lesssim 2$ with a dark photon mass above \SI{1}{\mega\eV}, dark photons produced at beam experiments decay visibly into SM particles. 
The observational signatures of visibly decaying dark photons have been studied extensively in the literature~\cite{Bergsma:1985is,Bergsma:1985qz,Konaka:1986cb,Bjorken:1988as,Davier:1989wz,Blumlein:1990ay,Blumlein:1991xh,Banerjee:2018vgk,Batley:2015lha,Tsai:2019mtm}.
In \cref{fig:const}, we plot the region of parameter space excluded by these experiments. This excluded region covers considerable parameter space, extending down to $\epsilon \sim 10^{-7}-10^{-8}$ for $m_\chi \lesssim \SI{100}{\mega\eV}$. 

Direct-detection experiments can probe the scattering of the DM on both electrons and nucleons (including the Migdal effect~\cite{Aprile:2019xxb,Barak:2020fql}) via dark photon exchange. 
In \cref{fig:const}, 
we consider the constraints from DarkSide, Xenon 1T, SuperCDMS, and SENSEI \cite{Agnes:2018oej,Aprile:2019jmx,Baxter:2019pnz,Aprile:2019xxb,Barak:2020fql,Amaral:2020ryn}. 
In the parameter space we consider, nuclear scattering limits derived by exploiting the Migdal effect set the strongest bound. These limits are primarily sensitive to the high-mass, high-$\epsilon$ corner of our parameter space.

\subsection{Supernova Constraints}

The production and escape of dark sector particles during a core-collapse supernova can lead to cooling of the proto-neutron star that differs from the SM prediction~\cite{Raffelt:1987yt,Raffelt1996}. Such anomalous cooling is constrained by our observation of SN1987A~\cite{Burrows1986,Burrows:1987zz}.\footnote{Alternative cooling models have also been proposed that cast doubt on the SN1987A bounds (see, e.g., Ref.~\cite{Bar:2019ifz}).}

Ref.~\cite{Chang:2018rso} carefully derived constraints on the $m_\chi$--$\epsilon$ plane in the vector-portal DM model using the SN1987A result for $m_\chi = 3 m_{A'}$, and for two discrete $\alpha_D$ values, together with constraints for models with only $A'$ and no DM. For fixed $\alpha_D$, $m_\chi$ and $m_{A'}$, the excluded region is generally enclosed by two boundary values of $\epsilon$. The lower boundary in $\epsilon$ is determined by the rate of production of the dark-sector particles from the SN core: models with smaller values of $\epsilon$ are allowed because they do not lead to enough production of dark-sector particles to modify the supernova evolution significantly. The upper boundary on $\epsilon$ is determined by whether the dark-sector particles will thermalize with the SM material in the proto-neutron star before escaping the SN, leaving these particles trapped; in models with larger values of $\epsilon$, the dark-sector particles are thermalized efficiently and do not escape and cool the proto-neutron star, and hence these scenarios are unconstrained. 

We now discuss how to recast the bounds in Ref.~\cite{Chang:2018rso} for different values of $\alpha_D$. The maximum value of $m_\chi$ is independent of $\alpha_D$, being set by the kinematics of the supernova. The behavior of the lower bound is determined by the DM mass with respect to the plasma frequency of the interior, $\omega_{p}\sim$ \SI{15}{\mega\eV}. For 2$m_\chi$ > $\omega_p$, the off-shell DM production via bremsstrahlung through virtual dark photons during neutron-proton collisions is suppressed, and the direct production of $A'$ is more important. Consequently, the lower bound in $\epsilon$ is very similar to that in the dark-photon-only case, and is roughly independent of $\alpha_D$. For 2$m_\chi$ < $\omega_p$, however, $\chi \overline{\chi}$-pairs can be produced through an on-shell $A'$, and the production rate is fixed by the value of $\alpha_D \epsilon^2$. For a lower bound given at a reference value $\alpha_{D,\text{ref}}$, we can therefore rescale to a new value of $\alpha_D$ by leaving the part of the bound where $2 m_\chi > \omega_p$ constant, and rescaling the $\epsilon$ limit where $2 m_\chi < \omega_p$ by $\sqrt{\alpha_{D,\text{ref}}/\alpha_D}$. 

The upper boundary of the limit on $\epsilon$ is determined by the dark-matter-proton scattering cross-section, and consequently varying $\alpha_D$ changes the asymptotically flat part of the upper boundary in $\epsilon$ such that $\alpha_D \epsilon^2$ is kept fixed, i.e.\ from a reference upper limit given for $\alpha_{D,\text{ref}}$, we rescale by $\sqrt{\alpha_{D,\text{ref}}/\alpha_D}$. 

We find that for $\epsilon \gtrsim 10^{-9}$, the DM rate of production in the supernova in our model is always large enough for a significant amount to be produced; our limits are therefore set by the upper limit on $\epsilon$, as determined by the thermalization condition. Note that this also happens for the lower boundary of our curves since there $\alpha_D$ is very large. The SN1987A constraints cover the low-$\epsilon$ and low-$m_\chi$ part of the parameter space, and generally lie entirely within the self-interaction constraints that we will describe next (albeit with different model-dependence).

\subsection{DM Self-Interactions}

The cross section for elastic DM-DM scattering is constrained by cluster mergers and halo shapes to satisfy $\sigma_\text{SI}/m_{\chi} \leq \SI{1}{\centi\meter\squared\per\gram} \sim \SI{5e3}{\per\giga\eV\cubed}$~\cite{Bondarenko:2020mpf}.
The DM self-interaction rates for $\chi \chi \to \chi \chi$  and $\chi \bar{\chi} \to \chi \bar{\chi}$ (and their conjugate processes) are determined in Refs.~\cite{PhysRevLett.115.061301,Cline:2017tka}. Including both $s$ and $t$-channel tree level diagrams, the averaged cross section $\sigma_\text{SI}$ is given by:
\begin{alignat}{1}
    \frac{\sigma_\text{SI}}{m_{\chi}} &= 3 \pi h(r) \frac{ \alpha_D^{2} }{ m_{\chi}^{3} } \nonumber \\
    &= \SI{1}{\centi\meter\squared\per\gram} \left(\frac{h(r)}{1.2}\right)\left( \frac{ \SI{10}{\mega\eV} }{ m_{\chi} } \right)^{3} \left( \frac{ \alpha_D }{0.02} \right)^{2} \,,
\end{alignat}
where
\begin{alignat}{1}
    h(r) \equiv \frac{16 - 16r^2 + 5r^4}{r^4(r^2 - 4)^2} \,.
\end{alignat}
Typical values of $h(r)$ are $h(1.3) = 0.2$ and $h(1.8) = 2.7$. 

As shown in Fig.~\ref{fig:const}, this constraint rules out a large fraction of the parameter space especially at low $\epsilon$, generically excluding $\epsilon$ as high as $ 10^{-6}$--$10^{-5}$ depending on $r$; this behavior occurs because the values of $\alpha_D$ required to obtain the correct relic density are higher at small $\epsilon$. In this sense the self-interaction bound is complementary to limits on the interactions with the SM, which are suppressed by small $\epsilon$. 

One possible way to evade this constraint is to consider a scenario where only some subdominant fraction of the DM is produced by the mechanisms we have considered in this work, as this limit is rather sensitive to the fraction of DM that is self-interacting. For example, Ref.~\cite{Pollack:2012hbv} shows that if the self-interacting component is less than 1$\%$ of the DM, these constraints become inapplicable. However, a full self-consistent treatment of fractionally abundance self-interacting dark matter constraints would require recalculation of the cosmological evolution in order to obtain a lower relic density, and is beyond the scope of this work.

\subsection{CMB Constraints on DM Annihilation}

During the post-recombination epoch, DM annihilation to $e^+ e^-$ leads to energy deposition into the baryonic gas; the resulting extra ionization can be constrained based on observations of the CMB anisotropy. We compare the annihilation cross section for $\chi  \bar{\chi}  \rightarrow f\bar{f}$ (see App. B) to the limits derived in Ref.~\cite{Slatyer:2015jla} and updated in Ref.~\cite{Aghanim:2018eyx}. We plot the region excluded by this constraint in~\cref{fig:const}.

We observe that these CMB constraints provide some of the strongest bounds on models of this type for $r$ close to 2, excluding most of the available parameter space. Even for smaller values of $r$, the CMB constraints provide stringent limits for models with low $m_\chi$ and high $\epsilon$. 

These limits could be lifted or relaxed if the dark-sector model were adjusted in order to suppress the DM annihilation to SM particles at low velocities. For example, this could be achieved if the DM was a scalar rather than a fermion, as then the leading-order annihilation through the dark photon would be $p$-wave and scale as $\langle \sigma v \rangle \propto v^2$.

\subsection{Cosmological Constraints on Light Relics}

Electromagnetically coupled DM with a mass of around \SI{1}{\mega\eV} can significantly affect the process of Big Bang Nucleosynthesis (BBN) by \textit{(i)} directly increasing the expansion rate as a contribution to the energy density of the universe, and \textit{(ii)} injecting entropy into the SM sector and changing the relative energy density of the electromagnetic sector as compared to the neutrino sector, altering the temperature evolution of both sectors with respect to standard cosmology. These changes in turn alter the predicted abundance of light nuclei like deuterium and helium-4, which can then be compared with existing measurements of the abundances of these nuclei (see e.g.\ Refs.~\cite{Izotov:2014fga,Aver:2015iza,Cooke:2017cwo,Zavarygin:2018ara,Valerdi:2019beb} for deuterium and helium-4). The injection of entropy from electromagnetically coupled DM can also decrease $N_\text{eff}$~\cite{Boehm:2013jpa}, the effective number of degrees of freedom, during the CMB epoch, which can then be constrained by the CMB anisotropy power spectrum~\cite{Aghanim:2018eyx}.

Ref.~\cite{Sabti:2019mhn} modelled the predicted primordial elemental abundances in the presence of an electromagnetically coupled dark matter particle; we adopt their results for our BBN constraints. They presented two constraints, depending on whether a prior was imposed on $\Omega_b h^2$ in the BBN calculations. When no prior was imposed, the bound is relatively weak, $m_\chi \gtrsim \SI{0.7}{\mega\eV}$ for Dirac fermion DM. With a prior based on CMB observations, $\Omega_b h^2 = 0.02225 \pm 0.00066$~\cite{Aghanim:2018eyx}, this bound improves to $m_\chi \gtrsim \SI{7}{\mega\eV}$, since the effect of entropy injection into the SM from the DM cannot be compensated for by lowering $\Omega_b h^2$ arbitrarily.

We note however that assuming the central value of $\Omega_b h^2$ from Planck leads to a standard BBN theoretical prediction of D/H that is roughly $2\sigma$ below the central measured value. This discrepancy may indicate an incomplete understanding of the process of BBN even in standard cosmology, which may therefore affect the bound given above. 

As mentioned above, one can also consider the impact of electromagnetically coupled DM particles on the CMB anisotropy power spectrum. Electromagnetically coupled DM particles heat the electromagnetic sector as they become nonrelativistic, effectively decreasing the number of relativistic degrees of freedom at late times by increasing the ratio of photon to neutrino temperatures. The Planck 2018 measurement~\cite{Aghanim:2018eyx} sets a constraint on electrophilic Dirac fermions of $m_\chi \gtrsim \SI{7.4}{\mega\eV}$. A joint constraint using both primordial elemental abundance and CMB data strengthens the constraint on electrophilic Dirac fermion to $m_\chi \gtrsim \SI{10}{\mega\eV}$. However, CMB $N_\text{eff}$ bounds are less robust than the BBN constraint, and can be overcome by e.g.\ adding dark, relativistic degrees of freedom to compensate for the effect of the electromagnetically coupled DM~\cite{Nollett:2013pwa}. 

Given the above consideration, we set a tentative constraint of $m_\chi > \SI{7}{\mega\eV}$ to indicate the potential constraint from BBN and CMB. Since the region with $m_\chi < \SI{10}{\mega\eV}$ is already strongly constrained by the CMB limits on DM $s$-wave annihilation, beam dump experiments and SN1987A, this constraint is not particularly important to understanding the viability of the model.

\subsection{Warm Dark Matter}

In the $1.5 \lesssim r \lesssim 2$ KINDER regime, the dark sector undergoes an early kinetic decoupling from the SM, after which the dark sector temperature $T'$ evolves only logarithmically with respect to the SM temperature $T$ until the $3 \to 2$ process freezes out. As a result, the dark sector temperature can be much higher than in the standard WIMP paradigm, where $T' = T$ until kinetic decoupling, after which $T' \propto (1+z)^2$. Models of warm dark matter (WDM) typically have suppressed structure on small scales~\cite{Abazajian:2017tcc,Adhikari:2016bei}, and can be constrained by measurements of the matter power spectrum from the Lyman-$\alpha$ forest~\cite{Irsic:2017ixq,Palanque-Delabrouille:2019iyz}, which are sensitive to modes with comoving wavenumber as large as $k_{\max} \sim \SI{3}{\h \per \mega \parsec}$. 

To get an estimate for how important the WDM Lyman-$\alpha$ bounds are to the KINDER regime, we estimate the comoving Jeans length $\lambda_J$ of DM, and compare this with $2\pi / k_{\max} \sim \SI{2}{\per\h\mega\parsec}$; for model parameters where $\lambda_J \ll 2\pi/k_{\max} $, the model is unlikely to leave a significant imprint on the matter power spectrum on scales currently probed by experiments. We leave a detailed analysis of such potential WDM constraints for future work. 

The comoving Jeans length for the DM is given by~\cite{Schneider:2013ria}
\begin{alignat}{1}
    \lambda_J(z) = (1+z) \sqrt{\frac{T'(z)}{m_\chi}} \frac{2 \sqrt{2} \pi}{H(z)} \,.
\end{alignat}
After the dark sector completely freezes out, $T' \propto (1+z)^2$; in the radiation dominated era, $\lambda_J$ stays roughly constant, while $\lambda_J \propto (1+z)^{1/2}$ during matter domination, decreasing with time. To make a conservative estimate, we therefore want to compare $\lambda_J(z_\text{eq})$ with $2 \pi/k_{\max}$ at the redshift of matter-radiation equality, $z_\text{eq}$.\footnote{The Jeans scale at matter-radiation equality is on the same order as the free-streaming length of warm dark matter at matter-radiation equality, another common method of determining the length scale below which structure is damped~\cite{Schneider:2013ria}.} We can estimate the temperature of the dark sector at $z_\text{eq}$ as
 \begin{alignat}{1}
     T'(z_\text{eq}) \simeq T'_{3} \frac{(1+z_{\text{eq}})^2}{(1 + z_3)^2} \simeq T_3' \frac{T_\text{eq}^2}{T_3^2}\,,
 \end{alignat}
 where $z_3$ and $T'_3$ are the redshift and dark sector temperature at $3 \to 2$ freezeout respectively. With this approximation, we have
 \begin{alignat}{1}
     \lambda_J(z_\text{eq}) \simeq (1 + z_\text{eq}) \frac{T_\text{eq}}{T_3 \sqrt{x_3'}} \frac{2 \sqrt{2} \pi}{H(z_\text{eq})} \,.
 \end{alignat}
 Taking $z_\text{eq} = 3402$ and assuming a $\Lambda$CDM cosmology, we can obtain the following estimate for the Jeans length at matter-radiation equality: 
 \begin{alignat}{1}
     \lambda_J(z_\text{eq}) \simeq \SI{0.2}{\per\h\mega\parsec} \left(\frac{x_3}{10^3}\right) \left(\frac{\SI{}{\mega\eV}}{m_\chi}\right) \left(\frac{10}{x_3'}\right)^{1/2} \,.
     \label{eq:Jeans_scale_value}
 \end{alignat}
In the $1.5 \lesssim r \lesssim 2$ KINDER regime, we know that $x_3' \sim x_d$, since $x'$ evolves logarithmically with respect to $x$ in thermodynamic phases B and C, while $x_3$ is largest when the $3 \leftrightarrow 2$ process freezes out at the latest possible time. We therefore find that $\lambda_J(z_\text{eq})$ is largest at \textit{(i)} small $\epsilon$, so that decoupling occurs early, minimizing $x_d$ and thus $x_3'$, and \textit{(ii)} large $\alpha_D$ with small $m_\chi$, so that the $3 \leftrightarrow 2$ cross section is large, and the process freezes out as late as possible, maximizing $x_3$. To maximize the impact on small-scale structure, we therefore take the smallest mass we consider $m_\chi = \SI{1}{\mega\eV}$, choose the largest perturbative value of $\alpha_D = 4 \pi$, giving $\epsilon = 3.5 \times 10^{-9}$ to achieve the observed relic abundance for $r = 1.8$. We find that $x_3 = 5500$ and $x_3' = 45$, leading to $\lambda_J(z_\text{eq}) \simeq \SI{0.5}{\per\h\mega\parsec}$, which is still small enough to be consistent with probes of small-scale structure. Other parameter combinations that obtain the observed relic abundance lead to smaller values of $\lambda_J(z_\text{eq})$. 

For $1 \lesssim r \lesssim 1.5$, Regime I has $T' = T$ until freezeout of the dark sector, while in Regimes II and III, the dark sector is actually colder than a dark sector that is thermally coupled to the SM until freezeout, easily avoiding these warm DM constraints. In Regime IV, a similar argument as above shows that $\lambda_J(z_\text{eq})$ is given by Eq.~\eqref{eq:Jeans_scale_value} with $x_3,x_3'$ replaced by $x_2,x_2'$. Once again, large values of $\alpha_D$, small values of $\epsilon$ and small $m_\chi$ would lead to the largest impact on small-scale structure. SN1987A constraints and the requirement of a perturbative value of $\alpha_D < 4 \pi$, however, are enough to constrain $\epsilon \gtrsim 10^{-9}$. Choosing $r = 1.4$, $\epsilon = 10^{-9}$, $m_\chi = \SI{1}{\mega\eV}$ and $\alpha_D = 0.19$, we find $x_2 = 630$, $x_2' = 89$ and $\lambda_J(z_\text{eq}) \simeq \SI{5e-3}{\per\h\mega\parsec}$, much smaller than would be observable. Larger values of $m_\chi$ require larger values of $\epsilon$ to meet the relic abundance criterion, and lead to even smaller values of $\lambda_J(z_\text{eq})$. Similar results hold for $r = 1.3$ as well. 

We therefore find that $\lambda_J(z_\text{eq}) \ll 2 \pi / k_\text{max}$ is satisfied throughout all relevant parameter space, leaving our model unconstrained by small-scale structure observations. However, parts of the KINDER regime are close to being constrained by existing power spectrum measurements; future improvements in WDM constraints could potentially probe these models. 

\subsection{Summary of Constraints}

Fig.~\ref{fig:const} shows a plot of the constraints on the $m_\chi$--$\epsilon$ plane with four different values of $r$, with $\alpha_D$ chosen at every point in parameter space such that the observed relic abundance of DM is attained, $\Omega_\chi h^2 = 0.12$. Regions ruled out by the constraints discussed above are marked in color; parts of the space that require $\alpha_D > 4\pi$ to obtain the correct relic abundance are also shaded gray, since perturbative control of our model breaks down there. The contour of $\alpha_D = 1$ is also shown for reference.

For $1.5 \lesssim r \lesssim 2$, we show the constraints for two representative values, $r = 1.6$ and $r = 1.8$. In both cases, a small region of open parameter space exists near $\epsilon \sim 10^{-6}$ and with DM masses of a few hundred MeV. For these values of $r$, the vector-portal DM model is bounded from below by the nonperturbative region, and is strongly constrained by the CMB $s$-wave annihilation bound and self-interaction limits. The available parameter space sits in the NFDM regime for $r = 1.6$, and in the KINDER regime for $r = 1.8$. The unconstrained regions are similar to those obtained in Ref.~\cite{Cline:2017tka} at the high-$\epsilon$ end, but differ at the low-$\epsilon$ end due to the KINDER regime that we have found in this paper. 

For $1 \lesssim r \lesssim 1.5$, we show the constraints for $r = 1.3$ and $r = 1.4$. Here, there are two viable regions of parameter space: both are in the range $m_\chi \gtrsim \SI{100}{\mega\eV}$, and are separated by the beam dump constraints: one region in Regime III is in the range $\epsilon \sim 10^{-8}$ -- $10^{-7}$, while the other is in Regime II and I in the range $\epsilon \sim 10^{-7}$ -- $5 \times 10^{-5}$. In this range of $r$-values, both the self-interaction and CMB $s$-wave annihilation limits are less constraining, allowing more open parameter space than for $1.5 \lesssim r \lesssim 2$. These new limits represent an improved calculation over those found in Ref.~\cite{D_Agnolo_2015}. In particular, most of the available parameter space is not in Regime I, as assumed by Ref.~\cite{D_Agnolo_2015}. In contrast to that work, we find that there is a lower limit of $\epsilon \gtrsim 10^{-8}$ imposed by perturbativity and self-interaction constraints, since (in Regime IV) $\alpha_D$ needs to become very large at such small values of $\epsilon$ in order to achieve the correct DM relic abundance. 

We emphasize that these constraints are derived assuming that the dark sector is in thermal equilibrium with the SM at $T \sim m_\chi$, which may not be a valid assumption for values smaller than $\epsilon_\text{eq}$ as defined in Eq.~\eqref{eq:epsilon_thermal_equilibrium}. For $\epsilon \sim 10^{-9}$ and below, other mechanisms such as freeze-in can potentially achieve the correct relic abundance without the dark sector ever being in thermal equilibrium with the SM. 

\subsection{Lifting CMB and Self-Interaction Constraints with Pseudo-Dirac DM}

In the previous subsections, we have demonstrated that the bulk of the parameter space for this class of models with $1.2 < r  < 1.8$ has been tested by existing observations and experiments, for the baseline scenario where the DM is a Dirac fermion. Narrow regions of parameter space remain open, but for example, Regime IV for $1 \lesssim r \lesssim 1.5$ appears to be fully excluded. 

However, these exclusions rely critically on constraints from the CMB and from self-interactions, both of which probe the behavior of the DM long after freezeout. This exclusion is model-dependent; it is possible to perturb our baseline model in ways that dramatically alleviate these constraints while leaving the cosmology during the freezeout epoch essentially unchanged.

As a specific example, suppose that the DM is a pseudo-Dirac fermion, where at low energies the DM is split into two nearly-degenerate Majorana mass eigenstates $\chi_1$, $\chi_2$ (see e.g.\ Refs.~\cite{Finkbeiner:2007kk,Finkbeiner:2010sm, Elor:2018xku} for specific models). The gauge interaction between the DM and the $A^\prime$ ($\bar{\chi} \slashed{A}^\prime \chi$) then gives rise to interactions of the form $\bar{\chi}_i \slashed{A}^\prime \chi_j$, $i \ne j$. There is no $\bar{\chi}_i \slashed{A}^\prime \chi_i$ vertex as Majorana fermions cannot carry a conserved dark charge. The heavier mass eigenstate $\chi_2$ can thus decay to the lighter eigenstate $\chi_1$ via emission of an off-shell $A^\prime$.

When the temperature of the dark sector exceeds the mass splitting between the states, the DM will behave as a Dirac fermion, and thus for a mass splitting $\Delta m_\chi \ll T'$ throughout freezeout, our previous cosmological results will still hold. However, once $T^\prime \ll \Delta m_\chi$, the DM will convert into the lighter mass eigenstate provided the lifetime of the heavier eigenstate is sufficiently short (even if the lifetime is long, DM-DM scattering can also efficiently deplete the heavier eigenstate). Thus during the recombination epoch and in galaxies at late times, any process requiring the presence of both mass eigenstates will be strongly suppressed. 

This suppression applies to both the annihilation $\bar{\chi} \chi \rightarrow e^+ e^-$ through an $s$-channel $A^\prime$, which determines the CMB constraint,\footnote{The relaxation of CMB bounds for pseudo-Dirac DM is well-known, see e.g. Ref.~\cite{Battaglieri:2017aum}.} and to the contribution to the tree-level self-interaction cross section $\bar{\chi} \chi \rightarrow \bar{\chi} \chi$ from an $s$-channel $A^\prime$. The contribution to the tree-level self-interaction cross sections from a $t$-channel $A^\prime$ exchange is suppressed for a related reason; if the initial state is $\chi_1 \chi_1$ then the final state (at tree level) can only be $\chi_2 \chi_2$, which is kinematically forbidden provided the kinetic energy of DM particles in the halo is much smaller than the mass splitting. There will still be a contribution to the self-interaction cross section at 1-loop order, and a CMB signal via $t$-channel annihilation of $\chi_1$'s to the 3-body final state $ A^\prime + e^+ + e^-$~\cite{Rizzo:2020jsm} (as well as possible contributions from the residual $\chi_2$ abundance), but these rates are parametrically suppressed compared to those relevant for the Dirac case.

Thus we expect both the CMB and self-interaction limits to be dramatically relaxed in the pseudo-Dirac case without changing the freezeout history, for mass splittings that are small compared to $T^\prime$ at freezeout, but large compared to the DM temperature during recombination and the kinetic energy of DM particles in present-day halos. This modification opens up allowed parameter space spanning all the freezeout regimes we have studied; we will present a detailed computation of the modified constraints in future work~\cite{KINDER_short_paper}.

\section{Conclusion}
\label{sec:conclusion}

We have fully characterized the possible freezeout histories of the vector-portal DM model in Eq.~\eqref{eq:Lagrangian}, in the region of parameter space in which the DM is a thermal relic, and $1 \lesssim r \lesssim 2$. In this region, the $\chi \chi \bar{\chi} \leftrightarrow \chi A^{\prime}$ ($3 \leftrightarrow 2$) and kinematically suppressed $\chi \bar{\chi} \leftrightarrow A^{\prime} A^{\prime}$ ($2 \leftrightarrow 2$) processes play important roles in the thermal freezeout of the DM. Extending beyond the scope of previous studies \cite{D_Agnolo_2015,Cline_2017}, we explored this model for values of the kinetic mixing parameter $\epsilon$ where the dark and SM sectors do not remain in kinetic equilibrium  throughout the process of DM thermal freezeout. Doing so reveals a rich set of novel thermal histories, leading to very different dependences of the DM relic abundance on the model parameters. 

We have identified four novel pathways by which thermal freezeout of the dark sector can proceed, in addition to those identified in previous studies. Two of these pathways share key features, and represent a general class of freezeout histories that we dub the ``KINetically DEcoupling Relic'' (KINDER). In the KINDER scenario, the DM relic abundance is determined primarily by the kinetic decoupling of the dark and SM sectors. KINDER is realized through a process of dark sector cannibalization, which was previously invoked in the ELDER scenario~\cite{Kuflik:2015isi,Kuflik:2017iqs}. In this work, we have demonstrated that cannibalization can be supported by a $3 \rightarrow 2$ annihilation process involving multiple dark sector species, and can proceed even in the presence of nonzero dark sector chemical potentials. ELDER DM can be regarded as an example of a KINDER scenario where the kinetic decoupling is controlled by elastic scattering between the DM and SM. 

We have presented detailed numerical results for the thermal history of the dark sector in each of these new regimes. Additionally, in a number of cases we were able to analytically derive the evolution of the dark sector temperature $T^{\prime}$ and dark matter abundance $Y_\chi$, throughout the freezeout of the DM; this allows us to analytically demonstrate the dependence of the DM relic abundance on the model parameters in much of parameter space.

The novel freezeout mechanisms we have characterized, and their corresponding distinct regimes of parameter space, can be separated into two main parameter regions in $r$. In the region $1.5 \lesssim r \lesssim 2$, in addition to the ``classic not-forbidden'' regime studied in Ref.~\cite{Cline_2017}, we have identified a realization of KINDER at low values of $\epsilon$. 

In the region $1 \lesssim r \lesssim 1.5$, in addition to the ``classic forbidden'' regime studied in Ref.~\cite{D_Agnolo_2015} (Regime I), which is valid at high $\epsilon$, we identify a second variation of KINDER at very low $\epsilon$ (Regime IV). At intermediate values of $\epsilon$, we find two previously unrecognized parameter regimes with distinct freezeout histories (Regimes II and III). In Regimes II and III the $A^{\prime} \rightarrow f \bar{f}$ process is fast enough to maintain $n_{A^{\prime}} \approx n_{A^{\prime},0} \left( T \right)$ until all number-changing processes have frozen out. However, during the period after $3 \rightarrow 2$ freezeout and before $2 \rightarrow 2$ freezeout, this process cannot maintain thermal equilibrium between the DM and SM sectors due to number and energy conservation requirements enforced by the Boltzmann equations. In these regimes the elastic scattering $\chi f \rightarrow \chi f$ process controls the heat exchange between the DM and SM sectors after the freezeout of the $3 \leftrightarrow 2$ process and before the freezeout of the $2 \leftrightarrow 2$ process, while the $2 \leftrightarrow 2$ process cools the dark sector. 

The distinguishing feature between Regimes II and III is the efficiency with which the elastic scattering process heats the dark sector. In Regime III, elastic scattering is inefficient, the dark sector is cooled by the kinematically forbidden $\chi \overline{\chi} \to A'A'$ ($2 \to 2$) process, and the chemical potential of the dark sector is such that the $\chi$ abundance no longer evolves appreciably after the $3 \leftrightarrow 2$ process freezes out. This leads to a DM relic abundance determined only by the freezeout of the $3 \leftrightarrow 2$ process, even though the $2 \leftrightarrow 2$ process is significantly faster. In Regime II, in contrast, elastic scattering remains efficient after the freezeout of the $3 \leftrightarrow 2$ process, and can counteract the cooling of the dark sector, allowing continued evolution of the DM density. This leads to a DM relic abundance determined by the interplay of elastic scattering and dark sector processes.

The two variations of KINDER we have identified differ in their evolution at late times, after the slower dark sector process freezes out. For $1.5 \lesssim r \lesssim 2$, cannibalization continues through the $3 \to 2$ process until all number-changing processes have frozen out, ensuring a slow evolution of the DM number density after kinetic decoupling. In contrast, for $1 \lesssim r \lesssim 1.5$, the cannibalization is halted once the $3 \leftrightarrow 2$ process freezes out. The number-changing $2 \leftrightarrow 2$ process is still active at this point, and cools the dark sector; however, the chemical potential evolves such that the $\chi$ abundance remains constant regardless.

We have calculated the relevant experimental constraints on our model. Our results drastically modify those of Ref.~\cite{D_Agnolo_2015} for $\epsilon \lesssim 10^{-5}$ (below Regime I) and those of Ref.~\cite{Cline_2017} for $\epsilon \lesssim 10^{-7}$ (the NFDM and KINDER Regimes). The KINDER mechanism realized in our model implies large self-interaction rates, and a large $s$-wave annihilation signal in the CMB, for symmetric Dirac fermion DM; these limits are in tension with the KINDER regime, although a small window of open parameter space remains for $r = 1.8$. There is also available parameter space in Regimes II and III for DM masses $\sim \left( 0.1 - 1 \right)$ GeV where experiments have not yet explored. In these allowed regions of parameter space, self-interactions can be in the correct range ($0.1 \text{ cm}^{2}/\text{g} \lesssim \sigma_\text{SI}/m_{\chi} \lesssim 1 \text{ cm}^{2}/\text{g}$) to have observable consequences for the small-scale structure of galaxies without being currently excluded. Our new calculations provide target regions that can be tested by future sub-GeV direct detection experiments and dark photon searches.

In this paper we have presented the baseline scenario of this vector-portal model in which the DM $\chi$ is a Dirac fermion. In a forthcoming paper~\cite{KINDER_short_paper}, we will present an alternative to this baseline scenario in which the DM is a pseudo-Dirac fermion which at low energies splits into two nearly-degenerate Majorana mass eigenstates. For the correct range of values of the mass splitting this scenario shares essentially the same cosmology as the Dirac case, while modifying the late-time cosmology in a way that relaxes both CMB and self-interaction constraints, thus opening windows of parameter space spanning all the novel freezeout regimes we have presented.

\textbf{Acknowledgments.} We thank James Cline, Yonit Hochberg, Eric Kuflik, Cristina Mondino, Nadav Outmezguine, Tom Rizzo, Joshua Ruderman, Martin Schmaltz, Oren Slone, and Wei Xue for useful discussions. We especially thank Jae Hyeok Chang for explaining how to recast the SN1987A bounds on vector-portal DM in Ref.~\cite{Chang:2018rso}. 

This material is partially based upon PF's work supported by the National Science Foundation Graduate Research Fellowship under Grant No.~1745302. TRS and PF are partially supported by the U.S. Department of Energy, Office of Science, Office of High Energy Physics, under grant Contract Number DE-SC0012567. HL is supported by the DOE under contract DESC0007968 and the NSF under award PHY-1915409.

Part of this document was prepared by Y.-D.T. using the resources of the Fermi National Accelerator Laboratory (Fermilab), a U.S. Department of Energy, Office of Science, HEP User Facility. Fermilab is managed by Fermi Research Alliance, LLC (FRA), acting under Contract No. DE-AC02-07CH11359. Part of this work was performed by Y.-D.T. at the Aspen Center for Physics, which is supported by the National Science Foundation grant PHY-1607611.

\bibliography{kinder}

%merlin.mbs apsrev4-1.bst 2010-07-25 4.21a (PWD, AO, DPC) hacked
%Control: key (0)
%Control: author (0) dotless jnrlst
%Control: editor formatted (1) identically to author
%Control: production of article title (0) allowed
%Control: page (1) range
%Control: year (0) verbatim
%Control: production of eprint (0) enabled
\begin{thebibliography}{68}%
\makeatletter
\providecommand \@ifxundefined [1]{%
 \@ifx{#1\undefined}
}%
\providecommand \@ifnum [1]{%
 \ifnum #1\expandafter \@firstoftwo
 \else \expandafter \@secondoftwo
 \fi
}%
\providecommand \@ifx [1]{%
 \ifx #1\expandafter \@firstoftwo
 \else \expandafter \@secondoftwo
 \fi
}%
\providecommand \natexlab [1]{#1}%
\providecommand \enquote  [1]{``#1''}%
\providecommand \bibnamefont  [1]{#1}%
\providecommand \bibfnamefont [1]{#1}%
\providecommand \citenamefont [1]{#1}%
\providecommand \href@noop [0]{\@secondoftwo}%
\providecommand \href [0]{\begingroup \@sanitize@url \@href}%
\providecommand \@href[1]{\@@startlink{#1}\@@href}%
\providecommand \@@href[1]{\endgroup#1\@@endlink}%
\providecommand \@sanitize@url [0]{\catcode `\\12\catcode `\$12\catcode
  `\&12\catcode `\#12\catcode `\^12\catcode `\_12\catcode `\%12\relax}%
\providecommand \@@startlink[1]{}%
\providecommand \@@endlink[0]{}%
\providecommand \url  [0]{\begingroup\@sanitize@url \@url }%
\providecommand \@url [1]{\endgroup\@href {#1}{\urlprefix }}%
\providecommand \urlprefix  [0]{URL }%
\providecommand \Eprint [0]{\href }%
\providecommand \doibase [0]{http://dx.doi.org/}%
\providecommand \selectlanguage [0]{\@gobble}%
\providecommand \bibinfo  [0]{\@secondoftwo}%
\providecommand \bibfield  [0]{\@secondoftwo}%
\providecommand \translation [1]{[#1]}%
\providecommand \BibitemOpen [0]{}%
\providecommand \bibitemStop [0]{}%
\providecommand \bibitemNoStop [0]{.\EOS\space}%
\providecommand \EOS [0]{\spacefactor3000\relax}%
\providecommand \BibitemShut  [1]{\csname bibitem#1\endcsname}%
\let\auto@bib@innerbib\@empty
%</preamble>
\bibitem [{\citenamefont {Battaglieri}\ \emph {et~al.}(2017)\citenamefont
  {Battaglieri} \emph {et~al.}}]{Battaglieri:2017aum}%
  \BibitemOpen
  \bibfield  {author} {\bibinfo {author} {\bibfnamefont {Marco}\ \bibnamefont
  {Battaglieri}} \emph {et~al.},\ }\bibfield  {title} {\enquote {\bibinfo
  {title} {{US Cosmic Visions: New Ideas in Dark Matter 2017: Community
  Report}},}\ }\href@noop {} {\  (\bibinfo {year} {2017})},\ \Eprint
  {http://arxiv.org/abs/1707.04591} {arXiv:1707.04591 [hep-ph]} \BibitemShut
  {NoStop}%
%%CITATION = ARXIV:1707.04591;%%
\bibitem [{\citenamefont {Akerib}\ \emph {et~al.}(2017)\citenamefont {Akerib}
  \emph {et~al.}}]{Akerib:2016vxi}%
  \BibitemOpen
  \bibfield  {author} {\bibinfo {author} {\bibfnamefont {D.S.}\ \bibnamefont
  {Akerib}} \emph {et~al.} (\bibinfo {collaboration} {LUX}),\ }\bibfield
  {title} {\enquote {\bibinfo {title} {{Results from a search for dark matter
  in the complete LUX exposure}},}\ }\href {\doibase
  10.1103/PhysRevLett.118.021303} {\bibfield  {journal} {\bibinfo  {journal}
  {Phys. Rev. Lett.}\ }\textbf {\bibinfo {volume} {118}},\ \bibinfo {pages}
  {021303} (\bibinfo {year} {2017})},\ \Eprint
  {http://arxiv.org/abs/1608.07648} {arXiv:1608.07648 [astro-ph.CO]}
  \BibitemShut {NoStop}%
\bibitem [{\citenamefont {Aprile}\ \emph {et~al.}(2018)\citenamefont {Aprile}
  \emph {et~al.}}]{Aprile:2018dbl}%
  \BibitemOpen
  \bibfield  {author} {\bibinfo {author} {\bibfnamefont {E.}~\bibnamefont
  {Aprile}} \emph {et~al.} (\bibinfo {collaboration} {XENON}),\ }\bibfield
  {title} {\enquote {\bibinfo {title} {{Dark Matter Search Results from a One
  Ton-Year Exposure of XENON1T}},}\ }\href {\doibase
  10.1103/PhysRevLett.121.111302} {\bibfield  {journal} {\bibinfo  {journal}
  {Phys. Rev. Lett.}\ }\textbf {\bibinfo {volume} {121}},\ \bibinfo {pages}
  {111302} (\bibinfo {year} {2018})},\ \Eprint
  {http://arxiv.org/abs/1805.12562} {arXiv:1805.12562 [astro-ph.CO]}
  \BibitemShut {NoStop}%
\bibitem [{\citenamefont {Agnese}\ \emph {et~al.}(2018)\citenamefont {Agnese}
  \emph {et~al.}}]{Agnese:2017njq}%
  \BibitemOpen
  \bibfield  {author} {\bibinfo {author} {\bibfnamefont {R.}~\bibnamefont
  {Agnese}} \emph {et~al.} (\bibinfo {collaboration} {SuperCDMS}),\ }\bibfield
  {title} {\enquote {\bibinfo {title} {{Results from the Super Cryogenic Dark
  Matter Search Experiment at Soudan}},}\ }\href {\doibase
  10.1103/PhysRevLett.120.061802} {\bibfield  {journal} {\bibinfo  {journal}
  {Phys. Rev. Lett.}\ }\textbf {\bibinfo {volume} {120}},\ \bibinfo {pages}
  {061802} (\bibinfo {year} {2018})},\ \Eprint
  {http://arxiv.org/abs/1708.08869} {arXiv:1708.08869 [hep-ex]} \BibitemShut
  {NoStop}%
\bibitem [{\citenamefont {Wang}\ \emph {et~al.}(2020)\citenamefont {Wang} \emph
  {et~al.}}]{Wang:2020coa}%
  \BibitemOpen
  \bibfield  {author} {\bibinfo {author} {\bibfnamefont {Qiuhong}\ \bibnamefont
  {Wang}} \emph {et~al.} (\bibinfo {collaboration} {PandaX-II}),\ }\bibfield
  {title} {\enquote {\bibinfo {title} {{Results of Dark Matter Search using the
  Full PandaX-II Exposure}},}\ }\href@noop {} {\  (\bibinfo {year} {2020})},\
  \Eprint {http://arxiv.org/abs/2007.15469} {arXiv:2007.15469 [astro-ph.CO]}
  \BibitemShut {NoStop}%
\bibitem [{\citenamefont {Bergsma}\ \emph {et~al.}(1986)\citenamefont {Bergsma}
  \emph {et~al.}}]{Bergsma:1985is}%
  \BibitemOpen
  \bibfield  {author} {\bibinfo {author} {\bibfnamefont {F.}~\bibnamefont
  {Bergsma}} \emph {et~al.} (\bibinfo {collaboration} {CHARM}),\ }\bibfield
  {title} {\enquote {\bibinfo {title} {{A search for decays of heavy neutrinos
  in the mass range 0.5 GeV to 2.8 GeV}},}\ }\href {\doibase
  10.1016/0370-2693(86)91601-1} {\bibfield  {journal} {\bibinfo  {journal}
  {Phys. Lett.}\ }\textbf {\bibinfo {volume} {B166}},\ \bibinfo {pages} {473}
  (\bibinfo {year} {1986})}\BibitemShut {NoStop}%
%%CITATION = PHLTA,B166,473;%%
\bibitem [{\citenamefont {Bergsma}\ \emph {et~al.}(1985)\citenamefont {Bergsma}
  \emph {et~al.}}]{Bergsma:1985qz}%
  \BibitemOpen
  \bibfield  {author} {\bibinfo {author} {\bibfnamefont {F.}~\bibnamefont
  {Bergsma}} \emph {et~al.} (\bibinfo {collaboration} {CHARM}),\ }\bibfield
  {title} {\enquote {\bibinfo {title} {{Search for Axion Like Particle
  Production in 400-{GeV} Proton - Copper Interactions}},}\ }\href {\doibase
  10.1016/0370-2693(85)90400-9} {\bibfield  {journal} {\bibinfo  {journal}
  {Phys. Lett.}\ }\textbf {\bibinfo {volume} {157B}},\ \bibinfo {pages}
  {458--462} (\bibinfo {year} {1985})}\BibitemShut {NoStop}%
%%CITATION = PHLTA,157B,458;%%
\bibitem [{\citenamefont {Konaka}\ \emph {et~al.}(1986)\citenamefont {Konaka}
  \emph {et~al.}}]{Konaka:1986cb}%
  \BibitemOpen
  \bibfield  {author} {\bibinfo {author} {\bibfnamefont {A.}~\bibnamefont
  {Konaka}} \emph {et~al.},\ }\bibfield  {title} {\enquote {\bibinfo {title}
  {{Search for neutral particles in electron-beam-dump experiment}},}\
  }\bibfield  {booktitle} {\emph {\bibinfo {booktitle} {{Proceedings, 23RD
  International Conference on High Energy Physics, JULY 16-23, 1986, Berkeley,
  CA}}},\ }\href {\doibase 10.1103/PhysRevLett.57.659} {\bibfield  {journal}
  {\bibinfo  {journal} {Phys. Rev. Lett.}\ }\textbf {\bibinfo {volume} {57}},\
  \bibinfo {pages} {659} (\bibinfo {year} {1986})}\BibitemShut {NoStop}%
%%CITATION = PRLTA,57,659;%%
\bibitem [{\citenamefont {Bjorken}\ \emph {et~al.}(1988)\citenamefont
  {Bjorken}, \citenamefont {Ecklund}, \citenamefont {Nelson}, \citenamefont
  {Abashian}, \citenamefont {Church}, \citenamefont {Lu}, \citenamefont {Mo},
  \citenamefont {Nunamaker},\ and\ \citenamefont {Rassmann}}]{Bjorken:1988as}%
  \BibitemOpen
  \bibfield  {author} {\bibinfo {author} {\bibfnamefont {J.~D.}\ \bibnamefont
  {Bjorken}}, \bibinfo {author} {\bibfnamefont {S.}~\bibnamefont {Ecklund}},
  \bibinfo {author} {\bibfnamefont {W.~R.}\ \bibnamefont {Nelson}}, \bibinfo
  {author} {\bibfnamefont {A.}~\bibnamefont {Abashian}}, \bibinfo {author}
  {\bibfnamefont {C.}~\bibnamefont {Church}}, \bibinfo {author} {\bibfnamefont
  {B.}~\bibnamefont {Lu}}, \bibinfo {author} {\bibfnamefont {L.~W.}\
  \bibnamefont {Mo}}, \bibinfo {author} {\bibfnamefont {T.~A.}\ \bibnamefont
  {Nunamaker}}, \ and\ \bibinfo {author} {\bibfnamefont {P.}~\bibnamefont
  {Rassmann}},\ }\bibfield  {title} {\enquote {\bibinfo {title} {{Search for
  neutral metastable penetrating particles produced in the SLAC beam dump}},}\
  }\href {\doibase 10.1103/PhysRevD.38.3375} {\bibfield  {journal} {\bibinfo
  {journal} {Phys. Rev.}\ }\textbf {\bibinfo {volume} {D38}},\ \bibinfo {pages}
  {3375} (\bibinfo {year} {1988})}\BibitemShut {NoStop}%
%%CITATION = PHRVA,D38,3375;%%
\bibitem [{\citenamefont {Davier}\ and\ \citenamefont
  {Nguyen~Ngoc}(1989)}]{Davier:1989wz}%
  \BibitemOpen
  \bibfield  {author} {\bibinfo {author} {\bibfnamefont {M.}~\bibnamefont
  {Davier}}\ and\ \bibinfo {author} {\bibfnamefont {H.}~\bibnamefont
  {Nguyen~Ngoc}},\ }\bibfield  {title} {\enquote {\bibinfo {title} {{An
  unambiguous search for a light Higgs boson}},}\ }\href {\doibase
  10.1016/0370-2693(89)90174-3} {\bibfield  {journal} {\bibinfo  {journal}
  {Phys. Lett.}\ }\textbf {\bibinfo {volume} {B229}},\ \bibinfo {pages} {150}
  (\bibinfo {year} {1989})}\BibitemShut {NoStop}%
%%CITATION = PHLTA,B229,150;%%
\bibitem [{\citenamefont {Bl{\"u}mlein}\ \emph {et~al.}(1991)\citenamefont
  {Bl{\"u}mlein} \emph {et~al.}}]{Blumlein:1990ay}%
  \BibitemOpen
  \bibfield  {author} {\bibinfo {author} {\bibfnamefont {J.}~\bibnamefont
  {Bl{\"u}mlein}} \emph {et~al.},\ }\bibfield  {title} {\enquote {\bibinfo
  {title} {{Limits on neutral light scalar and pseudoscalar particles in a
  proton beam dump experiment}},}\ }\href {\doibase 10.1007/BF01548556}
  {\bibfield  {journal} {\bibinfo  {journal} {Z. Phys.}\ }\textbf {\bibinfo
  {volume} {C51}},\ \bibinfo {pages} {341--350} (\bibinfo {year}
  {1991})}\BibitemShut {NoStop}%
%%CITATION = ZEPYA,C51,341;%%
\bibitem [{\citenamefont {Bl{\"u}mlein}\ \emph {et~al.}(1992)\citenamefont
  {Bl{\"u}mlein} \emph {et~al.}}]{Blumlein:1991xh}%
  \BibitemOpen
  \bibfield  {author} {\bibinfo {author} {\bibfnamefont {J.}~\bibnamefont
  {Bl{\"u}mlein}} \emph {et~al.},\ }\bibfield  {title} {\enquote {\bibinfo
  {title} {{Limits on the mass of light (pseudo)scalar particles from
  Bethe-Heitler $e^+ e^-$ and $\mu^+\mu^-$ pair production in a proton-iron
  beam dump experiment}},}\ }\href {\doibase 10.1142/S0217751X9200171X}
  {\bibfield  {journal} {\bibinfo  {journal} {Int. J. Mod. Phys.}\ }\textbf
  {\bibinfo {volume} {A7}},\ \bibinfo {pages} {3835--3850} (\bibinfo {year}
  {1992})}\BibitemShut {NoStop}%
%%CITATION = IMPAE,A7,3835;%%
\bibitem [{\citenamefont {Banerjee}\ \emph {et~al.}(2018)\citenamefont
  {Banerjee} \emph {et~al.}}]{Banerjee:2018vgk}%
  \BibitemOpen
  \bibfield  {author} {\bibinfo {author} {\bibfnamefont {D.}~\bibnamefont
  {Banerjee}} \emph {et~al.} (\bibinfo {collaboration} {NA64}),\ }\bibfield
  {title} {\enquote {\bibinfo {title} {{Search for a Hypothetical 16.7 MeV
  Gauge Boson and Dark Photons in the NA64 Experiment at CERN}},}\ }\href
  {\doibase 10.1103/PhysRevLett.120.231802} {\bibfield  {journal} {\bibinfo
  {journal} {Phys. Rev. Lett.}\ }\textbf {\bibinfo {volume} {120}},\ \bibinfo
  {pages} {231802} (\bibinfo {year} {2018})},\ \Eprint
  {http://arxiv.org/abs/1803.07748} {arXiv:1803.07748 [hep-ex]} \BibitemShut
  {NoStop}%
\bibitem [{\citenamefont {Batley}\ \emph {et~al.}(2015)\citenamefont {Batley}
  \emph {et~al.}}]{Batley:2015lha}%
  \BibitemOpen
  \bibfield  {author} {\bibinfo {author} {\bibfnamefont {J.R.}\ \bibnamefont
  {Batley}} \emph {et~al.} (\bibinfo {collaboration} {NA48/2}),\ }\bibfield
  {title} {\enquote {\bibinfo {title} {{Search for the dark photon in $\pi^0$
  decays}},}\ }\href {\doibase 10.1016/j.physletb.2015.04.068} {\bibfield
  {journal} {\bibinfo  {journal} {Phys. Lett. B}\ }\textbf {\bibinfo {volume}
  {746}},\ \bibinfo {pages} {178--185} (\bibinfo {year} {2015})},\ \Eprint
  {http://arxiv.org/abs/1504.00607} {arXiv:1504.00607 [hep-ex]} \BibitemShut
  {NoStop}%
\bibitem [{\citenamefont {Tsai}\ \emph {et~al.}(2019)\citenamefont {Tsai},
  \citenamefont {deNiverville},\ and\ \citenamefont {Liu}}]{Tsai:2019mtm}%
  \BibitemOpen
  \bibfield  {author} {\bibinfo {author} {\bibfnamefont {Yu-Dai}\ \bibnamefont
  {Tsai}}, \bibinfo {author} {\bibfnamefont {Patrick}\ \bibnamefont
  {deNiverville}}, \ and\ \bibinfo {author} {\bibfnamefont {Ming~Xiong}\
  \bibnamefont {Liu}},\ }\bibfield  {title} {\enquote {\bibinfo {title} {{The
  High-Energy Frontier of the Intensity Frontier: Closing the Dark Photon,
  Inelastic Dark Matter, and Muon g-2 Windows}},}\ }\href@noop {} {\  (\bibinfo
  {year} {2019})},\ \Eprint {http://arxiv.org/abs/1908.07525} {arXiv:1908.07525
  [hep-ph]} \BibitemShut {NoStop}%
\bibitem [{\citenamefont {Hochberg}\ \emph {et~al.}(2014)\citenamefont
  {Hochberg}, \citenamefont {Kuflik}, \citenamefont {Volansky},\ and\
  \citenamefont {Wacker}}]{Hochberg_2014}%
  \BibitemOpen
  \bibfield  {author} {\bibinfo {author} {\bibfnamefont {Yonit}\ \bibnamefont
  {Hochberg}}, \bibinfo {author} {\bibfnamefont {Eric}\ \bibnamefont {Kuflik}},
  \bibinfo {author} {\bibfnamefont {Tomer}\ \bibnamefont {Volansky}}, \ and\
  \bibinfo {author} {\bibfnamefont {Jay~G.}\ \bibnamefont {Wacker}},\
  }\bibfield  {title} {\enquote {\bibinfo {title} {Mechanism for thermal relic
  dark matter of strongly interacting massive particles},}\ }\href {\doibase
  10.1103/physrevlett.113.171301} {\bibfield  {journal} {\bibinfo  {journal}
  {Physical Review Letters}\ }\textbf {\bibinfo {volume} {113}} (\bibinfo
  {year} {2014}),\ 10.1103/physrevlett.113.171301}\BibitemShut {NoStop}%
\bibitem [{\citenamefont {Kuflik}\ \emph {et~al.}(2016)\citenamefont {Kuflik},
  \citenamefont {Perelstein}, \citenamefont {Lorier},\ and\ \citenamefont
  {Tsai}}]{Kuflik:2015isi}%
  \BibitemOpen
  \bibfield  {author} {\bibinfo {author} {\bibfnamefont {Eric}\ \bibnamefont
  {Kuflik}}, \bibinfo {author} {\bibfnamefont {Maxim}\ \bibnamefont
  {Perelstein}}, \bibinfo {author} {\bibfnamefont {Nicolas Rey-Le}\
  \bibnamefont {Lorier}}, \ and\ \bibinfo {author} {\bibfnamefont {Yu-Dai}\
  \bibnamefont {Tsai}},\ }\bibfield  {title} {\enquote {\bibinfo {title}
  {{Elastically Decoupling Dark Matter}},}\ }\href {\doibase
  10.1103/PhysRevLett.116.221302} {\bibfield  {journal} {\bibinfo  {journal}
  {Phys. Rev. Lett.}\ }\textbf {\bibinfo {volume} {116}},\ \bibinfo {pages}
  {221302} (\bibinfo {year} {2016})},\ \Eprint
  {http://arxiv.org/abs/1512.04545} {arXiv:1512.04545 [hep-ph]} \BibitemShut
  {NoStop}%
\bibitem [{\citenamefont {Kuflik}\ \emph {et~al.}(2017)\citenamefont {Kuflik},
  \citenamefont {Perelstein}, \citenamefont {Lorier},\ and\ \citenamefont
  {Tsai}}]{Kuflik:2017iqs}%
  \BibitemOpen
  \bibfield  {author} {\bibinfo {author} {\bibfnamefont {Eric}\ \bibnamefont
  {Kuflik}}, \bibinfo {author} {\bibfnamefont {Maxim}\ \bibnamefont
  {Perelstein}}, \bibinfo {author} {\bibfnamefont {Nicolas Rey-Le}\
  \bibnamefont {Lorier}}, \ and\ \bibinfo {author} {\bibfnamefont {Yu-Dai}\
  \bibnamefont {Tsai}},\ }\bibfield  {title} {\enquote {\bibinfo {title}
  {{Phenomenology of ELDER Dark Matter}},}\ }\href {\doibase
  10.1007/JHEP08(2017)078} {\bibfield  {journal} {\bibinfo  {journal} {JHEP}\
  }\textbf {\bibinfo {volume} {08}},\ \bibinfo {pages} {078} (\bibinfo {year}
  {2017})},\ \Eprint {http://arxiv.org/abs/1706.05381} {arXiv:1706.05381
  [hep-ph]} \BibitemShut {NoStop}%
\bibitem [{\citenamefont {{Carlson}}\ \emph {et~al.}(1992)\citenamefont
  {{Carlson}}, \citenamefont {{Machacek}},\ and\ \citenamefont
  {{Hall}}}]{1992ApJ...398...43C}%
  \BibitemOpen
  \bibfield  {author} {\bibinfo {author} {\bibfnamefont {Eric~D.}\ \bibnamefont
  {{Carlson}}}, \bibinfo {author} {\bibfnamefont {Marie~E.}\ \bibnamefont
  {{Machacek}}}, \ and\ \bibinfo {author} {\bibfnamefont {Lawrence~J.}\
  \bibnamefont {{Hall}}},\ }\bibfield  {title} {\enquote {\bibinfo {title}
  {{Self-interacting Dark Matter}},}\ }\href {\doibase 10.1086/171833}
  {\bibfield  {journal} {\bibinfo  {journal} {\apj}\ }\textbf {\bibinfo
  {volume} {398}},\ \bibinfo {pages} {43} (\bibinfo {year} {1992})}\BibitemShut
  {NoStop}%
\bibitem [{\citenamefont {Griest}\ and\ \citenamefont
  {Seckel}(1991)}]{PhysRevD.43.3191}%
  \BibitemOpen
  \bibfield  {author} {\bibinfo {author} {\bibfnamefont {Kim}\ \bibnamefont
  {Griest}}\ and\ \bibinfo {author} {\bibfnamefont {David}\ \bibnamefont
  {Seckel}},\ }\bibfield  {title} {\enquote {\bibinfo {title} {Three exceptions
  in the calculation of relic abundances},}\ }\href {\doibase
  10.1103/PhysRevD.43.3191} {\bibfield  {journal} {\bibinfo  {journal} {Phys.
  Rev. D}\ }\textbf {\bibinfo {volume} {43}},\ \bibinfo {pages} {3191--3203}
  (\bibinfo {year} {1991})}\BibitemShut {NoStop}%
\bibitem [{\citenamefont {D’Agnolo}\ and\ \citenamefont
  {Ruderman}(2015)}]{D_Agnolo_2015}%
  \BibitemOpen
  \bibfield  {author} {\bibinfo {author} {\bibfnamefont {Raffaele~Tito}\
  \bibnamefont {D’Agnolo}}\ and\ \bibinfo {author} {\bibfnamefont
  {Joshua~T.}\ \bibnamefont {Ruderman}},\ }\bibfield  {title} {\enquote
  {\bibinfo {title} {Light dark matter from forbidden channels},}\ }\href
  {\doibase 10.1103/physrevlett.115.061301} {\bibfield  {journal} {\bibinfo
  {journal} {Physical Review Letters}\ }\textbf {\bibinfo {volume} {115}}
  (\bibinfo {year} {2015}),\ 10.1103/physrevlett.115.061301}\BibitemShut
  {NoStop}%
\bibitem [{\citenamefont {Cline}\ \emph
  {et~al.}(2017{\natexlab{a}})\citenamefont {Cline}, \citenamefont {Liu},
  \citenamefont {Slatyer},\ and\ \citenamefont {Xue}}]{Cline_2017}%
  \BibitemOpen
  \bibfield  {author} {\bibinfo {author} {\bibfnamefont {James~M.}\
  \bibnamefont {Cline}}, \bibinfo {author} {\bibfnamefont {Hongwan}\
  \bibnamefont {Liu}}, \bibinfo {author} {\bibfnamefont {Tracy~R.}\
  \bibnamefont {Slatyer}}, \ and\ \bibinfo {author} {\bibfnamefont {Wei}\
  \bibnamefont {Xue}},\ }\bibfield  {title} {\enquote {\bibinfo {title}
  {Enabling forbidden dark matter},}\ }\href {\doibase
  10.1103/physrevd.96.083521} {\bibfield  {journal} {\bibinfo  {journal}
  {Physical Review D}\ }\textbf {\bibinfo {volume} {96}} (\bibinfo {year}
  {2017}{\natexlab{a}}),\ 10.1103/physrevd.96.083521}\BibitemShut {NoStop}%
\bibitem [{\citenamefont {Aghanim}\ \emph {et~al.}(2020)\citenamefont {Aghanim}
  \emph {et~al.}}]{Aghanim:2018eyx}%
  \BibitemOpen
  \bibfield  {author} {\bibinfo {author} {\bibfnamefont {N.}~\bibnamefont
  {Aghanim}} \emph {et~al.} (\bibinfo {collaboration} {Planck}),\ }\bibfield
  {title} {\enquote {\bibinfo {title} {{Planck 2018 results. VI. Cosmological
  parameters}},}\ }\href {\doibase 10.1051/0004-6361/201833910} {\bibfield
  {journal} {\bibinfo  {journal} {Astron. Astrophys.}\ }\textbf {\bibinfo
  {volume} {641}},\ \bibinfo {pages} {A6} (\bibinfo {year} {2020})},\ \Eprint
  {http://arxiv.org/abs/1807.06209} {arXiv:1807.06209 [astro-ph.CO]}
  \BibitemShut {NoStop}%
\bibitem [{\citenamefont {Gherghetta}\ \emph {et~al.}(2019)\citenamefont
  {Gherghetta}, \citenamefont {Kersten}, \citenamefont {Olive},\ and\
  \citenamefont {Pospelov}}]{Gherghetta:2019coi}%
  \BibitemOpen
  \bibfield  {author} {\bibinfo {author} {\bibfnamefont {Tony}\ \bibnamefont
  {Gherghetta}}, \bibinfo {author} {\bibfnamefont {J\"orn}\ \bibnamefont
  {Kersten}}, \bibinfo {author} {\bibfnamefont {Keith}\ \bibnamefont {Olive}},
  \ and\ \bibinfo {author} {\bibfnamefont {Maxim}\ \bibnamefont {Pospelov}},\
  }\bibfield  {title} {\enquote {\bibinfo {title} {{Evaluating the price of
  tiny kinetic mixing}},}\ }\href {\doibase 10.1103/PhysRevD.100.095001}
  {\bibfield  {journal} {\bibinfo  {journal} {Phys. Rev. D}\ }\textbf {\bibinfo
  {volume} {100}},\ \bibinfo {pages} {095001} (\bibinfo {year} {2019})},\
  \Eprint {http://arxiv.org/abs/1909.00696} {arXiv:1909.00696 [hep-ph]}
  \BibitemShut {NoStop}%
\bibitem [{\citenamefont {Cirelli}\ \emph {et~al.}(2017)\citenamefont
  {Cirelli}, \citenamefont {Panci}, \citenamefont {Petraki}, \citenamefont
  {Sala},\ and\ \citenamefont {Taoso}}]{Cirelli_2017}%
  \BibitemOpen
  \bibfield  {author} {\bibinfo {author} {\bibfnamefont {Marco}\ \bibnamefont
  {Cirelli}}, \bibinfo {author} {\bibfnamefont {Paolo}\ \bibnamefont {Panci}},
  \bibinfo {author} {\bibfnamefont {Kalliopi}\ \bibnamefont {Petraki}},
  \bibinfo {author} {\bibfnamefont {Filippo}\ \bibnamefont {Sala}}, \ and\
  \bibinfo {author} {\bibfnamefont {Marco}\ \bibnamefont {Taoso}},\ }\bibfield
  {title} {\enquote {\bibinfo {title} {Dark matter's secret liaisons:
  phenomenology of a dark u(1) sector with bound states},}\ }\href {\doibase
  10.1088/1475-7516/2017/05/036} {\bibfield  {journal} {\bibinfo  {journal}
  {Journal of Cosmology and Astroparticle Physics}\ }\textbf {\bibinfo {volume}
  {2017}},\ \bibinfo {pages} {036--036} (\bibinfo {year} {2017})}\BibitemShut
  {NoStop}%
\bibitem [{\citenamefont {Edsjo}\ and\ \citenamefont
  {Gondolo}(1997)}]{Edsjo:1997bg}%
  \BibitemOpen
  \bibfield  {author} {\bibinfo {author} {\bibfnamefont {Joakim}\ \bibnamefont
  {Edsjo}}\ and\ \bibinfo {author} {\bibfnamefont {Paolo}\ \bibnamefont
  {Gondolo}},\ }\bibfield  {title} {\enquote {\bibinfo {title} {{Neutralino
  relic density including coannihilations}},}\ }\href {\doibase
  10.1103/PhysRevD.56.1879} {\bibfield  {journal} {\bibinfo  {journal} {Phys.
  Rev. D}\ }\textbf {\bibinfo {volume} {56}},\ \bibinfo {pages} {1879--1894}
  (\bibinfo {year} {1997})},\ \Eprint {http://arxiv.org/abs/hep-ph/9704361}
  {arXiv:hep-ph/9704361} \BibitemShut {NoStop}%
\bibitem [{\citenamefont {D'Agnolo}\ and\ \citenamefont
  {Ruderman}(2015)}]{PhysRevLett.115.061301}%
  \BibitemOpen
  \bibfield  {author} {\bibinfo {author} {\bibfnamefont {Raffaele~Tito}\
  \bibnamefont {D'Agnolo}}\ and\ \bibinfo {author} {\bibfnamefont {Joshua~T.}\
  \bibnamefont {Ruderman}},\ }\bibfield  {title} {\enquote {\bibinfo {title}
  {Light dark matter from forbidden channels},}\ }\href {\doibase
  10.1103/PhysRevLett.115.061301} {\bibfield  {journal} {\bibinfo  {journal}
  {Phys. Rev. Lett.}\ }\textbf {\bibinfo {volume} {115}},\ \bibinfo {pages}
  {061301} (\bibinfo {year} {2015})}\BibitemShut {NoStop}%
\bibitem [{\citenamefont {Cline}\ \emph
  {et~al.}(2017{\natexlab{b}})\citenamefont {Cline}, \citenamefont {Liu},
  \citenamefont {Slatyer},\ and\ \citenamefont {Xue}}]{Cline:2017tka}%
  \BibitemOpen
  \bibfield  {author} {\bibinfo {author} {\bibfnamefont {James~M.}\
  \bibnamefont {Cline}}, \bibinfo {author} {\bibfnamefont {Hongwan}\
  \bibnamefont {Liu}}, \bibinfo {author} {\bibfnamefont {Tracy}\ \bibnamefont
  {Slatyer}}, \ and\ \bibinfo {author} {\bibfnamefont {Wei}\ \bibnamefont
  {Xue}},\ }\bibfield  {title} {\enquote {\bibinfo {title} {{Enabling Forbidden
  Dark Matter}},}\ }\href {\doibase 10.1103/PhysRevD.96.083521} {\bibfield
  {journal} {\bibinfo  {journal} {Phys. Rev.}\ }\textbf {\bibinfo {volume}
  {D96}},\ \bibinfo {pages} {083521} (\bibinfo {year} {2017}{\natexlab{b}})},\
  \Eprint {http://arxiv.org/abs/1702.07716} {arXiv:1702.07716 [hep-ph]}
  \BibitemShut {NoStop}%
%%CITATION = ARXIV:1702.07716;%%
\bibitem [{\citenamefont {Carlson}\ \emph {et~al.}(1992)\citenamefont
  {Carlson}, \citenamefont {Machacek},\ and\ \citenamefont
  {Hall}}]{Carlson:1992fn}%
  \BibitemOpen
  \bibfield  {author} {\bibinfo {author} {\bibfnamefont {Eric~D.}\ \bibnamefont
  {Carlson}}, \bibinfo {author} {\bibfnamefont {Marie~E.}\ \bibnamefont
  {Machacek}}, \ and\ \bibinfo {author} {\bibfnamefont {Lawrence~J.}\
  \bibnamefont {Hall}},\ }\bibfield  {title} {\enquote {\bibinfo {title}
  {{Self-interacting dark matter}},}\ }\href {\doibase 10.1086/171833}
  {\bibfield  {journal} {\bibinfo  {journal} {Astrophys. J.}\ }\textbf
  {\bibinfo {volume} {398}},\ \bibinfo {pages} {43--52} (\bibinfo {year}
  {1992})}\BibitemShut {NoStop}%
\bibitem [{\citenamefont {Pappadopulo}\ \emph {et~al.}(2016)\citenamefont
  {Pappadopulo}, \citenamefont {Ruderman},\ and\ \citenamefont
  {Trevisan}}]{Pappadopulo:2016pkp}%
  \BibitemOpen
  \bibfield  {author} {\bibinfo {author} {\bibfnamefont {Duccio}\ \bibnamefont
  {Pappadopulo}}, \bibinfo {author} {\bibfnamefont {Joshua~T.}\ \bibnamefont
  {Ruderman}}, \ and\ \bibinfo {author} {\bibfnamefont {Gabriele}\ \bibnamefont
  {Trevisan}},\ }\bibfield  {title} {\enquote {\bibinfo {title} {{Dark matter
  freeze-out in a nonrelativistic sector}},}\ }\href {\doibase
  10.1103/PhysRevD.94.035005} {\bibfield  {journal} {\bibinfo  {journal} {Phys.
  Rev. D}\ }\textbf {\bibinfo {volume} {94}},\ \bibinfo {pages} {035005}
  (\bibinfo {year} {2016})},\ \Eprint {http://arxiv.org/abs/1602.04219}
  {arXiv:1602.04219 [hep-ph]} \BibitemShut {NoStop}%
\bibitem [{\citenamefont {Farina}\ \emph {et~al.}(2016)\citenamefont {Farina},
  \citenamefont {Pappadopulo}, \citenamefont {Ruderman},\ and\ \citenamefont
  {Trevisan}}]{Farina:2016llk}%
  \BibitemOpen
  \bibfield  {author} {\bibinfo {author} {\bibfnamefont {Marco}\ \bibnamefont
  {Farina}}, \bibinfo {author} {\bibfnamefont {Duccio}\ \bibnamefont
  {Pappadopulo}}, \bibinfo {author} {\bibfnamefont {Joshua~T.}\ \bibnamefont
  {Ruderman}}, \ and\ \bibinfo {author} {\bibfnamefont {Gabriele}\ \bibnamefont
  {Trevisan}},\ }\bibfield  {title} {\enquote {\bibinfo {title} {{Phases of
  Cannibal Dark Matter}},}\ }\href {\doibase 10.1007/JHEP12(2016)039}
  {\bibfield  {journal} {\bibinfo  {journal} {JHEP}\ }\textbf {\bibinfo
  {volume} {12}},\ \bibinfo {pages} {039} (\bibinfo {year} {2016})},\ \Eprint
  {http://arxiv.org/abs/1607.03108} {arXiv:1607.03108 [hep-ph]} \BibitemShut
  {NoStop}%
\bibitem [{\citenamefont {Evans}\ \emph {et~al.}(2018)\citenamefont {Evans},
  \citenamefont {Gori},\ and\ \citenamefont {Shelton}}]{Evans:2017kti}%
  \BibitemOpen
  \bibfield  {author} {\bibinfo {author} {\bibfnamefont {Jared~A.}\
  \bibnamefont {Evans}}, \bibinfo {author} {\bibfnamefont {Stefania}\
  \bibnamefont {Gori}}, \ and\ \bibinfo {author} {\bibfnamefont {Jessie}\
  \bibnamefont {Shelton}},\ }\bibfield  {title} {\enquote {\bibinfo {title}
  {{Looking for the WIMP Next Door}},}\ }\href {\doibase
  10.1007/JHEP02(2018)100} {\bibfield  {journal} {\bibinfo  {journal} {JHEP}\
  }\textbf {\bibinfo {volume} {02}},\ \bibinfo {pages} {100} (\bibinfo {year}
  {2018})},\ \Eprint {http://arxiv.org/abs/1712.03974} {arXiv:1712.03974
  [hep-ph]} \BibitemShut {NoStop}%
\bibitem [{\citenamefont {Aprile}\ \emph
  {et~al.}(2019{\natexlab{a}})\citenamefont {Aprile} \emph
  {et~al.}}]{Aprile:2019xxb}%
  \BibitemOpen
  \bibfield  {author} {\bibinfo {author} {\bibfnamefont {E.}~\bibnamefont
  {Aprile}} \emph {et~al.} (\bibinfo {collaboration} {XENON}),\ }\bibfield
  {title} {\enquote {\bibinfo {title} {{Light Dark Matter Search with
  Ionization Signals in XENON1T}},}\ }\href {\doibase
  10.1103/PhysRevLett.123.251801} {\bibfield  {journal} {\bibinfo  {journal}
  {Phys. Rev. Lett.}\ }\textbf {\bibinfo {volume} {123}},\ \bibinfo {pages}
  {251801} (\bibinfo {year} {2019}{\natexlab{a}})},\ \Eprint
  {http://arxiv.org/abs/1907.11485} {arXiv:1907.11485 [hep-ex]} \BibitemShut
  {NoStop}%
\bibitem [{\citenamefont {Barak}\ \emph {et~al.}(2020)\citenamefont {Barak}
  \emph {et~al.}}]{Barak:2020fql}%
  \BibitemOpen
  \bibfield  {author} {\bibinfo {author} {\bibfnamefont {Liron}\ \bibnamefont
  {Barak}} \emph {et~al.} (\bibinfo {collaboration} {SENSEI}),\ }\bibfield
  {title} {\enquote {\bibinfo {title} {{SENSEI: Direct-Detection Results on
  sub-GeV Dark Matter from a New Skipper-CCD}},}\ }\href@noop {} {\  (\bibinfo
  {year} {2020})},\ \Eprint {http://arxiv.org/abs/2004.11378} {arXiv:2004.11378
  [astro-ph.CO]} \BibitemShut {NoStop}%
\bibitem [{\citenamefont {Agnes}\ \emph {et~al.}(2018)\citenamefont {Agnes}
  \emph {et~al.}}]{Agnes:2018oej}%
  \BibitemOpen
  \bibfield  {author} {\bibinfo {author} {\bibfnamefont {P.}~\bibnamefont
  {Agnes}} \emph {et~al.} (\bibinfo {collaboration} {DarkSide}),\ }\bibfield
  {title} {\enquote {\bibinfo {title} {{Constraints on Sub-GeV
  Dark-Matter\textendash{}Electron Scattering from the DarkSide-50
  Experiment}},}\ }\href {\doibase 10.1103/PhysRevLett.121.111303} {\bibfield
  {journal} {\bibinfo  {journal} {Phys. Rev. Lett.}\ }\textbf {\bibinfo
  {volume} {121}},\ \bibinfo {pages} {111303} (\bibinfo {year} {2018})},\
  \Eprint {http://arxiv.org/abs/1802.06998} {arXiv:1802.06998 [astro-ph.CO]}
  \BibitemShut {NoStop}%
\bibitem [{\citenamefont {Aprile}\ \emph
  {et~al.}(2019{\natexlab{b}})\citenamefont {Aprile} \emph
  {et~al.}}]{Aprile:2019jmx}%
  \BibitemOpen
  \bibfield  {author} {\bibinfo {author} {\bibfnamefont {E.}~\bibnamefont
  {Aprile}} \emph {et~al.} (\bibinfo {collaboration} {XENON}),\ }\bibfield
  {title} {\enquote {\bibinfo {title} {{Search for Light Dark Matter
  Interactions Enhanced by the Migdal Effect or Bremsstrahlung in XENON1T}},}\
  }\href {\doibase 10.1103/PhysRevLett.123.241803} {\bibfield  {journal}
  {\bibinfo  {journal} {Phys. Rev. Lett.}\ }\textbf {\bibinfo {volume} {123}},\
  \bibinfo {pages} {241803} (\bibinfo {year} {2019}{\natexlab{b}})},\ \Eprint
  {http://arxiv.org/abs/1907.12771} {arXiv:1907.12771 [hep-ex]} \BibitemShut
  {NoStop}%
\bibitem [{\citenamefont {Baxter}\ \emph {et~al.}(2020)\citenamefont {Baxter},
  \citenamefont {Kahn},\ and\ \citenamefont {Krnjaic}}]{Baxter:2019pnz}%
  \BibitemOpen
  \bibfield  {author} {\bibinfo {author} {\bibfnamefont {Daniel}\ \bibnamefont
  {Baxter}}, \bibinfo {author} {\bibfnamefont {Yonatan}\ \bibnamefont {Kahn}},
  \ and\ \bibinfo {author} {\bibfnamefont {Gordan}\ \bibnamefont {Krnjaic}},\
  }\bibfield  {title} {\enquote {\bibinfo {title} {{Electron Ionization via
  Dark Matter-Electron Scattering and the Migdal Effect}},}\ }\href {\doibase
  10.1103/PhysRevD.101.076014} {\bibfield  {journal} {\bibinfo  {journal}
  {Phys. Rev. D}\ }\textbf {\bibinfo {volume} {101}},\ \bibinfo {pages}
  {076014} (\bibinfo {year} {2020})},\ \Eprint
  {http://arxiv.org/abs/1908.00012} {arXiv:1908.00012 [hep-ph]} \BibitemShut
  {NoStop}%
\bibitem [{\citenamefont {Amaral}\ \emph {et~al.}(2020)\citenamefont {Amaral}
  \emph {et~al.}}]{Amaral:2020ryn}%
  \BibitemOpen
  \bibfield  {author} {\bibinfo {author} {\bibfnamefont {D.W.}\ \bibnamefont
  {Amaral}} \emph {et~al.} (\bibinfo {collaboration} {SuperCDMS}),\ }\bibfield
  {title} {\enquote {\bibinfo {title} {{Constraints on low-mass, relic dark
  matter candidates from a surface-operated SuperCDMS single-charge sensitive
  detector}},}\ }\href@noop {} {\  (\bibinfo {year} {2020})},\ \Eprint
  {http://arxiv.org/abs/2005.14067} {arXiv:2005.14067 [hep-ex]} \BibitemShut
  {NoStop}%
\bibitem [{\citenamefont {Raffelt}\ and\ \citenamefont
  {Seckel}(1988)}]{Raffelt:1987yt}%
  \BibitemOpen
  \bibfield  {author} {\bibinfo {author} {\bibfnamefont {Georg}\ \bibnamefont
  {Raffelt}}\ and\ \bibinfo {author} {\bibfnamefont {David}\ \bibnamefont
  {Seckel}},\ }\bibfield  {title} {\enquote {\bibinfo {title} {{Bounds on
  Exotic Particle Interactions from SN 1987a}},}\ }\href {\doibase
  10.1103/PhysRevLett.60.1793} {\bibfield  {journal} {\bibinfo  {journal}
  {Phys. Rev. Lett.}\ }\textbf {\bibinfo {volume} {60}},\ \bibinfo {pages}
  {1793} (\bibinfo {year} {1988})}\BibitemShut {NoStop}%
\bibitem [{\citenamefont {Raffelt}(1996)}]{Raffelt1996}%
  \BibitemOpen
  \bibfield  {author} {\bibinfo {author} {\bibfnamefont {G.~G.}\ \bibnamefont
  {Raffelt}},\ }\href@noop {} {\emph {\bibinfo {title} {{Stars as laboratories
  for fundamental physics}}}}\ (\bibinfo  {publisher} {University of Chicago
  Press},\ \bibinfo {year} {1996})\BibitemShut {NoStop}%
%%CITATION = INSPIRE-430034;%%
\bibitem [{\citenamefont {{Burrows}}\ and\ \citenamefont
  {{Lattimer}}(1986)}]{Burrows1986}%
  \BibitemOpen
  \bibfield  {author} {\bibinfo {author} {\bibfnamefont {A.}~\bibnamefont
  {{Burrows}}}\ and\ \bibinfo {author} {\bibfnamefont {J.~M.}\ \bibnamefont
  {{Lattimer}}},\ }\bibfield  {title} {\enquote {\bibinfo {title} {{The birth
  of neutron stars}},}\ }\href {\doibase 10.1086/164405} {\bibfield  {journal}
  {\bibinfo  {journal} {\apj}\ }\textbf {\bibinfo {volume} {307}},\ \bibinfo
  {pages} {178--196} (\bibinfo {year} {1986})}\BibitemShut {NoStop}%
\bibitem [{\citenamefont {Burrows}\ and\ \citenamefont
  {Lattimer}(1987)}]{Burrows:1987zz}%
  \BibitemOpen
  \bibfield  {author} {\bibinfo {author} {\bibfnamefont {Adam}\ \bibnamefont
  {Burrows}}\ and\ \bibinfo {author} {\bibfnamefont {James~M.}\ \bibnamefont
  {Lattimer}},\ }\bibfield  {title} {\enquote {\bibinfo {title} {{Neutrinos
  from SN 1987A}},}\ }\href {\doibase 10.1086/184938} {\bibfield  {journal}
  {\bibinfo  {journal} {Astrophys. J. Lett.}\ }\textbf {\bibinfo {volume}
  {318}},\ \bibinfo {pages} {L63--L68} (\bibinfo {year} {1987})}\BibitemShut
  {NoStop}%
\bibitem [{\citenamefont {Bar}\ \emph {et~al.}(2020)\citenamefont {Bar},
  \citenamefont {Blum},\ and\ \citenamefont {D'Amico}}]{Bar:2019ifz}%
  \BibitemOpen
  \bibfield  {author} {\bibinfo {author} {\bibfnamefont {Nitsan}\ \bibnamefont
  {Bar}}, \bibinfo {author} {\bibfnamefont {Kfir}\ \bibnamefont {Blum}}, \ and\
  \bibinfo {author} {\bibfnamefont {Guido}\ \bibnamefont {D'Amico}},\
  }\bibfield  {title} {\enquote {\bibinfo {title} {{Is there a supernova bound
  on axions?}}}\ }\href {\doibase 10.1103/PhysRevD.101.123025} {\bibfield
  {journal} {\bibinfo  {journal} {Phys. Rev. D}\ }\textbf {\bibinfo {volume}
  {101}},\ \bibinfo {pages} {123025} (\bibinfo {year} {2020})},\ \Eprint
  {http://arxiv.org/abs/1907.05020} {arXiv:1907.05020 [hep-ph]} \BibitemShut
  {NoStop}%
\bibitem [{\citenamefont {Chang}\ \emph {et~al.}(2018)\citenamefont {Chang},
  \citenamefont {Essig},\ and\ \citenamefont {McDermott}}]{Chang:2018rso}%
  \BibitemOpen
  \bibfield  {author} {\bibinfo {author} {\bibfnamefont {Jae~Hyeok}\
  \bibnamefont {Chang}}, \bibinfo {author} {\bibfnamefont {Rouven}\
  \bibnamefont {Essig}}, \ and\ \bibinfo {author} {\bibfnamefont {Samuel~D.}\
  \bibnamefont {McDermott}},\ }\bibfield  {title} {\enquote {\bibinfo {title}
  {{Supernova 1987A Constraints on Sub-GeV Dark Sectors, Millicharged
  Particles, the QCD Axion, and an Axion-like Particle}},}\ }\href {\doibase
  10.1007/JHEP09(2018)051} {\bibfield  {journal} {\bibinfo  {journal} {JHEP}\
  }\textbf {\bibinfo {volume} {09}},\ \bibinfo {pages} {051} (\bibinfo {year}
  {2018})},\ \Eprint {http://arxiv.org/abs/1803.00993} {arXiv:1803.00993
  [hep-ph]} \BibitemShut {NoStop}%
%%CITATION = ARXIV:1803.00993;%%
\bibitem [{\citenamefont {Bondarenko}\ \emph {et~al.}(2020)\citenamefont
  {Bondarenko}, \citenamefont {Sokolenko}, \citenamefont {Boyarsky},
  \citenamefont {Robertson}, \citenamefont {Harvey},\ and\ \citenamefont
  {Revaz}}]{Bondarenko:2020mpf}%
  \BibitemOpen
  \bibfield  {author} {\bibinfo {author} {\bibfnamefont {Kyrylo}\ \bibnamefont
  {Bondarenko}}, \bibinfo {author} {\bibfnamefont {Anastasia}\ \bibnamefont
  {Sokolenko}}, \bibinfo {author} {\bibfnamefont {Alexey}\ \bibnamefont
  {Boyarsky}}, \bibinfo {author} {\bibfnamefont {Andrew}\ \bibnamefont
  {Robertson}}, \bibinfo {author} {\bibfnamefont {David}\ \bibnamefont
  {Harvey}}, \ and\ \bibinfo {author} {\bibfnamefont {Yves}\ \bibnamefont
  {Revaz}},\ }\bibfield  {title} {\enquote {\bibinfo {title} {{From dwarf
  galaxies to galaxy clusters: Self-Interacting Dark Matter over 7 orders of
  magnitude in halo mass}},}\ }\href@noop {} {\  (\bibinfo {year} {2020})},\
  \Eprint {http://arxiv.org/abs/2006.06623} {arXiv:2006.06623 [astro-ph.CO]}
  \BibitemShut {NoStop}%
\bibitem [{\citenamefont {Pollack}(2012)}]{Pollack:2012hbv}%
  \BibitemOpen
  \bibfield  {author} {\bibinfo {author} {\bibfnamefont {Jason}\ \bibnamefont
  {Pollack}},\ }\emph {\bibinfo {title} {{Supermassive Black Holes from
  Gravothermal Collapse of Fractional Self-Interacting Dark Matter halos}}},\
  \href@noop {} {\bibinfo {type} {Bachelor thesis}},\ \bibinfo  {school}
  {Princeton U.} (\bibinfo {year} {2012})\BibitemShut {NoStop}%
\bibitem [{\citenamefont {Slatyer}(2016)}]{Slatyer:2015jla}%
  \BibitemOpen
  \bibfield  {author} {\bibinfo {author} {\bibfnamefont {Tracy~R.}\
  \bibnamefont {Slatyer}},\ }\bibfield  {title} {\enquote {\bibinfo {title}
  {{Indirect dark matter signatures in the cosmic dark ages. I. Generalizing
  the bound on s-wave dark matter annihilation from Planck results}},}\ }\href
  {\doibase 10.1103/PhysRevD.93.023527} {\bibfield  {journal} {\bibinfo
  {journal} {Phys. Rev. D}\ }\textbf {\bibinfo {volume} {93}},\ \bibinfo
  {pages} {023527} (\bibinfo {year} {2016})},\ \Eprint
  {http://arxiv.org/abs/1506.03811} {arXiv:1506.03811 [hep-ph]} \BibitemShut
  {NoStop}%
\bibitem [{\citenamefont {Izotov}\ \emph {et~al.}(2014)\citenamefont {Izotov},
  \citenamefont {Thuan},\ and\ \citenamefont {Guseva}}]{Izotov:2014fga}%
  \BibitemOpen
  \bibfield  {author} {\bibinfo {author} {\bibfnamefont {Y.I.}\ \bibnamefont
  {Izotov}}, \bibinfo {author} {\bibfnamefont {T.X.}\ \bibnamefont {Thuan}}, \
  and\ \bibinfo {author} {\bibfnamefont {N.G.}\ \bibnamefont {Guseva}},\
  }\bibfield  {title} {\enquote {\bibinfo {title} {{A new determination of the
  primordial He abundance using the He i $\lambda$10830 \r{A} emission line:
  cosmological implications}},}\ }\href {\doibase 10.1093/mnras/stu1771}
  {\bibfield  {journal} {\bibinfo  {journal} {Mon. Not. Roy. Astron. Soc.}\
  }\textbf {\bibinfo {volume} {445}},\ \bibinfo {pages} {778--793} (\bibinfo
  {year} {2014})},\ \Eprint {http://arxiv.org/abs/1408.6953} {arXiv:1408.6953
  [astro-ph.CO]} \BibitemShut {NoStop}%
\bibitem [{\citenamefont {Aver}\ \emph {et~al.}(2015)\citenamefont {Aver},
  \citenamefont {Olive},\ and\ \citenamefont {Skillman}}]{Aver:2015iza}%
  \BibitemOpen
  \bibfield  {author} {\bibinfo {author} {\bibfnamefont {Erik}\ \bibnamefont
  {Aver}}, \bibinfo {author} {\bibfnamefont {Keith~A.}\ \bibnamefont {Olive}},
  \ and\ \bibinfo {author} {\bibfnamefont {Evan~D.}\ \bibnamefont {Skillman}},\
  }\bibfield  {title} {\enquote {\bibinfo {title} {{The effects of He I
  \ensuremath{\lambda}10830 on helium abundance determinations}},}\ }\href
  {\doibase 10.1088/1475-7516/2015/07/011} {\bibfield  {journal} {\bibinfo
  {journal} {JCAP}\ }\textbf {\bibinfo {volume} {07}},\ \bibinfo {pages} {011}
  (\bibinfo {year} {2015})},\ \Eprint {http://arxiv.org/abs/1503.08146}
  {arXiv:1503.08146 [astro-ph.CO]} \BibitemShut {NoStop}%
\bibitem [{\citenamefont {Cooke}\ \emph {et~al.}(2018)\citenamefont {Cooke},
  \citenamefont {Pettini},\ and\ \citenamefont {Steidel}}]{Cooke:2017cwo}%
  \BibitemOpen
  \bibfield  {author} {\bibinfo {author} {\bibfnamefont {Ryan~J.}\ \bibnamefont
  {Cooke}}, \bibinfo {author} {\bibfnamefont {Max}\ \bibnamefont {Pettini}}, \
  and\ \bibinfo {author} {\bibfnamefont {Charles~C.}\ \bibnamefont {Steidel}},\
  }\bibfield  {title} {\enquote {\bibinfo {title} {{One Percent Determination
  of the Primordial Deuterium Abundance}},}\ }\href {\doibase
  10.3847/1538-4357/aaab53} {\bibfield  {journal} {\bibinfo  {journal}
  {Astrophys. J.}\ }\textbf {\bibinfo {volume} {855}},\ \bibinfo {pages} {102}
  (\bibinfo {year} {2018})},\ \Eprint {http://arxiv.org/abs/1710.11129}
  {arXiv:1710.11129 [astro-ph.CO]} \BibitemShut {NoStop}%
\bibitem [{\citenamefont {Zavarygin}\ \emph {et~al.}(2018)\citenamefont
  {Zavarygin}, \citenamefont {Webb}, \citenamefont {Riemer-S\o{}rensen},\ and\
  \citenamefont {Dumont}}]{Zavarygin:2018ara}%
  \BibitemOpen
  \bibfield  {author} {\bibinfo {author} {\bibfnamefont {E.O.}\ \bibnamefont
  {Zavarygin}}, \bibinfo {author} {\bibfnamefont {J.K.}\ \bibnamefont {Webb}},
  \bibinfo {author} {\bibfnamefont {S.}~\bibnamefont {Riemer-S\o{}rensen}}, \
  and\ \bibinfo {author} {\bibfnamefont {V.}~\bibnamefont {Dumont}},\
  }\bibfield  {title} {\enquote {\bibinfo {title} {{Primordial deuterium
  abundance at $z_{abs}$ = 2:504 towards Q1009+2956}},}\ }\href {\doibase
  10.1088/1742-6596/1038/1/012012} {\bibfield  {journal} {\bibinfo  {journal}
  {J. Phys. Conf. Ser.}\ }\textbf {\bibinfo {volume} {1038}},\ \bibinfo {pages}
  {012012} (\bibinfo {year} {2018})},\ \Eprint
  {http://arxiv.org/abs/1801.04704} {arXiv:1801.04704 [astro-ph.CO]}
  \BibitemShut {NoStop}%
\bibitem [{\citenamefont {Valerdi}\ \emph {et~al.}(2019)\citenamefont
  {Valerdi}, \citenamefont {Peimbert}, \citenamefont {Peimbert},\ and\
  \citenamefont {Sixtos}}]{Valerdi:2019beb}%
  \BibitemOpen
  \bibfield  {author} {\bibinfo {author} {\bibfnamefont {Mabel}\ \bibnamefont
  {Valerdi}}, \bibinfo {author} {\bibfnamefont {Antonio}\ \bibnamefont
  {Peimbert}}, \bibinfo {author} {\bibfnamefont {Manuel}\ \bibnamefont
  {Peimbert}}, \ and\ \bibinfo {author} {\bibfnamefont {Andr\'es}\ \bibnamefont
  {Sixtos}},\ }\bibfield  {title} {\enquote {\bibinfo {title} {{Determination
  of the Primordial Helium Abundance Based on NGC 346, an H ii Region of the
  Small Magellanic Cloud}},}\ }\href {\doibase 10.3847/1538-4357/ab14e4}
  {\bibfield  {journal} {\bibinfo  {journal} {Astrophys. J.}\ }\textbf
  {\bibinfo {volume} {876}},\ \bibinfo {pages} {98} (\bibinfo {year} {2019})},\
  \Eprint {http://arxiv.org/abs/1904.01594} {arXiv:1904.01594 [astro-ph.GA]}
  \BibitemShut {NoStop}%
\bibitem [{\citenamefont {Boehm}\ \emph {et~al.}(2013)\citenamefont {Boehm},
  \citenamefont {Dolan},\ and\ \citenamefont {McCabe}}]{Boehm:2013jpa}%
  \BibitemOpen
  \bibfield  {author} {\bibinfo {author} {\bibfnamefont {C\'eline}\
  \bibnamefont {Boehm}}, \bibinfo {author} {\bibfnamefont {Matthew~J.}\
  \bibnamefont {Dolan}}, \ and\ \bibinfo {author} {\bibfnamefont {Christopher}\
  \bibnamefont {McCabe}},\ }\bibfield  {title} {\enquote {\bibinfo {title} {{A
  Lower Bound on the Mass of Cold Thermal Dark Matter from Planck}},}\ }\href
  {\doibase 10.1088/1475-7516/2013/08/041} {\bibfield  {journal} {\bibinfo
  {journal} {JCAP}\ }\textbf {\bibinfo {volume} {08}},\ \bibinfo {pages} {041}
  (\bibinfo {year} {2013})},\ \Eprint {http://arxiv.org/abs/1303.6270}
  {arXiv:1303.6270 [hep-ph]} \BibitemShut {NoStop}%
\bibitem [{\citenamefont {Sabti}\ \emph {et~al.}(2020)\citenamefont {Sabti},
  \citenamefont {Alvey}, \citenamefont {Escudero}, \citenamefont {Fairbairn},\
  and\ \citenamefont {Blas}}]{Sabti:2019mhn}%
  \BibitemOpen
  \bibfield  {author} {\bibinfo {author} {\bibfnamefont {Nashwan}\ \bibnamefont
  {Sabti}}, \bibinfo {author} {\bibfnamefont {James}\ \bibnamefont {Alvey}},
  \bibinfo {author} {\bibfnamefont {Miguel}\ \bibnamefont {Escudero}}, \bibinfo
  {author} {\bibfnamefont {Malcolm}\ \bibnamefont {Fairbairn}}, \ and\ \bibinfo
  {author} {\bibfnamefont {Diego}\ \bibnamefont {Blas}},\ }\bibfield  {title}
  {\enquote {\bibinfo {title} {{Refined Bounds on MeV-scale Thermal Dark
  Sectors from BBN and the CMB}},}\ }\href {\doibase
  10.1088/1475-7516/2020/01/004} {\bibfield  {journal} {\bibinfo  {journal}
  {JCAP}\ }\textbf {\bibinfo {volume} {01}},\ \bibinfo {pages} {004} (\bibinfo
  {year} {2020})},\ \Eprint {http://arxiv.org/abs/1910.01649} {arXiv:1910.01649
  [hep-ph]} \BibitemShut {NoStop}%
\bibitem [{\citenamefont {Nollett}\ and\ \citenamefont
  {Steigman}(2014)}]{Nollett:2013pwa}%
  \BibitemOpen
  \bibfield  {author} {\bibinfo {author} {\bibfnamefont {Kenneth~M.}\
  \bibnamefont {Nollett}}\ and\ \bibinfo {author} {\bibfnamefont {Gary}\
  \bibnamefont {Steigman}},\ }\bibfield  {title} {\enquote {\bibinfo {title}
  {{BBN And The CMB Constrain Light, Electromagnetically Coupled WIMPs}},}\
  }\href {\doibase 10.1103/PhysRevD.89.083508} {\bibfield  {journal} {\bibinfo
  {journal} {Phys. Rev. D}\ }\textbf {\bibinfo {volume} {89}},\ \bibinfo
  {pages} {083508} (\bibinfo {year} {2014})},\ \Eprint
  {http://arxiv.org/abs/1312.5725} {arXiv:1312.5725 [astro-ph.CO]} \BibitemShut
  {NoStop}%
\bibitem [{\citenamefont {Abazajian}(2017)}]{Abazajian:2017tcc}%
  \BibitemOpen
  \bibfield  {author} {\bibinfo {author} {\bibfnamefont {Kevork~N.}\
  \bibnamefont {Abazajian}},\ }\bibfield  {title} {\enquote {\bibinfo {title}
  {{Sterile neutrinos in cosmology}},}\ }\href {\doibase
  10.1016/j.physrep.2017.10.003} {\bibfield  {journal} {\bibinfo  {journal}
  {Phys. Rept.}\ }\textbf {\bibinfo {volume} {711-712}},\ \bibinfo {pages}
  {1--28} (\bibinfo {year} {2017})},\ \Eprint {http://arxiv.org/abs/1705.01837}
  {arXiv:1705.01837 [hep-ph]} \BibitemShut {NoStop}%
%%CITATION = ARXIV:1705.01837;%%
\bibitem [{\citenamefont {Drewes}\ \emph {et~al.}(2017)\citenamefont {Drewes}
  \emph {et~al.}}]{Adhikari:2016bei}%
  \BibitemOpen
  \bibfield  {author} {\bibinfo {author} {\bibfnamefont {M.}~\bibnamefont
  {Drewes}} \emph {et~al.},\ }\bibfield  {title} {\enquote {\bibinfo {title}
  {{A White Paper on keV Sterile Neutrino Dark Matter}},}\ }\href {\doibase
  10.1088/1475-7516/2017/01/025} {\bibfield  {journal} {\bibinfo  {journal}
  {JCAP}\ }\textbf {\bibinfo {volume} {01}},\ \bibinfo {pages} {025} (\bibinfo
  {year} {2017})},\ \Eprint {http://arxiv.org/abs/1602.04816} {arXiv:1602.04816
  [hep-ph]} \BibitemShut {NoStop}%
\bibitem [{\citenamefont {Ir\v{s}i\v{c}}\ \emph {et~al.}(2017)\citenamefont
  {Ir\v{s}i\v{c}} \emph {et~al.}}]{Irsic:2017ixq}%
  \BibitemOpen
  \bibfield  {author} {\bibinfo {author} {\bibfnamefont {Vid}\ \bibnamefont
  {Ir\v{s}i\v{c}}} \emph {et~al.},\ }\bibfield  {title} {\enquote {\bibinfo
  {title} {{New Constraints on the free-streaming of warm dark matter from
  intermediate and small scale Lyman-$\alpha$ forest data}},}\ }\href {\doibase
  10.1103/PhysRevD.96.023522} {\bibfield  {journal} {\bibinfo  {journal} {Phys.
  Rev. D}\ }\textbf {\bibinfo {volume} {96}},\ \bibinfo {pages} {023522}
  (\bibinfo {year} {2017})},\ \Eprint {http://arxiv.org/abs/1702.01764}
  {arXiv:1702.01764 [astro-ph.CO]} \BibitemShut {NoStop}%
\bibitem [{\citenamefont {Palanque-Delabrouille}\ \emph
  {et~al.}(2020)\citenamefont {Palanque-Delabrouille}, \citenamefont {Y\`eche},
  \citenamefont {Sch\"oneberg}, \citenamefont {Lesgourgues}, \citenamefont
  {Walther}, \citenamefont {Chabanier},\ and\ \citenamefont
  {Armengaud}}]{Palanque-Delabrouille:2019iyz}%
  \BibitemOpen
  \bibfield  {author} {\bibinfo {author} {\bibfnamefont {Nathalie}\
  \bibnamefont {Palanque-Delabrouille}}, \bibinfo {author} {\bibfnamefont
  {Christophe}\ \bibnamefont {Y\`eche}}, \bibinfo {author} {\bibfnamefont
  {Nils}\ \bibnamefont {Sch\"oneberg}}, \bibinfo {author} {\bibfnamefont
  {Julien}\ \bibnamefont {Lesgourgues}}, \bibinfo {author} {\bibfnamefont
  {Michael}\ \bibnamefont {Walther}}, \bibinfo {author} {\bibfnamefont
  {Sol\`ene}\ \bibnamefont {Chabanier}}, \ and\ \bibinfo {author}
  {\bibfnamefont {Eric}\ \bibnamefont {Armengaud}},\ }\bibfield  {title}
  {\enquote {\bibinfo {title} {{Hints, neutrino bounds and WDM constraints from
  SDSS DR14 Lyman-$\alpha$ and Planck full-survey data}},}\ }\href {\doibase
  10.1088/1475-7516/2020/04/038} {\bibfield  {journal} {\bibinfo  {journal}
  {JCAP}\ }\textbf {\bibinfo {volume} {04}},\ \bibinfo {pages} {038} (\bibinfo
  {year} {2020})},\ \Eprint {http://arxiv.org/abs/1911.09073} {arXiv:1911.09073
  [astro-ph.CO]} \BibitemShut {NoStop}%
\bibitem [{\citenamefont {Schneider}\ \emph {et~al.}(2013)\citenamefont
  {Schneider}, \citenamefont {Smith},\ and\ \citenamefont
  {Reed}}]{Schneider:2013ria}%
  \BibitemOpen
  \bibfield  {author} {\bibinfo {author} {\bibfnamefont {Aurel}\ \bibnamefont
  {Schneider}}, \bibinfo {author} {\bibfnamefont {Robert~E.}\ \bibnamefont
  {Smith}}, \ and\ \bibinfo {author} {\bibfnamefont {Darren}\ \bibnamefont
  {Reed}},\ }\bibfield  {title} {\enquote {\bibinfo {title} {{Halo Mass
  Function and the Free Streaming Scale}},}\ }\href {\doibase
  10.1093/mnras/stt829} {\bibfield  {journal} {\bibinfo  {journal} {Mon. Not.
  Roy. Astron. Soc.}\ }\textbf {\bibinfo {volume} {433}},\ \bibinfo {pages}
  {1573} (\bibinfo {year} {2013})},\ \Eprint {http://arxiv.org/abs/1303.0839}
  {arXiv:1303.0839 [astro-ph.CO]} \BibitemShut {NoStop}%
\bibitem [{\citenamefont {Finkbeiner}\ and\ \citenamefont
  {Weiner}(2007)}]{Finkbeiner:2007kk}%
  \BibitemOpen
  \bibfield  {author} {\bibinfo {author} {\bibfnamefont {Douglas~P.}\
  \bibnamefont {Finkbeiner}}\ and\ \bibinfo {author} {\bibfnamefont {Neal}\
  \bibnamefont {Weiner}},\ }\bibfield  {title} {\enquote {\bibinfo {title}
  {{Exciting Dark Matter and the INTEGRAL/SPI 511 keV signal}},}\ }\href
  {\doibase 10.1103/PhysRevD.76.083519} {\bibfield  {journal} {\bibinfo
  {journal} {Phys. Rev. D}\ }\textbf {\bibinfo {volume} {76}},\ \bibinfo
  {pages} {083519} (\bibinfo {year} {2007})},\ \Eprint
  {http://arxiv.org/abs/astro-ph/0702587} {arXiv:astro-ph/0702587} \BibitemShut
  {NoStop}%
\bibitem [{\citenamefont {Finkbeiner}\ \emph {et~al.}(2011)\citenamefont
  {Finkbeiner}, \citenamefont {Goodenough}, \citenamefont {Slatyer},
  \citenamefont {Vogelsberger},\ and\ \citenamefont
  {Weiner}}]{Finkbeiner:2010sm}%
  \BibitemOpen
  \bibfield  {author} {\bibinfo {author} {\bibfnamefont {Douglas~P.}\
  \bibnamefont {Finkbeiner}}, \bibinfo {author} {\bibfnamefont {Lisa}\
  \bibnamefont {Goodenough}}, \bibinfo {author} {\bibfnamefont {Tracy~R.}\
  \bibnamefont {Slatyer}}, \bibinfo {author} {\bibfnamefont {Mark}\
  \bibnamefont {Vogelsberger}}, \ and\ \bibinfo {author} {\bibfnamefont {Neal}\
  \bibnamefont {Weiner}},\ }\bibfield  {title} {\enquote {\bibinfo {title}
  {{Consistent Scenarios for Cosmic-Ray Excesses from Sommerfeld-Enhanced Dark
  Matter Annihilation}},}\ }\href {\doibase 10.1088/1475-7516/2011/05/002}
  {\bibfield  {journal} {\bibinfo  {journal} {JCAP}\ }\textbf {\bibinfo
  {volume} {05}},\ \bibinfo {pages} {002} (\bibinfo {year} {2011})},\ \Eprint
  {http://arxiv.org/abs/1011.3082} {arXiv:1011.3082 [hep-ph]} \BibitemShut
  {NoStop}%
\bibitem [{\citenamefont {Elor}\ \emph {et~al.}(2018)\citenamefont {Elor},
  \citenamefont {Liu}, \citenamefont {Slatyer},\ and\ \citenamefont
  {Soreq}}]{Elor:2018xku}%
  \BibitemOpen
  \bibfield  {author} {\bibinfo {author} {\bibfnamefont {Gilly}\ \bibnamefont
  {Elor}}, \bibinfo {author} {\bibfnamefont {Hongwan}\ \bibnamefont {Liu}},
  \bibinfo {author} {\bibfnamefont {Tracy~R.}\ \bibnamefont {Slatyer}}, \ and\
  \bibinfo {author} {\bibfnamefont {Yotam}\ \bibnamefont {Soreq}},\ }\bibfield
  {title} {\enquote {\bibinfo {title} {{Complementarity for Dark Sector Bound
  States}},}\ }\href {\doibase 10.1103/PhysRevD.98.036015} {\bibfield
  {journal} {\bibinfo  {journal} {Phys. Rev. D}\ }\textbf {\bibinfo {volume}
  {98}},\ \bibinfo {pages} {036015} (\bibinfo {year} {2018})},\ \Eprint
  {http://arxiv.org/abs/1801.07723} {arXiv:1801.07723 [hep-ph]} \BibitemShut
  {NoStop}%
\bibitem [{\citenamefont {Rizzo}(2020)}]{Rizzo:2020jsm}%
  \BibitemOpen
  \bibfield  {author} {\bibinfo {author} {\bibfnamefont {Thomas~G.}\
  \bibnamefont {Rizzo}},\ }\bibfield  {title} {\enquote {\bibinfo {title}
  {{Dark Initial State Radiation and the Kinetic Mixing Portal}},}\ }\href@noop
  {} {\  (\bibinfo {year} {2020})},\ \Eprint {http://arxiv.org/abs/2006.08502}
  {arXiv:2006.08502 [hep-ph]} \BibitemShut {NoStop}%
\bibitem [{\citenamefont {Fitzpatrick}\ \emph {et~al.}()\citenamefont
  {Fitzpatrick}, \citenamefont {Liu}, \citenamefont {Slatyer},\ and\
  \citenamefont {Tsai}}]{KINDER_short_paper}%
  \BibitemOpen
  \bibfield  {author} {\bibinfo {author} {\bibfnamefont {Patrick}\ \bibnamefont
  {Fitzpatrick}}, \bibinfo {author} {\bibfnamefont {Hongwan}\ \bibnamefont
  {Liu}}, \bibinfo {author} {\bibfnamefont {Tracy~R.}\ \bibnamefont {Slatyer}},
  \ and\ \bibinfo {author} {\bibfnamefont {Yu-Dai}\ \bibnamefont {Tsai}},\
  }\href@noop {} {\bibinfo  {journal} {to appear}\ }\BibitemShut {NoStop}%
\bibitem [{\citenamefont {Bringmann}\ and\ \citenamefont
  {Hofmann}(2007)}]{Bringmann:2006mu}%
  \BibitemOpen
\bibfield  {journal} {  }\bibfield  {author} {\bibinfo {author} {\bibfnamefont
  {Torsten}\ \bibnamefont {Bringmann}}\ and\ \bibinfo {author} {\bibfnamefont
  {Stefan}\ \bibnamefont {Hofmann}},\ }\bibfield  {title} {\enquote {\bibinfo
  {title} {{Thermal decoupling of WIMPs from first principles}},}\ }\href
  {\doibase 10.1088/1475-7516/2007/04/016} {\bibfield  {journal} {\bibinfo
  {journal} {JCAP}\ }\textbf {\bibinfo {volume} {04}},\ \bibinfo {pages} {016}
  (\bibinfo {year} {2007})},\ \bibinfo {note} {[Erratum: JCAP 03, E02
  (2016)]},\ \Eprint {http://arxiv.org/abs/hep-ph/0612238}
  {arXiv:hep-ph/0612238} \BibitemShut {NoStop}%
\bibitem [{\citenamefont {Bringmann}(2009)}]{Bringmann:2009vf}%
  \BibitemOpen
  \bibfield  {author} {\bibinfo {author} {\bibfnamefont {Torsten}\ \bibnamefont
  {Bringmann}},\ }\bibfield  {title} {\enquote {\bibinfo {title} {{Particle
  Models and the Small-Scale Structure of Dark Matter}},}\ }\href {\doibase
  10.1088/1367-2630/11/10/105027} {\bibfield  {journal} {\bibinfo  {journal}
  {New J. Phys.}\ }\textbf {\bibinfo {volume} {11}},\ \bibinfo {pages} {105027}
  (\bibinfo {year} {2009})},\ \Eprint {http://arxiv.org/abs/0903.0189}
  {arXiv:0903.0189 [astro-ph.CO]} \BibitemShut {NoStop}%
\bibitem [{\citenamefont {Liu}\ \emph {et~al.}(2020)\citenamefont {Liu},
  \citenamefont {Ridgway},\ and\ \citenamefont {Slatyer}}]{Liu:2019bbm}%
  \BibitemOpen
  \bibfield  {author} {\bibinfo {author} {\bibfnamefont {Hongwan}\ \bibnamefont
  {Liu}}, \bibinfo {author} {\bibfnamefont {Gregory~W.}\ \bibnamefont
  {Ridgway}}, \ and\ \bibinfo {author} {\bibfnamefont {Tracy~R.}\ \bibnamefont
  {Slatyer}},\ }\bibfield  {title} {\enquote {\bibinfo {title} {{Code package
  for calculating modified cosmic ionization and thermal histories with dark
  matter and other exotic energy injections}},}\ }\href {\doibase
  10.1103/PhysRevD.101.023530} {\bibfield  {journal} {\bibinfo  {journal}
  {Phys. Rev. D}\ }\textbf {\bibinfo {volume} {101}},\ \bibinfo {pages}
  {023530} (\bibinfo {year} {2020})},\ \Eprint
  {http://arxiv.org/abs/1904.09296} {arXiv:1904.09296 [astro-ph.CO]}
  \BibitemShut {NoStop}%
\end{thebibliography}%
\newpage

\appendix

\section{Cross Sections and Decay Widths}
\label{sec:Cross}

The thermally averaged cross sections and decay widths are computed using the same conventions as in Ref.~\cite{Cline:2017tka}. Table~\ref{tab:xsec} gives a list of the relevant cross sections and decay widths used throughout this paper. 

\renewcommand{\arraystretch}{2} 

\setlength{\tabcolsep}{15pt}

\begin{table*}
\begin{tabular}{c c} 
\toprule
Process & Cross Section or Decay Width \\[1ex]
\hline
$\chi \overline{\chi} \chi \to A' \chi$ & $\langle \sigma v^2 \rangle = $\mbox{\fontsize{12}{2}\selectfont \( \frac{g_D^6 (r-4)(r+4)(-32r^8 + 167r^6 - 534r^4 + 668r^2 - 512)}{36 m_\chi^2 (r^2 - 4)^4 (r^2 + 2)^2} \frac{\sqrt{r^4 - 20r^2 + 64}}{96 \pi m_\chi^3}\)} \\
$A' A' \to \chi \overline{\chi}$ & $\langle \sigma v \rangle = $\mbox{\fontsize{12}{2}\selectfont \( \frac{32g_D^4(r^4 - 1)}{9r^4} \frac{\sqrt{r^2 - 1}}{8 \pi m_\chi^2 r^3}\)}\\
$\chi \overline{\chi} \to e^+e^-$ & $ \langle \sigma v \rangle = $ \mbox{\fontsize{12}{2}\selectfont \(\frac{4 e^2 \epsilon^2 g_D^2 \left(2 + m_e^2/m_\chi^2\right)}{(r^2 - 4)^2} \frac{\sqrt{1 - m_e^2/m_\chi^2}}{8 \pi m_\chi^2}\)} \\
$A' \to f \overline{f}$ & $\Gamma =$ \mbox{\fontsize{12}{2}\selectfont \( \frac{\epsilon^2 e^2}{12\pi} \left(1 + \frac{2 m_f^2}{m_{A'}^2}\right) \)}$\sqrt{m_{A'}^2 - 4 m_f^2}$ \\[2ex]
\botrule
\end{tabular}
\caption{List of cross sections and decay widths for the dark sector processes considered in this paper. All quantities are evaluated at the kinematic threshold.}
\label{tab:xsec}
\end{table*}

\section{Elastic Scattering Energy Transfer Rate}
\label{sec:Rate}

In this section, we outline the derivation of the elastic scattering energy transfer cross section $\langle \sigma v \delta E \rangle_{\chi f \to \chi f}$ that appears in the energy density Boltzmann equation, Eq.~\eqref{eq:Boltz_rho}. For consistency with existing literature, we switch notations within this section so that a subscript $\chi,\overline{\chi}$ denotes quantities for both the DM particle and antiparticle, while a single subscript $\chi$ denotes a quantity associated only with the DM particle alone. 

Following Ref.~\cite{Kuflik:2017iqs}, the Boltzmann equation for the phase space distribution $f_{\chi,\overline{\chi}} (\vec{p}_1,t)$ of DM (both $\chi$ and $\overline{\chi}$) is
\begin{alignat}{1}
    \frac{\partial f_{\chi,\overline{\chi}}}{\partial t} - H \frac{\vec{p}_1^2}{E_1} \frac{\partial f_{\chi,\overline{\chi}}}{\partial E_1} = \frac{C[f_\chi]}{E_1} \,,
    \label{eq:true_boltzmann_eqn}
\end{alignat}
where $C[f_\chi]$ is the collision term, and $E_1^2 = \vec{p}_1^2 + m_\chi^2$. Here, we will focus on the elastic scattering collision term, which includes $\chi f \to \chi f$, $\overline{\chi} f \to \overline{\chi} f$, $\chi \overline{f} \to \chi \overline{f}$ and $\overline{\chi} \overline{f} \to \overline{\chi} \overline{f}$ scatterings, where $f$ is an SM fermion, which we take to be the electron throughout this paper for simplicity. From here on, $\chi f \to \chi f$ should be taken as shorthand for all four of these processes. Explicitly, taking the indices 1 and 3 for incoming and outgoing dark sector particles, and the indices 2 and 4 for incoming and outgoing SM particles, the collision operator is
\begin{multline}
    C[f_\chi] = \frac{1}{2} \int \frac{d^3 \vec{p}_2}{(2\pi)^3 2E_2} \int \frac{d^3 \vec{p}_3}{(2\pi)^3 2E_3} \int \frac{d^3 \vec{p}_4}{(2\pi)^3 2E_4} \\
    \times (2\pi)^4 \delta^4(p_1 + p_2 - p_3 - p_4) \overline{|\mathcal{M}|^2} \\ 
    \times \big[f_3 f_4(1 \mp f_1)(1 \mp f_2) \\
    - f_1 f_2(1 \mp f_3)(1 \mp f_4)\big] \,,
\end{multline}
with $f_i$ denoting the phase space distribution of the particle indexed by $i$, We follow the conventions of Ref.~\cite{Bringmann:2006mu}, where the number density of a particle $i$ is related to its phase space distribution via
\begin{alignat}{1}
    n_i = \int \frac{d^3 \vec{p}}{(2\pi)^3} f_i(\vec{p}) \,,
\end{alignat}
with the number of degrees of freedom of particle $i$ absorbed into the definition of $f_i$. Furthermore, the matrix element squared is summed over final states but averaged over initial states, and is a sum of all four matrix elements squared for the four conjugate processes. The number of degrees of freedom for Dirac fermions $\chi$ is $g_\chi = 2$, and likewise for the SM fermion $f$; the degrees of freedom of particles and antiparticles are always counted separately. We note that all of these conventions are different from those used in Ref.~\cite{Kuflik:2017iqs}, but the final results are equivalent. 

Following Ref.~\cite{Kuflik:2017iqs} Eq.~(B1), we can multiply Eq.~\eqref{eq:true_boltzmann_eqn} by $E$ and integrate over all momenta to obtain
\begin{alignat}{1}
    \dot{\rho}_{\chi,\overline{\chi}} + 3H(\rho_{\chi,\overline{\chi}} + P_{\chi,\overline{\chi}}) &\simeq \int \frac{d^3 \vec{p}_1}{(2\pi)^3 2E_1} \frac{\vec{p}_1^2}{m_\chi} C[f_\chi]
    \label{eq:energy_boltzmann_for_elastic_scattering}
\end{alignat}
after taking the nonrelativistic approximation for $\chi$, as found in Ref.~\cite{Kuflik:2017iqs} Eq.~(B1).

Ref.~\cite{Kuflik:2017iqs} provides an expression assuming that $f$ is relativistic; however, for $m_\chi \sim \mathcal{O}(\SI{}{\mega\eV})$, electrons are nonrelativistic before the dark sector decouples from the SM. We present here a compact expression for $C[f_\chi]$ that applies for electrons in all regimes. Following Ref.~\cite{Bringmann:2009vf}, we can write the collision term as
\begin{alignat}{1}
    C[f_\chi] \simeq c(T) m_\chi^2 \left[m_\chi T \Delta_{\vec{p}_1} + \vec{p}_1 \cdot \nabla_{\vec{p}_1} + 3 \right] f_\chi \,,
    \label{eq:elastic_scattering_collision_term}
\end{alignat}
where $\Delta_{\vec{p}}$ and $\nabla_{\vec{p}}$ are the Laplacian and del operators with respect to $\vec{p}$ respectively. Again, the only approximation made is that $\chi$ is nonrelativistic. The expression for $c(T)$ is given as~\cite{Bringmann:2009vf}
\begin{multline}
    c(T) = \frac{g_f}{12(2\pi)^3 m_\chi^4 T} \\
    \times \int dp_2 \, p_2^5 E_2^{-1} g^{\pm} \left(1 \mp g^\pm \right) \overline{|\mathcal{M}|^2}_{\substack{t=0 \\ s = s_0}} \,,
    \label{eq:c(T)}
\end{multline}
where $g^\pm \equiv [\exp(E_2/T) \pm 1]^{-1}$, taking the plus sign when $f$ is a fermion,\footnote{We do not choose a sign for $g^\pm$ to be as general as possible, since the calculation follows equally easily for scattering off a boson.} and the matrix element squared is to be evaluated at $t = 0$ and $s = s_0 \equiv m_\chi^2 + 2 m_\chi E_2 + m_f^2$. Ref.~\cite{Bringmann:2006mu} found an analytic expression for $c(T)$ by making the relativistic approximation for $f$, i.e. $E_2 \simeq p_2$, and writing
\begin{alignat}{1}
    \overline{|\mathcal{M}|^2}_{\substack{t=0 \\ s = s_0}} = c_n \left(\frac{E_2}{m_\chi}\right)^n + \mathcal{O} \left[ \left(\frac{E_2}{m_\chi}\right)^{n+1}\right]
\end{alignat}
and keeping only the leading order term. For $\chi f \to \chi f$ in our model, i.e.\ with $\chi$ and $f$ both Dirac fermions mediated by $A'$, this approximation is actually exact, with $n = 2$ being the only term in the expansion, 
\begin{alignat}{1}
    c_2 = \frac{16(4\pi)^2 \alpha_\text{EM} \alpha_D \epsilon^2}{r^4}  \qquad (\chi f \to \chi f)\,.
\end{alignat}

In fact, the integral in Eq.~\eqref{eq:c(T)} can be performed analytically without making the relativistic approximation for the SM fermions; the result is
\begin{alignat}{1}
    c(T) = \frac{g_f c_n T^{4+n}}{12(2\pi)^3 m_\chi^{4+n}} R_\pm(n+3,\xi)  \,,
     \label{eq:elastic_scattering_c(T)}
\end{alignat}
where $\xi = m_f/T$. The function $R_\pm$ is defined as
\begin{multline}
    R_\pm (q, \xi) \equiv \big[(q+1) Q_\pm (q, \xi) - 2(q-1) \xi^2 Q_\pm (q-2, \xi) \\
    + (q-3) \xi^4 Q_\pm(q-4, \xi) \big] \,,
\end{multline}
with
\begin{alignat}{1}
    Q_\pm (q, \xi) \equiv \mp q! \sum_{s=0}^q \frac{\xi^s}{s!} \text{Li}_{q-s+1} \left(\mp e^{-\xi}\right) \,,
\end{alignat}
where $\text{Li}_m(z)$ is the polylogarithm~\cite{Liu:2019bbm}. We can check that in the relativistic limit for the fermion $\xi \to 0$, we get 
\begin{alignat}{1}
    R_+(q,0) &\equiv N^+_q = (q+1)! (1 - 2^{-q})\zeta(q+1) \nonumber \\
    R_-(q,0) &\equiv N^-_q= (q+1)! \zeta(q+1) \,,
\end{alignat}
where $\zeta(n)$ is the Riemann zeta function. This recovers the expression presented in Eq.~(B22) of Ref.~\cite{Bringmann:2006mu} after substituting these expressions into Eq.~\eqref{eq:elastic_scattering_c(T)}. 

In the nonrelativistic limit, $\xi \to \infty$, we obtain
\begin{alignat}{1}
    Q_\pm (q,\xi \gg 1) \simeq \Gamma(q+1, \xi) \mp 2^{-(q+1)} \Gamma(q+1, 2\xi)\,,
\end{alignat}
where $\Gamma(n,y)$ is the incomplete Gamma function, 
\begin{alignat}{1}
    \Gamma(n,y) \equiv \int_y^\infty dt \, t^{n-1} e^{-t} \,.
\end{alignat}

With the expression for $c(T)$ in Eq.~\eqref{eq:elastic_scattering_c(T)}, we can substitute Eq.~\eqref{eq:elastic_scattering_collision_term} into Eq.~\eqref{eq:energy_boltzmann_for_elastic_scattering}, and perform the momentum integral, assuming the dark matter follows a Maxwell-Boltzmann phase space distribution. The result is
\begin{multline}
    \int \frac{d^3 \vec{p}_1}{(2\pi)^3 2E_1} \frac{\vec{p}_1^2}{m_\chi} C[f_\chi] \simeq \frac{g_f c_n}{64 \pi^3} m_\chi  n_{\chi,\overline{\chi}}\\
    \times \left(\frac{T}{m_\chi}\right)^{4+n} R_{\pm}(n+3, \xi) (T - T') \,.
\end{multline}
We can now define the right-hand side of Eq.~\eqref{eq:energy_boltzmann_for_elastic_scattering} as $-n_{\chi,\overline{\chi}} n_{f,\overline{f}} \langle \sigma v \delta E \rangle_{\chi f \to \chi f}$, giving
\begin{multline}
    \langle \sigma v \delta E \rangle_{\chi f \to \chi f} \simeq \frac{g_f c_n m_\chi}{64 \pi ^3 n_{f,\overline{f}}} \left(\frac{T}{m_\chi}\right)^{4+n} \\
    \times R_{\pm}(n+3,\xi) (T' - T) \,.
    \label{eq:elastic_scattering_rate_full}
\end{multline}
To our knowledge, this result is new, and is accurate for SM fermions in both the relativistic and nonrelativistic regimes, assuming nonrelativistic DM. In the limit where SM fermions are relativistic, we find for $\chi f \to \chi f$ mediated by $A'$
\begin{alignat}{1}
    \langle \sigma v \delta E \rangle_{\chi f \to \chi f} \simeq \frac{8}{\pi} \frac{\alpha_\text{EM}\alpha_D \epsilon^2}{r^4 x^6} 6! (1 - 2^{-5}) \zeta(6) (T' - T) \,.
    \label{eq:elastic_scattering_rate_approx}
\end{alignat}

\end{document}